\documentclass[twocolumn,secnumarabic,amssymb, nobibnotes, superscriptaddress, nofootinbib,aps,prd]{revtex4-2}
\usepackage[colorlinks,citecolor=blue,linkcolor=blue,anchorcolor=blue,filecolor=blue, urlcolor=blue]{hyperref}
\usepackage{natbib}

\usepackage{amsmath}
\usepackage{adjustbox}
\usepackage{graphicx}
\usepackage{booktabs}       
\usepackage{amsfonts}       
\usepackage{nicefrac}       
\usepackage{microtype}
\usepackage{cleveref}
\usepackage{xcolor}
\usepackage{multirow}
\usepackage{makecell}
\usepackage{comment}
\usepackage{todonotes}
\usepackage{array}
\usepackage{orcidlink}
\usepackage{float}
\usepackage[normalem]{ulem}

\begin{document}

\title{Radial and Non-Radial Oscillations of Protoneutron Stars with Hyperonic Composition} 
\author{Prashant Thakur~\orcidlink{0000-0001-5930-7179}}
\email{prashantthakur1921@gmail.com}
\affiliation{Department of Physics, Yonsei University, Seoul, 03722, South Korea}

\author{Adamu Issifu \orcidlink{0000-0002-2843-835X}} 
\email{ai@academico.ufpb.br}
\affiliation{Departamento de F\'isica, Instituto Tecnol\'ogico de Aeron\'autica, DCTA, 12228-900, S\~ao Jos\'e dos Campos, SP, Brazil} 
\affiliation{Laborat\'orio de Computa\c c\~ao Cient\'ifica Avan\c cada e Modelamento (Lab-CCAM)}

\author{Ishfaq Ahmad Rather~\orcidlink{0000-0001-5930-7179}}
\email{rather@astro.uni-frankfurt.de}
\affiliation{Institut f\"{u}r Theoretische Physik, Goethe Universit\"{a}t,
Max-von-Laue-Str.~1, D-60438 Frankfurt am Main, Germany}

\author{Y. Lim~\orcidlink{0000-0001-9252-823X}} 
\email{ylim@yonsei.ac.kr}
\affiliation{Department of Physics, Yonsei University, Seoul, 03722, South Korea}

\author{Tobias Frederico \orcidlink{0000-0002-5497-5490}} 
\email{tobias@ita.br}

\affiliation{Departamento de F\'isica, Instituto Tecnol\'ogico de Aeron\'autica, DCTA, 12228-900, S\~ao Jos\'e dos Campos, SP, Brazil} 
\affiliation{Laborat\'orio de Computa\c c\~ao Cient\'ifica Avan\c cada e Modelamento (Lab-CCAM)}


\begin{abstract}
This paper explores radial and non-radial oscillations of protoneutron stars (PNSs) as they evolve from hot, neutrino-rich configurations through deleptonization to cold, catalyzed states. The equation of state (EoS) is modeled using a density-dependent relativistic mean-field framework, with stellar evolution characterized by changes in entropy and lepton fraction. Both nucleonic and hyperonic compositions are considered. Non-radial $f$- and $p_1$-mode oscillations are computed using both the Cowling approximation and the full General Relativistic framework. Trapped neutrinos initially increase the error in the Cowling approximation for $f$-modes, which decreases during deleptonization and rises again in the cold phase. In contrast, $p_1$-mode errors peak during intermediate stages due to evolving pressure and density gradients. The emergence of hyperons modestly raises oscillation frequencies in both modes. Existing universal relations for $f$-mode frequency and damping time lack model independence for PNSs, motivating a more robust relation. In particular, our proposed universal relation involving the moment of inertia and $\tilde{\eta}$ shows strong agreement across all evolutionary phases, offering a temperature-sensitive, model-independent scaling for asteroseismology. Radial oscillations of a $1.4\,M_\odot$ PNS are also studied for different EoSs. Our results show that displacement ($\xi$) and pressure perturbation ($\eta$) profiles are highly sensitive to thermal state, composition, and compactness. Hyperonic stars show higher frequencies, altered node structures, and stronger pressure perturbations due to EoS softening. Differences in frequency separation $\Delta \nu_n$ and fundamental frequency $\nu_0$ between nucleonic and hyperonic models provide clear observational diagnostics for probing the interiors of PNSs and constraining the EoS of dense matter.
\end{abstract}
\maketitle

\section{Introduction}

{The study of core-collapse supernova (CCSNe) is a crucial topic in nuclear physics, as it is one of the few phenomena where all four fundamental forces of nature interact, shaping the dynamics of stellar collapse and explosion. Gravity drives the collapse when a massive star exhausts its nuclear fuel, the electromagnetic force governs interactions between charged particles, the weak nuclear force facilitates neutrino interactions, and the strong nuclear force dictates the behavior of nuclear matter at extreme densities \cite{Janka:2012wk, Kotake:2005zn}. Aside from that, CCSNe serve as a suitable phenomenon for probing the emission of light particles with masses \( \lesssim 100 \) MeV \cite{Raffelt:1990yz, Raffelt:1987yt}. Additionally, the observation of the neutrino burst from neutron star (NS) 1987A \cite{Kamiokande-II:1987idp, Hirata:1988ad} marked a turning point in particle physics, leading to further constraints on neutrino properties. 

The implosion of the massive star leads to the formation of a PNS when the core contracts and heats up. This collapse releases a vast amount of energy, causing an explosion and resulting in the creation of a hot PNS \cite{Pons:1998mm, Prakash:1996xs, Vartanyan:2018iah, Camelio:2017nka}. Over time, the PNS undergoes processes such as neutrino emission, deleptonization, thermal radiation, and neutrino diffusion, which cause it to cool and contract, ultimately forming a cold, catalyzed NS \cite{Nakazato:2019ojk}. To study the thermodynamic evolution of PNSs, the EoS is essential, as it relates pressure, energy density, and temperature, providing insight into the composition and properties of matter within the PNS \cite{Lattimer:2015nhk, Oertel:2016bki, Issifu:2023qyi, Issifu:2024fuw, Raduta:2020fdn, Kunkel:2024otq}.

In this work, we use the EoS initially published in \cite{Issifu:2023qyi}, which has been extensively applied in studies on quark core formation in PNSs \cite{Issifu:2024fuw} and the impact of mirror dark matter on PNS evolution \cite{Issifu:2024htq}. Our primary objective is to investigate the radial and non-radial oscillations of PNSs during their evolutionary stages and to examine how the emergence of hyperons influences these oscillations. This analysis provides deeper insights into the structure, stability, and composition of PNSs as they evolve. A key advantage of our approach is that radial and non-radial oscillations offer complementary perspectives. Radial oscillations primarily probe the stability, expansion, and contraction of PNSs, but they do not produce direct observational signatures in gravitational waves. In \cite{Sun:2024vdt}, the authors analyzed the radial oscillations of PNSs and determined that the oscillation frequency is lower in neutrino-trapped matter. They also observed a rapid decline in the fundamental frequencies near the maximum mass. Further studies on radial oscillations of PNSs are fairly limited because of the complex physical processes in PNS and can be found in \cite{Gondek:1997fd, Kokkotas:2000up, Sotani:2021kvj}.

In contrast, non-radial oscillations provide crucial insights into the internal composition of PNSs, including the role of exotic matter such as hyperons. Additionally, they serve as valuable probes of gravitational wave signals, offering rich astrophysical information. Interested readers may refer to Refs.~\cite{Rodriguez:2023nay, Barman:2024zuo, Pradhan:2022vdf,Kunjipurayil:2022zah,Roy:2023gzi, Zheng:2025xlr, Kumar:2024bvd}, and the references therein, for further details on the $f$- and $p_1$-mode oscillations. However, studying non-radial oscillations presents significant mathematical and computational challenges. By investigating both radial and non-radial modes together, we obtain a more comprehensive understanding of how temperature and the presence of hyperons influence the properties and dynamics of PNSs. More importantly, in the non-radial case, we compare full general relativity (GR) calculations with the Cowling approximation for the first time in the study of PNSs' evolution with exotic baryon composition. The present work extends the analysis of Ref.~\cite{Barman:2024zuo}, which focused on non-radial oscillations in PNSs using full GR but did not include neutrino effects or hyperonic matter.

These two approaches differ in key ways: the full GR method considers both fluid perturbations, which peak near the stellar surface, and metric perturbations, which peak at the stellar center. In contrast, the Cowling approximation neglects metric perturbations, introducing small errors, particularly in massive stars, where strong fluid perturbations dominate \cite{Kunjipurayil:2022zah}. Our goal is to quantify the deviation of the Cowling approximation from the full GR approach throughout stellar evolution and to assess the role of neutrinos, temperature, and hyperons in this deviation. Furthermore, \cite{Ferrari:2002ut} demonstrated that as PNSs evolve, changes in the frequencies and damping times of non-radial oscillations significantly impact their structural properties and gravitational wave emission.

This study presents several key novelties in the asteroseismology of PNSs. This study provides a first integrated and comprehensive picture where we perform a unified and systematic analysis of both radial and non-radial oscillations within a single framework, comparing the Cowling approximation with full general relativity across the complete evolutionary path from hot, neutrino-rich PNSs to cold, catalyzed NSs. This approach self-consistently incorporates hyperonic degrees of freedom along with the effects of neutrino trapping, deleptonization, and neutrino transparency, enabling us to quantify the evolving accuracy of the Cowling approximation under controlled conditions. We show that the non-barotropic nature of the hot PNS EoS leads to the breakdown of the well-established universal relations for the $f$-mode, and we introduce a new robust relation based on effective compactness that remains valid throughout evolution. 

In addition, we identify a distinct non-linear signature in radial oscillations, arising from avoided crossings induced by hyperons, which provides a potential diagnostic of dense matter composition. By simultaneously analyzing $f$-modes, $p$-modes, radial oscillations, and universal relations, our work offers the first integrated and comprehensive picture of the dynamic seismic fingerprints of PNSs. Additionally, the Cowling approximation error is smallest for the $f$-mode in hot, moderate-lepton PNSs, particularly when hyperons are present. In contrast, for the $p$-mode, it is lowest in cooler, high-lepton configurations, reflecting the opposite influence of stellar compactness on the two modes. Although several previous studies have explored specific aspects of neutron star oscillation \cite{Thapa:2023grg, Kumar:2024jky, Sun:2024vdt, Zheng:2025xlr}, to the
best of our knowledge, no comprehensive study has yet been performed that simultaneously investigates both radial and non-radial oscillations in PNSs with nucleonic and hyperonic compositions using both Cawling and GR formalism.

The work is organized as follows: In \Cref{NS1}, we present the microphysics governing the evolution of the PNSs. \Cref{oscillaltions} briefly describes the formalism for radial and non-radial oscillation modes. In \Cref{results}, we present the EoS, mass-radius relations, and analyze the $f$-mode frequency with several subsections discussing the gravitational-wave asteroseismology, highlighting universal relations and the behavior of $p$-mode frequencies as well as radial profiles, and Gravitational-wave detectability. Finally, we summarize our findings and conclusions in \Cref{summary}.
}

\section{Microphysics}\label{NS1}

{The EoS used for hadronic interactions is based on quantum hadrodynamics (QHD) \cite{Serot:1997xg}, which describes strong interactions between hadrons mediated by massive mesons. The full Lagrangian density for the relativistic mean-field approximation model \cite{Menezes:2021jmw} is given by:
\begin{equation}\label{rmf}
     \mathcal{L}_{\rm RMF}= \mathcal{L}_{H}+ \mathcal{L}_{\rm m}+ \mathcal{L}_{\rm l},
\end{equation}
where $\mathcal{L}_H$, $\mathcal{L}_m$, and $\mathcal{L}_{\rm l}$ represent the Lagrangian densities for the baryon octet, the meson fields, and the leptons, respectively. Explicitly:  
\begin{align}
 \mathcal{L}_{H}= {}& \sum_{b\in H}  \bar \psi_b \Big[  i \gamma^\mu\partial_\mu - \gamma^0  \big(g_{\omega b} \omega_0  +  g_{\phi b} \phi_0+ g_{\rho b} I_{3b} \rho_{03}  \big)
 - \Big( m_b- g_{\sigma b} \sigma_0 \Big)  \Big] \psi_b,\\
 \mathcal{L}_{\rm m}&= - \frac{1}{2} m_\sigma^2 \sigma_0^2  +\frac{1}{2} m_\omega^2 \omega_0^2 \label{lagrangian} +\frac{1}{2} m_\phi^2 \phi_0^2 +\frac{1}{2} m_\rho^2 \rho_{03}^2 ,\\
    \mathcal{L}_{\rm l} &= \sum_l\Bar{\psi}_l\left(i\gamma^\mu\partial_\mu-m_l\right)\psi_l \label{l1}.
\end{align}
Here, $\psi_b$ is the baryonic Dirac field with the subscript $b$ representing the individual baryons present, $i=\sigma,\; \omega,\; \phi$ and $\rho$ are the meson fields with $\omega$ and $\phi$ being the vector-isoscalar meson (where $\phi$ has a hidden strangeness), $\sigma$ is a scalar meson and $\rho$ is a vector-isovector mesons while the subscript `0' represents the mean-field approximation presentation. The corresponding masses of the mesons are given by; $m_i$ with their values provided in \Cref{T1}, $I_{3b}=\pm 1/2$ is the isospin projection and $\psi_l$ is the free lepton fields with the subscript `$l$' representing the different lepton species in the system. The baryon octet and the lepton species have a degeneracy of two, except in a situation where neutrinos are trapped in the stellar matter; at this stage, we consider electron neutrinos ($\nu_e$) with a degeneracy of one, in consistency with supernova physics \cite{Prakash:1996xs}. At the neutrino-trapped regime, we neglect the presence of muons in our analysis since their presence becomes only relevant after all the neutrinos have escaped from the stellar core \cite{Janka:2012wk}. Additionally, in the neutrino-transparent stage, when all the neutrinos have escaped from the stellar core, we consider the presence of electrons ($e$) and muons ($\mu$) in the stellar matter, while tau leptons are considered too heavy to be present \cite{Raduta:2020fdn, Pons:1998mm}. 

Moreover, we consider the presence of hyperons in the stellar core at each stage of the star's evolution, mediated by the $\phi$ meson with mass $m_\phi =1019.45$ MeV. The presence of hyperons in the stellar core is theoretically predicted as a result of energy minimization in high-density environments, where the increasing degeneracy pressure of nucleons makes hyperon formation energetically favorable. This process is governed by the Pauli exclusion principle, which dictates the behavior of fermionic particles. The emergence of hyperons introduces additional degrees of freedom, reducing the net energy of the system and, consequently, softening the EoS. This phenomenon, commonly referred to as the hyperon puzzle, has been extensively discussed in the literature \cite{Bombaci:2016xzl, Menezes:2021jmw}.


\begin{table}[t]
\begin{center}
\caption {DDME2 parameters for nucleonic matter in this work.}
\begin{tabular}{ c c c c c c c }
\hline
 meson($i$) & $m_i(\text{MeV})$ & $a_i$ & $b_i$ & $c_i$ & $d_i$ & $g_{i N} (n_0)$\\
 \hline
 \hline
 $\sigma$ & 550 & 1.3881 & 1.0943 & 1.7057 & 0.4421 & 10.5396 \\  
 $\omega$ & 783 & 1.3892 & 0.9240 & 1.4620 & 0.4775 & 13.0189  \\
 $\rho$ & 763 & 0.5647 & --- & --- & --- & 7.3672 \\
 \hline
 \hline
\end{tabular}
\label{T1}
\end{center}
\end{table}

We employ the density-dependent coupling adjusted by the DDME2 parameterization~\citep{ddme2} represented by:
{ \begin{equation}
    g_{i N} (n_B) = g_{iN} (n_0)a_i  \frac{1+b_i (\eta + d_i)^2}{1 +c_i (\eta + d_i)^2},
\end{equation}
with $i=\sigma, \omega, \phi$ and 
\begin{equation}
    g_{\rho N} (n_B) = g_{\rho N} (n_0) \exp\left[ - a_\rho \big( \eta -1 \big) \right].
\end{equation}}
Here, $\eta =n_B/n_0$, where $n_0=0.152$~fm$^{-3}$ is the baryon saturation density under this parameterization and $n_B$ is the baryon density. The model parameters ($a_i,\, b_i,\, c_i$ and $d_i$) are determined through fitting to experimental bulk nuclear matter properties around $n_0$. Other nuclear properties for this parameterization include the binding energy, compressibility modulus, symmetry energy, and its density slope, given by $B/A = -16.4$~MeV, $K_0=251.9$~MeV, $J=32.3$~MeV, and $L= 51.3$~MeV, respectively~\citep{Dutra, Lattimer:2012xj, ddme2,reed2021}. To incorporate hyperons within this framework, extensive work has been done to extend the couplings using various methods \cite{Glendenning:1991es, Weissenborn:2011kb}. These approaches aim to obtain a massive hyperonic star with a maximum mass consistent with the observed NS mass threshold of 2\,$M_\odot$. For this paper, we adopt the coupling constraints determined in \cite{Lopes:2022vjx}, which uses SU(3) and SU(6) symmetry arguments to extend the nucleon-meson couplings to include hyperons and $\Delta$-resonances in a unified framework. The advantage of this choice of coupling is that it leads to massive hyperonic stars within the 2\,$M_\odot$ threshold. The corresponding parameters for hyperons are listed in \Cref{T2}, represented as the ratio of the nucleon-meson couplings in the form $\chi_{ib}=g_{ib}/g_{iN}$.

This self-consistent treatment naturally softens the EoS by including hyperonic degrees of freedom ab initio within a unified theoretical framework. As a result, the mass–radius relation is softer than in a purely nucleonic model, even at densities below the hyperon production threshold \cite{Oertel:2016bki, Glendenning1985, Fortin:2016hny, Balberg:1997yw, Fortin:2017dsj}. This self-consistent framework fundamentally differs from phenomenological methods that add hyperons `by hand' to a pre-existing nucleonic EoS, for instance by treating them as a non-interacting gas or without modifying the underlying nucleon interactions \cite{Bednarek2012, Djapo:2008au}.

\subsection{Conditions Relevant for PNS Evolution}

The corresponding EoS is determined by solving the components of the energy-momentum tensor of the $\mathcal{L}_{\rm RMF}$ in Eq.~(\ref{rmf}).  To construct stable and potentially observable stars, we enforce charge neutrality and $\beta$-equilibrium conditions in calculating the EoS. Detailed formulations and applications of the EoS can be found in \cite{Issifu:2023qyi, Issifu:2024fuw, Raduta:2020fdn, Menezes:2021jmw}. Rather than repeating the derivation of these EoSs, we highlight only the key relations relevant to this work. The Helmholtz free energy of the system is given by: $\mathcal{F} = \varepsilon_t - Ts$, with $\varepsilon_t$ as the total energy of the system, $s$ as the entropy density, and $T$ as the temperature distribution of the system. The expression that connects $s,\, T,\, \varepsilon_t,\, P_t$(total pressure), $\mu_j$ (the chemical potential of the individual particles represented by $j$) and $n_j$(the number density of the individual particles) derived from $\mathcal{F}$ is given by,
\begin{align}
     sT= \varepsilon_t + P_t- \sum_{b} \mu_{b} n_{b} -\sum_{l} \mu_{l} n_{l}, 
\end{align}
where the sum; $b$ and $l$ are over all the baryons and leptons, respectively. Applying charge neutrality and $\beta$-equilibrium conditions to the above expression, we obtain 
\begin{equation}
    sT=P_t+\varepsilon_t -n_B\mu_B-\mu_{\nu_e}(n_{\nu_e} +n_e),
\end{equation}
for the neutrino-trapped regime, where the neutrino number density modifies the resulting equation, 
and
\begin{equation}
    sT=P_t+\varepsilon_t -n_B\mu_B,
\end{equation}
for the neutrino-transparent regime \cite{Issifu:2024fuw}.

To study the evolutionary stages, we divide our analysis into two broad phases: the neutrino-trapped and neutrino-transparent phases. We adopt the quasi-static approximation to study PNS evolution, assuming the star evolves through a sequence of equilibrium states. To simplify the analysis, we impose {\it ad hoc} thermodynamic conditions on the stellar configurations. This approach provides a reasonable description of the PNS structure, simplifies calculations, and facilitates comparison with certain simulation results \cite{Pons:1998mm, Prakash:1996xs, Vartanyan:2018iah, Camelio:2017nka}. The outcomes are examined in four distinct snapshots. he first stage corresponds to the early phase of PNS evolution, during which the star is lepton-rich. In this regime, the trapped leptons temporarily enhance the thermal pressure and significantly influence the stellar structure. We model this stage by fixing the lepton fraction at $Y_l = 0.4$ and by defining the entropy per baryon as $s_B = s/n_B=1$. The second stage is the deleptonization phase, occurring approximately 0.5 to 1 second after the explosion. During this period, the PNS rapidly loses neutrinos, leading to a reduction in the $Y_l$. In the model framework, we fix $Y_l = 0.2$ and $s_B = 2$ for this stage. 

The third stage occurs after nearly all neutrinos have escaped from the stellar core, approximately $50$s after the explosion, marking the transition to the cooling phase. At this point, the matter has reached its peak temperature and begins to cool down. In the model framework we set $Y_{\nu_e}=0$ and $s_B = 2$. The final stage occurs after the star has cooled over several decades to centuries ($\sim 50 - 100$ years or more), forming cold and catalyzed NS. The EoS is constrained by both experimental and observational data to ensure that, at the end of the star's evolution, its structure remains consistent with key astrophysical constraints. These constraints include the GW170817 event \cite{LIGOScientific:2017vwq, LIGOScientific:2018cki} and recent measurements from the NICER X-ray observatory, which provides mass and radius measurements for pulsars such as PSR J0740+6620 \cite{riley2021, Miller:2021qha} and PSR J0030+0451 \cite{riley2019, Miller:2019cac}.}

\begin{table}
\begin{center}
\caption {The ratio of the baryon coupling to the corresponding nucleon coupling for hyperons. 
}
\label{T2}
\begin{tabular}{ c c c c c } 
\hline
\hline
 b & $\chi_{\omega b}$ & $\chi_{\sigma b}$ & $\chi_{\rho b}$ & $\chi_{\phi b}$  \\
 \hline
 $\Lambda$ & 0.714 & 0.646 & 0 & -0.808  \\  
$\Sigma^0$ & 1 & 0.663 & 0 & -0.404  \\
  $\Sigma^{-}$, $\Sigma^{+}$ & 1 & 0.663 & 1 & -0.404  \\
$\Xi^-$, $\Xi^0$  & 0.571 & 0.453 & 0 & - 1.01 \\
  \hline
  \hline
\end{tabular}
\end{center}
\end{table}
\section{Oscillation modes}
\label{oscillaltions}
\subsection{Non-Radial}

In the full GR framework, non‐radial oscillations of NS are analyzed by introducing small perturbations to the static background spacetime metric. These perturbations lead to oscillation modes whose complex frequencies encode both the actual oscillation (through the real part) and the damping due to gravitational wave emission (through the imaginary part). In this approach, the complete set of Einstein's field equations is solved by considering gravitational waves as perturbations to the static metric of a non-rotating star. 

In contrast, the Cowling approximation simplifies the problem by neglecting the perturbations in the gravitational field and focusing solely on the fluid perturbations within the star. By disregarding the back-reaction on the gravitational potential, the Cowling approximation reduces the computational complexity of the problem. However, this simplification comes at the cost of slightly less accurate frequency estimates, since it does not fully capture the dynamic interplay between the metric and the fluid, which is intrinsic to the complete GR treatment~\cite{Pradhan:2022vdf,Kunjipurayil:2022zah,Roy:2023gzi}.  In the present study, we are using both formalisms to make a comparison of the $f$- and $p_1$-mode frequencies for different stages of NS evolution. For a detailed derivation and the underlying equations, see Ref.~\cite{Rather:2024nry}.  
Since buoyancy plays a negligible role in these modes, it is common to approximate the adiabatic and equilibrium sound speeds as equal, i.e., $c_s^2 \approx c_e^2$, in this formalism. This simplifies the analysis by effectively setting the Brunt--Väisälä frequency to zero, thereby excluding buoyancy-driven $g$-modes. More details can be found in Ref.~\cite{Zhao:2022toc}. This approximation is particularly justified in the quasi-static treatment of PNSs, where entropy and lepton number profiles are fixed. Similar simplifications have been used in earlier works employing both full GR~\cite{Barman:2024zuo,Kumar:2024jky} and the Cowling approximation~\cite{Thapa:2023grg}.

\subsection{Radial}
In a spherically symmetric system with radial motion, the metric function becomes explicitly time-dependent, allowing one to employ the Einstein field equations to analyze the radial oscillation characteristics of a static equilibrium configuration \cite{1966ApJ...145..505B}. In this framework, the radial displacement, denoted by $\Delta r$, and the pressure perturbation, $\Delta P$, are introduced to quantify the star’s response to small perturbations. By defining the dimensionless variables $\xi = \Delta r/r$ and $\eta = \Delta P/P$, the problem reduces to solving a set of perturbed differential equations. For a full description of the Radial oscillations, see Refs.~\cite{Rather:2023dom, Rather:2024hmo}.

Although the detailed derivation involves these coupled equations, the essential idea is to integrate them under appropriate boundary conditions. At the center of the star, regularity conditions are imposed (e.g., $\eta = -3\gamma\, \xi$), while at the stellar surface the requirement that the Lagrangian pressure perturbation vanishes selects discrete eigenvalues of $\omega^2$. These eigenvalues correspond to the natural modes of radial oscillation, ordered by the number of nodes in the radial eigenfunctions. In particular, the fundamental mode (with no nodes) is critical for assessing the stability of the star, as real eigenfrequencies indicate a stable configuration, whereas any imaginary component would signal instability.

The oscillation frequencies are commonly expressed in kHz, scaled by a dimensionless constant and a characteristic frequency $\omega_0\equiv\sqrt{M/R^3}$, where $M$ and $R$ represent the mass and radius of the star, respectively. Numerical methods such as the shooting method are typically employed to integrate the equations from the center to the surface, ensuring that all boundary conditions are met and yielding a discrete spectrum of radial eigenfrequencies.

  The analysis of radial oscillations for PNSs presented herein assumes an isentropic EoS throughout the star's evolutionary stages. The radial oscillations are mathematically modeled as infinitesimal adiabatic perturbations of the stellar equilibrium configuration, which fundamentally implies the absence of heat transfer and entropy modification during the oscillatory process. Furthermore, in deriving the isentropic EoS that governs PNS evolution, $s_B$ is fixed, which aligns fundamentally with the assumption of adiabatic perturbations. As a result, this approach eliminates the necessity for thermal correction terms, preserving the computational efficiency of stability and pulsation analyses while maintaining their theoretical integrity. The underlying mathematical framework governing this discussion remains valid and can be found in Ref.~\cite{Glendenning:1997wn}. Moreover, this analytical framework extends analogously to non-radial oscillations, as both modalities are predicated on the small adiabatic perturbation assumption and utilize the EoS as a fundamental input parameter.


\section{Results and Analysis}
\label{results}

\subsection{Equation of State and Mass-Radius relations}
\label{mr}

\begin{figure}[t!]
  \centering
    \includegraphics[width=\linewidth]{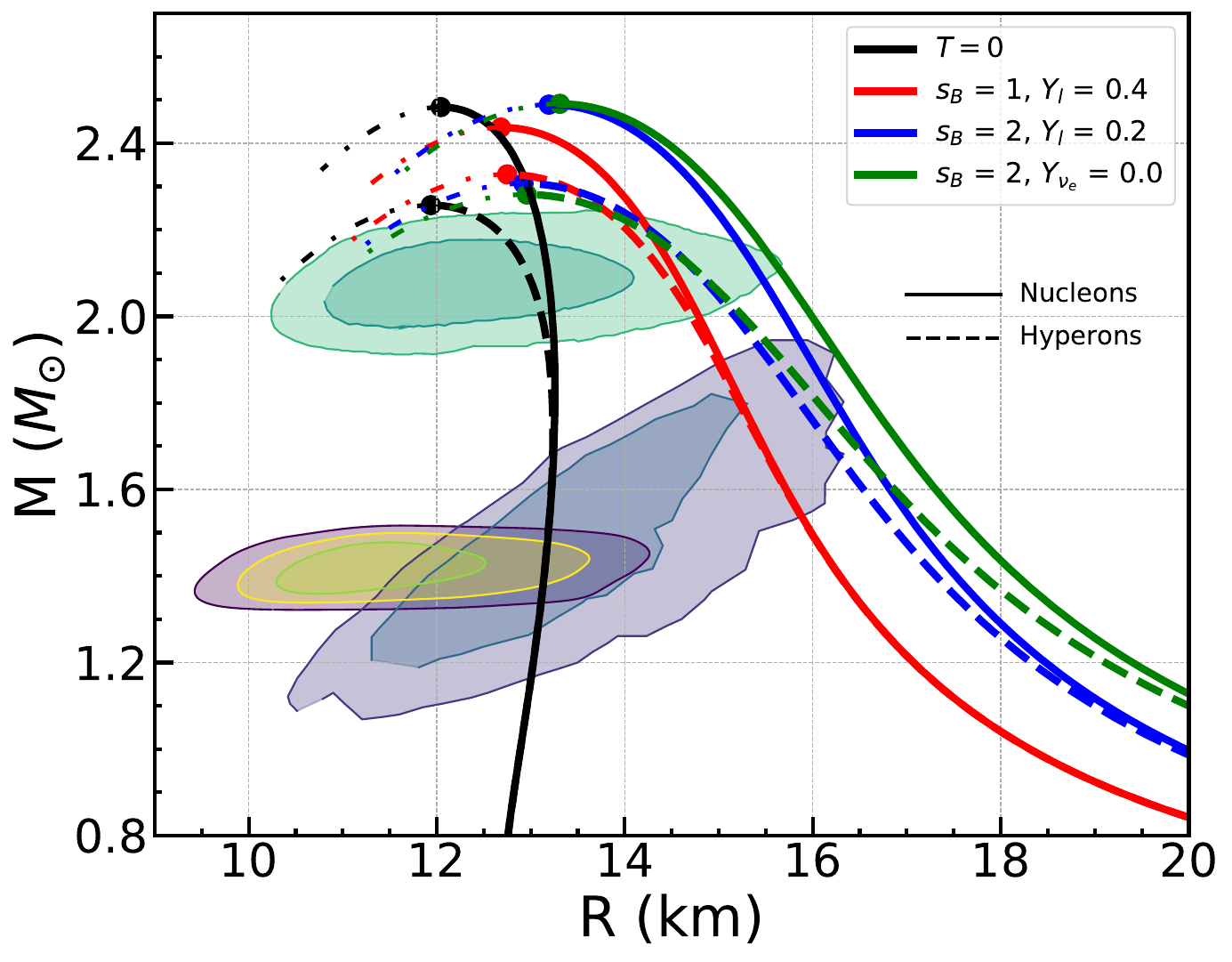}
  \caption{Total gravitational mass ($M$) of a cold NS (T = 0) as well as for a PNS capturing the evolutionary stages of neutrino-trapped, $\beta$-equilibrated stellar matter, as a function of its radius ($R$).  The different PNS stages are characterized by different entropy per baryon ($s_B$) and lepton fraction ($Y_l$), and are compared with a neutrino-transparent star where $s_B = 2$ and $Y_{\nu_e} = 0$. Solid (dashed) lines correspond to the nucleonic (hyperonic) EoSs. The 1$\sigma$ (68\%) confidence intervals for the 2D mass-radius posterior distributions of the millisecond pulsars PSR J0030+0451 \cite{Riley:2019yda, Miller:2019cac} and PSR J0740+6620 \cite{Riley:2021pdl, Miller:2021qha}, obtained from NICER X-ray observations, are included. Furthermore, the plot presents the most recent NICER constraints on the mass and radius of PSR J0437$-$4715 \cite{Choudhury:2024xbk}.}
  \label{mr}
\end{figure}

{\Cref{mr} illustrates the structure of PNSs as determined using the TOV equations \cite{PhysRev.55.374, Oppenheimer:1939ne}, showing their evolution from neutrino-rich objects at birth to cold, catalyzed NS at maturity. While PNSs undergo significant thermal and compositional evolution, their structure at any given time can be approximately described by the hydrostatic equilibrium configuration using the TOV framework \cite{Pons:1998mm, Camelio:2017nka}.
The results obtained from this process are then compared with those of observed pulsars, with the corresponding confidence contours shown in the right panel of \Cref{mr}. The full details of the thermodynamic conditions relevant for PNS evolution and how they impact the structure of the star can be found in Refs.~\cite{Pons:1998mm, Prakash:1996xs}.

To construct consistent stellar models, the high-density core EoS must be connected to an appropriate description of the low-density crust. For this regime, we employ the SFHo EoS~\cite{Hempel:2009mc}.  
The complete stellar EoS is obtained by matching the SFHo crust to the high-density core EoS discussed earlier. The crust--core transition is determined at the density where the pressures of the two models coincide, which occurs at approximately half the saturation density ($n_B$ $\approx$ 0.08 fm$^{-3}$), ensuring thermodynamic consistency \cite{rather2020effect, Fortin:2016hny}.

Aside from the leptons, whose composition depends on the star's evolutionary stage, we consider PNSs composed solely of nucleons, as well as those containing both nucleons and hyperons. In general, we find that the presence of hyperons lowers the star's maximum mass, as expected \cite{Bombaci:2016xzl}. This occurs because the onset of hyperons introduces additional degrees of freedom, weakening the pressure support at higher densities. Consequently, the EoS softens, and the star can support less mass against gravitational collapse \cite{Issifu:2024fuw, Raduta:2020fdn}. From \Cref{tab:model_frequencies_with_cowling}, comparing the maximum masses, we observe that in nucleon-only stellar matter, the neutrino-trapped phase allows the star to sustain about $0.05\,M_\odot$ more mass from the first to the second phase, with the star expanding due to heating. On the other hand, when hyperons are present, the star's mass decreases from the first to the second stage, even as it expands.  In the neutrino-transparent phase—the third and final stage—the star shrinks and becomes more compact, experiencing a slight reduction in mass and a significant decrease in radius for both matter compositions. 
}

\begin{figure}
    \centering
    \includegraphics[scale=0.45]{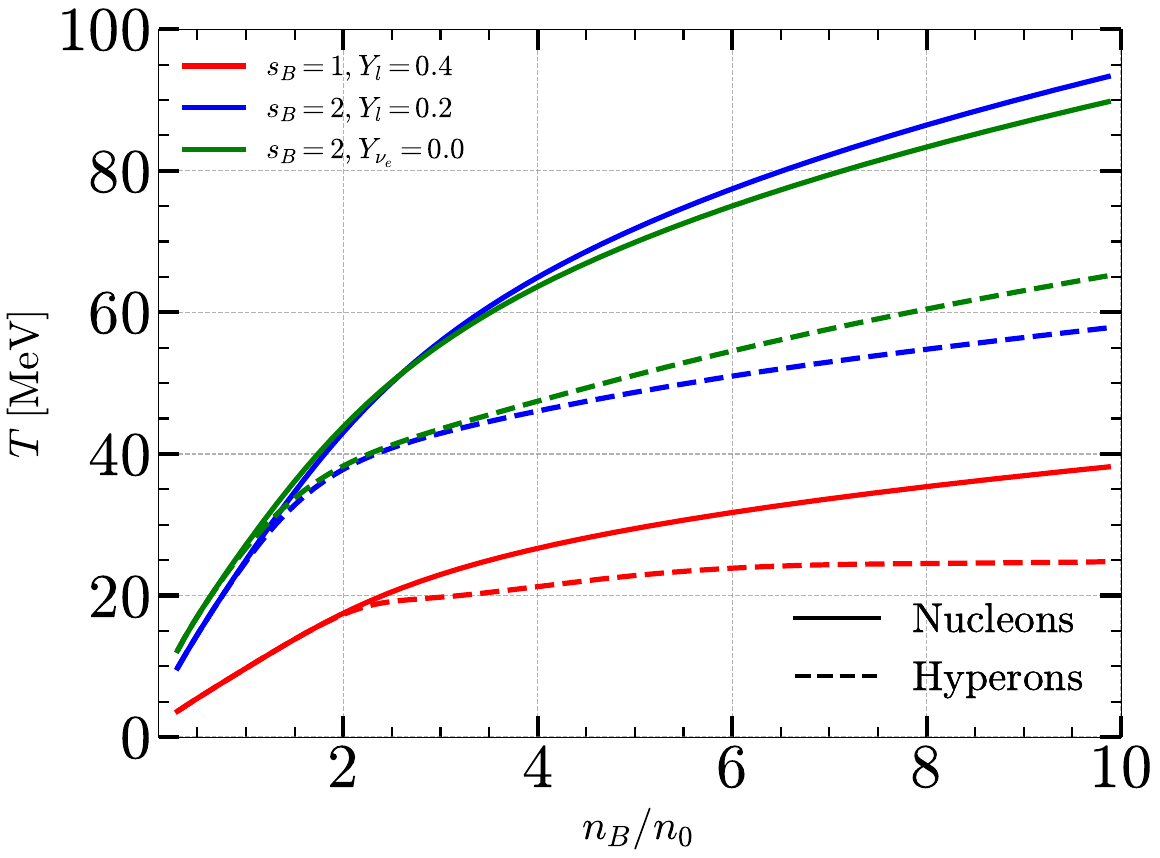}
   
   \caption{The temperature distribution in the stellar matter for nucleon-only and nucleon-hyperon admixed PNSs at different stages of the PNS evolution.}
   \label{Temperature_baryon_density}
\end{figure}

\Cref{Temperature_baryon_density} shows the temperature distribution in the stellar matter as a function of the baryon density in units of baryon saturation density. In the first stage, the temperature is lowest when the $s_B$ is minimal and the $Y_l$ is highest. In the second stage, the temperature reaches its peak in the core. In the third stage, where the stellar matter is expected to attain its maximum temperature before cooling, the temperature rises to $\sim 3n_0$ before gradually decreasing, eventually falling below the second-stage temperature in purely hadronic matter in the core. However, the inclusion of hyperons lowers the temperature throughout the stellar matter. This occurs because additional degrees of freedom redistribute the available thermal energy among more particle species, reducing the thermal energy per particle and consequently decreasing the overall temperature at a given density. Furthermore, the appearance of hyperons increases the specific heat capacity ($C_V \approx (\partial\varepsilon/\partial T)_V$) of the matter. As a result, more thermal energy is required to raise the temperature. However, since the total thermal energy of the PNSs remains nearly conserved, the temperature decreases to compensate for the increased $C_V$ \cite{Raduta:2020fdn, Prakash:1996xs, Pons:1998mm}.  Here, the temperature variation follows the pure nucleonic matter phase, except that the temperature is highest for the third stage towards the core.

\begin{figure*}[htbp]
  \centering
  \begin{minipage}[b]{0.47\linewidth}
    \centering
    \includegraphics[width=\linewidth]{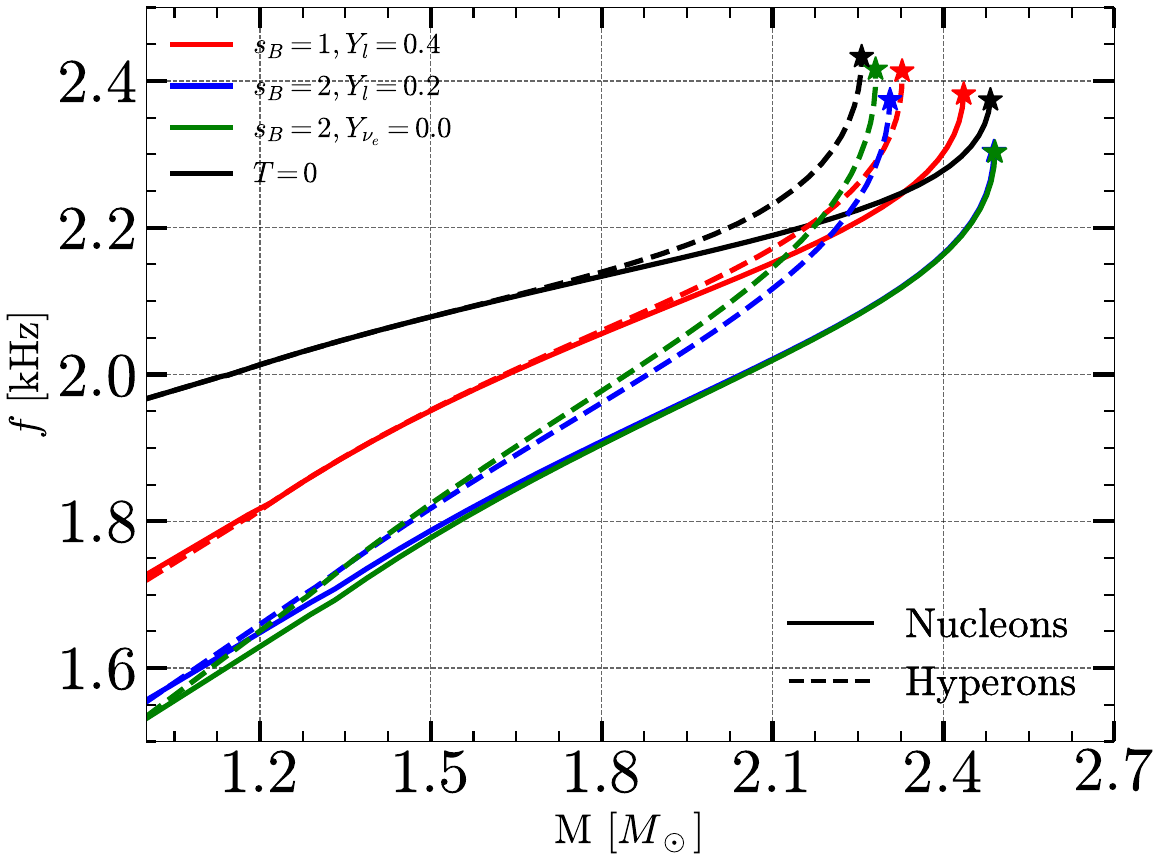}
  \end{minipage}\hspace{0.02\linewidth}%
  \begin{minipage}[b]{0.47\linewidth}
    \centering
    \includegraphics[width=\linewidth]{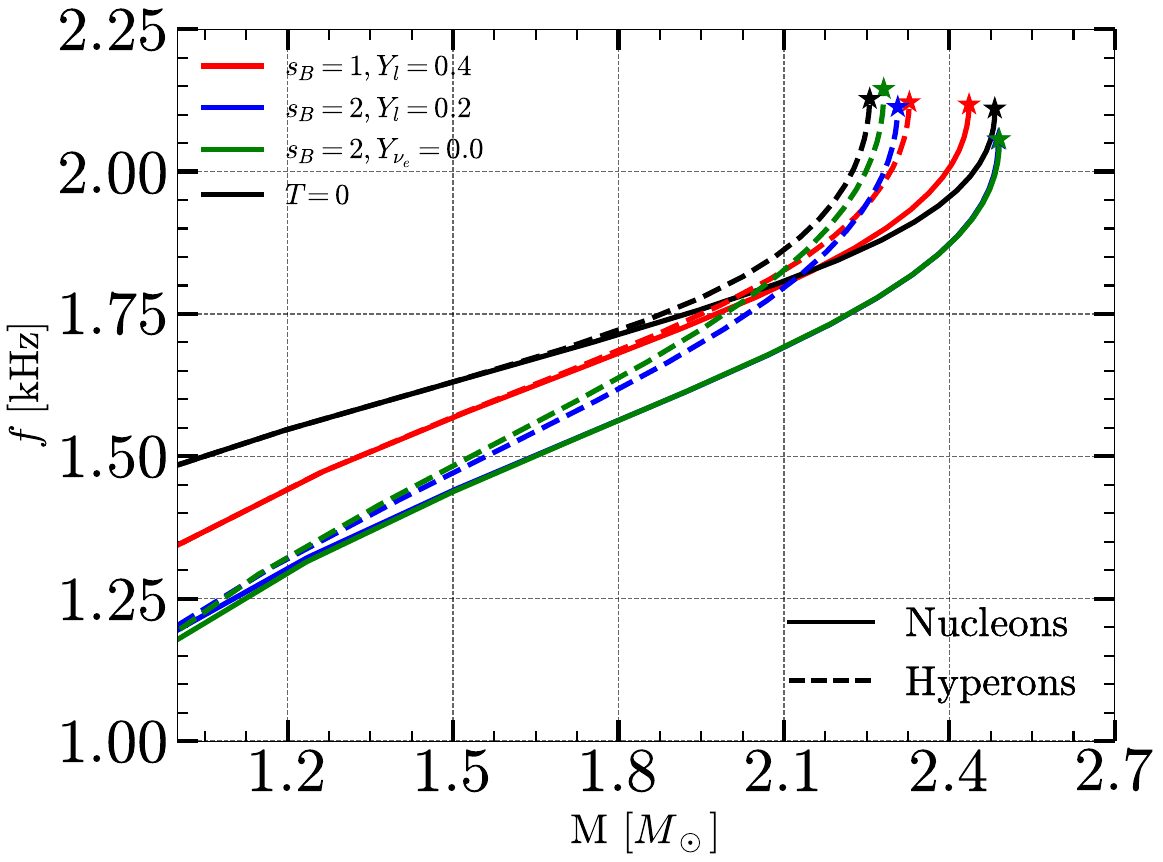}
  \end{minipage}
  \caption{Gravitational Mass ($M$) versus fundamental frequency ($f$) of non-radial oscillation modes for different stages of the PNS evolution with nucleons (solid lines) and with hyperons (dashed lines). The left panel represents results obtained using the Cowling approximation, while the right panel shows calculations based on the full General Relativity (GR) framework.}
  \label{frequency_mass_1}
\end{figure*}

\begin{table*}[t!]
\centering
\scriptsize
\caption{Stellar maximum mass $M_{max}$, corresponding radius $R_{max}$, and $f$-mode properties for different configurations studied. Values for hyperonic stars are shown in parentheses.}
\begin{tabular}{lccccccccc}
\toprule
Model & 
\makecell{$M_{max}$ \\ ($M_\odot$)} & 
\makecell{$R_{max}$ \\ (km)} & 
\makecell{$f_{1.4}$ [GR] \\ (kHz)} & 
\makecell{$f_{1.4}$ [Cow] \\ (kHz)} & 
\makecell{$f_{2.0}$ [GR] \\ (kHz)} & 
\makecell{$f_{2.0}$ [Cow] \\ (kHz)} & 
\makecell{P.E.$_{1.4}$ \\ (\%)} & 
\makecell{P.E.$_{2.0}$ \\ (\%)} & 
\makecell{$\tau_{1.4}$ \\ (s)} \\ 
\hline
$s_B=1;\; Y_l=0.4$       & 2.44 (2.33)   & 12.34 (12.41)   & 1.57 (1.57)  & 1.91 (1.91)  & 1.78 (1.80)  & 2.12 (2.13)  & 21.66 (21.66)  & 19.10 (18.33)  & 0.288 (0.287) \\
$s_B=2;\; Y_l=0.2$       & 2.49 (2.29)   & 12.83 (12.59)   & 1.45 (1.48)  & 1.75 (1.77)  & 1.67 (1.76)  & 1.98 (2.06)  & 20.69 (19.59)  & 18.56 (17.05)  & 0.339 (0.322) \\
$s_B=2;\; Y_{\nu_e}=0$   & 2.49 (2.28)   & 12.87 (12.51)   & 1.46 (1.50)  & 1.74 (1.78)  & 1.68 (1.79)  & 1.98 (2.08)  & 19.18 (18.67)  & 17.86 (16.20)  & 0.334 (0.314) \\
$T=0$                    & 2.48 (2.26)   & 12.03 (11.96)   & 1.60 (1.60)  & 2.05 (2.05)  & 1.77 (1.80)  & 2.17 (2.19)  & 28.13 (28.13)  & 22.60 (21.67)  & 0.285 (0.285) \\ 
\bottomrule
\end{tabular}
\label{tab:model_frequencies_with_cowling}
\end{table*}

\subsection{$f$-mode frequency: GR vs Cowling}
\label{frequency}
In the left panel of \Cref{frequency_mass_1}, the $f$-mode frequency (in kHz) is shown as a function of the gravitational mass $M$ (in $M_\odot$) for various thermodynamic conditions, including different combinations of $s_B$ and lepton fractions ($Y_l$ and $Y_{\nu_e}$) at $\beta$-equilibrium, using the Cowling approximation. This approach neglects metric perturbations, thereby simplifying the oscillation equations. The solid lines correspond to stars composed purely of nucleons, while the dashed lines represent stars with hyperonic degrees of freedom. The right panel presents the corresponding results from full GR calculations, where both fluid and spacetime perturbations are included. Each color represents a distinct thermodynamic scenario: $s_B = 1, Y_l = 0.4$ (red), $s_B = 2, Y_l = 0.2$ (blue), $s_B = 2, Y_{\nu_e} = 0$ (green), and cold matter at $T = 0$ (black). Both panels demonstrate that the Cowling approximation systematically overestimates the $f$-mode frequencies compared to full GR. At a canonical mass of $1.4\,M_\odot$, this overestimation reaches up to 28\%  for both nucleonic and hyperonic stars, as shown in Table~\ref{tab:model_frequencies_with_cowling}. 

Interestingly, the percentage error (P.E.) from the Cowling approximation decreases for the intermediate stages when the star is hot and enlarged, and rises again when the star is cold and catalyzed. 
This trend is consistent with the known behavior of the Cowling approximation in $f$-mode studies, where the P.E. tends to decrease for more massive stars~\cite{Kunjipurayil:2022zah,Pradhan:2022vdf,Rather:2024nry}. This behavior arises because $f$-mode oscillations involve fluid perturbations that peak near the stellar surface, while metric perturbations dominate near the core. In massive NSs, strong surface fluid perturbations couple only weakly to central metric perturbations, making the Cowling approximation more accurate. Furthermore, the same trend is evident at $2.0\,M_\odot$, where the P.E. continues to decrease for configurations such as $s_B = 2, Y_l = 0.2$ and $s_B = 2, Y_{\nu_e} = 0$, with the effect being more pronounced in hyperonic stars due to the additional softening of the EoS, increased compactness, and its impact on the gravitational field. The P.E. further decreases when evaluated at the maximum mass configurations for each model, reaching an average of approximately 11.53\%. The smallest deviation is observed for the $s_B = 2, Y_l = 0.2$ (hyperons) model, with a relative error of 7.83\%, indicating that the Cowling approximation performs well in this thermodynamic setup. This result highlights that, particularly during intermediate evolutionary stages of PNSs, the Cowling approximation can be suitably employed. 

Considering that the masses of the PNSs in the second and third stages are nearly identical, we infer that the presence of neutrinos in the second stage increases the P.E. in the $f$-mode frequency when using the Cowling approximation. In the first stage, where a larger number of neutrinos are trapped, the P.E. is also higher than in the intermediate stages, further supporting the role of neutrinos in increasing the error. This is because neutrinos contribute to the EoS through their pressure, influencing the star’s stability, gravitational field, and the restoring force responsible for fluid oscillations. Since the Cowling approximation neglects metric perturbations, it fails to capture the full impact of neutrino-induced effects, leading to less accurate predictions in neutrino-rich environments than full GR calculations \cite{Sotani:2020mwc}.

\begin{figure*}[t!]
  \centering
  \begin{minipage}[b]{0.47\linewidth}
    \centering
    \includegraphics[width=\linewidth]{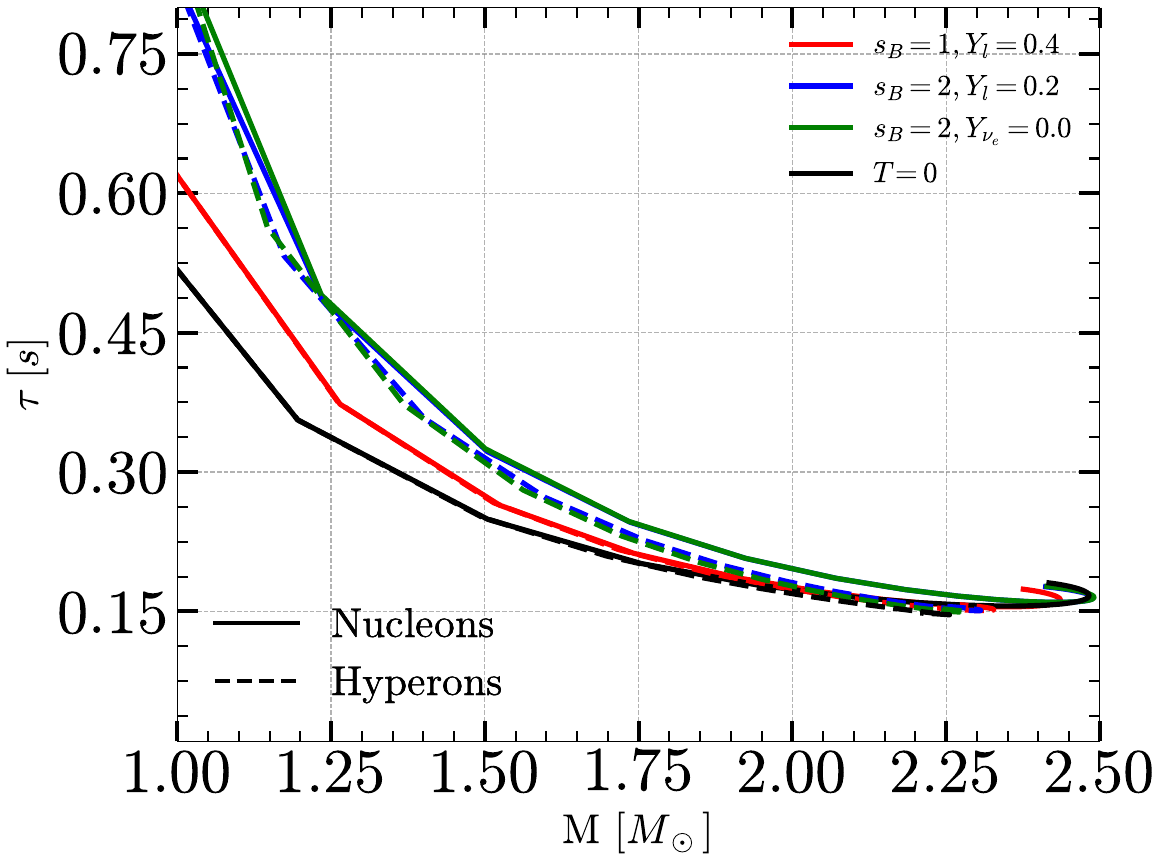}
  \end{minipage}\hspace{0.02\linewidth}%
  \begin{minipage}[b]{0.47\linewidth}
    \centering
    \includegraphics[width=\linewidth]{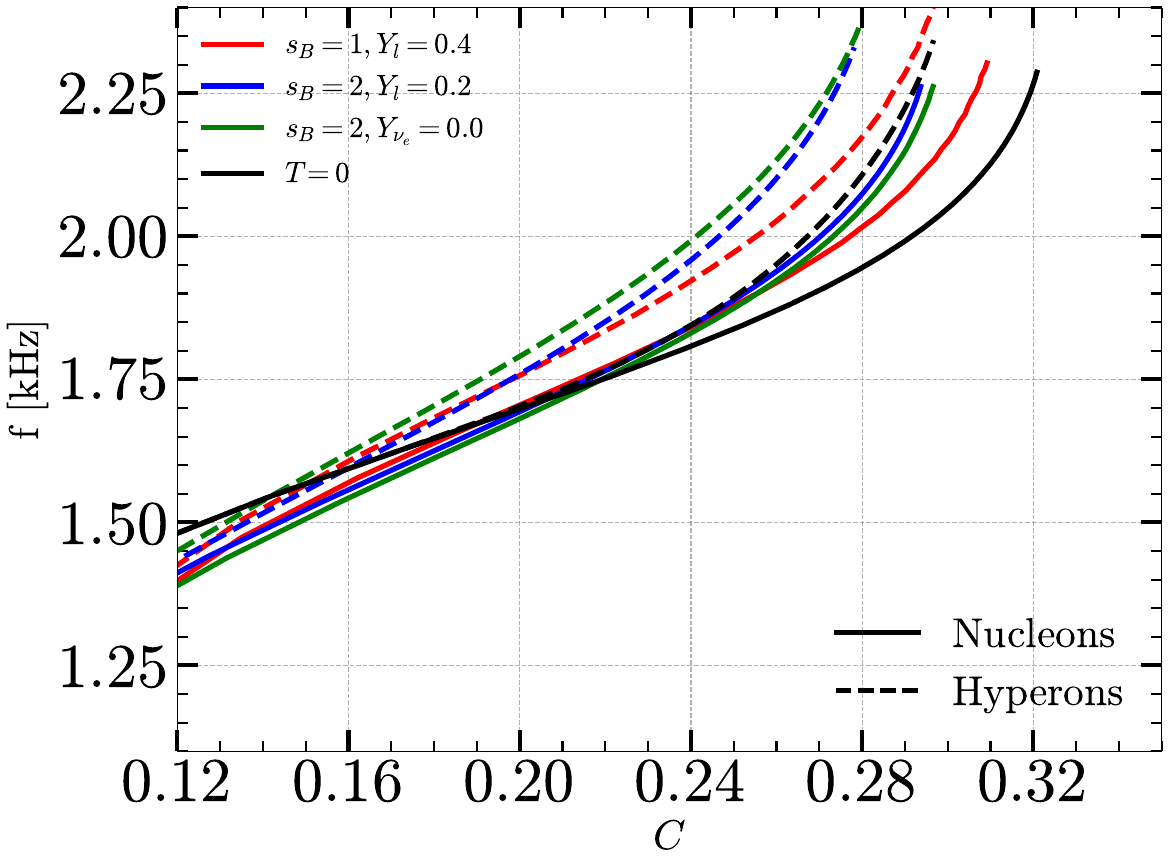}
  \end{minipage}
  \caption{Left: Damping time ($\tau$) as a function of gravitational mass ($M$) for non-radial $f$-mode oscillation. Right: Non-radial $f$-mode frequency as a function of Compactness ($M/R$) at different stages of the PNS evolution for nucleons (solid lines) and hyperons (dashed lines). }
  \label{damping_freq}
\end{figure*}

From \Cref{frequency_mass_1} and \Cref{tab:model_frequencies_with_cowling}, it is clear that at a mass of $1.4\,M_{\odot}$, the $f$-mode frequencies for nucleonic and hyperonic EoSs are quite close. However, at higher masses such as $2.0\,M_{\odot}$, the frequencies become significantly higher for hyperonic EoS, emphasizing the increased influence of EoS softening due to the presence of hyperons. An interesting observation can be made from the GR values in the $f_{1.4}$ column of \Cref{tab:model_frequencies_with_cowling}. The frequencies are lower for configurations with $s_B = 2, Y_l = 0.2$ and $s_B = 2, Y_{\nu_e} = 0$ compared to those with $s_B = 1, \, Y_l = 0.4$ and cold matter at $T = 0$. This effect can be attributed to changes in temperature and compactness during the star's evolutionary stages, resulting from variations in $s_B$ and $Y_l$. In the early neutrino-trapping phase (first stage), the high $Y_l$ and trapped neutrinos provide additional pressure support, keeping the EoS relatively stiff. 

As the star deleptonizes (second and third stages), neutrinos diffuse out, leading to thermal heating and expansion, which reduces the star’s compactness compared to the initial phase. This expansion results in a softened EoS, lowering the sound speed ($c^2_s = \partial P/\partial\varepsilon$) and reducing the restoring force for oscillations, thereby decreasing the $f$-mode frequencies. As deleptonization progresses, the star continues to cool, and once all neutrinos escape, it contracts to form a cold, catalyzed NS. At this stage, the EoS stiffens again due to the increased compactness and the loss of thermal pressure contributions. The stiffer EoS provides stronger pressure support and a higher restoring force, leading to an increase in $f$-mode frequencies. Thus, the EoS is stiffer in both the first stage (when the temperature is relatively low, see \Cref{Temperature_baryon_density}) and the final stage (when the star has cooled into a cold, catalyzed NS). Although the $T=0$ and $Y_l=0.4,\,s_B=1$ configurations have smaller radii due to higher compactness, the underlying EoS remains stiffer, 
that allow these stages to sustain comparable maximum masses. This results in higher $f$-mode frequencies in these phases, whereas the intermediate deleptonization stages, characterized by thermal expansion and a softer EoS, exhibit lower $f$-mode frequencies, as observed in Ref.~\cite{{Zhao:2022tcw}}.

This trend continues even at the maximum mass configuration for nucleonic stars, where the $s_B = 1, Y_l = 0.4$ and $T = 0$ cases have a frequency of 2.12 and 2.11\,kHz, respectively, while the $s_B = 2, Y_l = 0.2$ and $s_B = 2, Y_{\nu_e} = 0$ cases show slightly lower frequency of 2.06\,kHz. In contrast, for hyperonic stars, the highest frequency at maximum mass is observed for the $s_B = 2, Y_{\nu_e} = 0$ case at 2.16\,kHz, while the lowest is for the cold $T = 0$ case at 2.12\,kHz.

The left plot of \Cref{damping_freq} shows the damping time $\tau$ (in seconds) of $f$-mode oscillations as a function of gravitational mass $M$ (in $M_{\odot}$). The legends indicating nucleonic (solid lines) and hyperonic (dashed lines) compositions, as well as the color-coded models based on $s_B$, $Y_l, Y_{\nu_e}$, and the cold $T = 0$ cases, are the same as those described in the previous figures. 

The damping time $\tau$ represents the characteristic timescale over which the $f$-mode oscillation amplitude decays due to gravitational wave emission. As expected, $\tau$ decreases monotonically with increasing stellar mass for all compositions and thermodynamic conditions. For a typical NS, the $f$-mode frequency falls within the range of 1–3 kHz, while the corresponding damping time, $\tau_f$, is typically a few tenths of a second \cite{Kokkotas:1999bd,Pradhan:2020amo}. In our case, the damping time at $1.4\,M_{\odot}$ lies in the range of approximately 0.28 to 0.33 seconds and is nearly the same for both nucleonic and hyperonic compositions. At the maximum mass configuration, the damping time decreases to around 0.16 seconds for nucleonic and 0.14 seconds for hyperonic. The trends observed in the $f$-mode frequencies can be more clearly interpreted by analyzing the role of stellar compactness, as illustrated in the right plot of \Cref{damping_freq}, which is strongly influenced by the stiffness of the EoS.

Hyperonic EoSs, being softer than nucleonic ones, lead to more compact NSs with smaller radii. This increased compactness influences the dynamics of $f$-mode oscillations. Since $f$-mode oscillations involve fluid perturbations peaking near the surface and metric perturbations peaking at the center, the enhanced compactness in hyperonic stars tends to modify the frequency behavior compared to their nucleonic counterparts. For NSs with a nucleonic EoS, which is typically stiffer than its hyperonic admixture counterparts, the larger radius results in a lower mean density. This weakens the restoring force driving fluid oscillations, thereby lowering the $f$-mode frequencies. The reduced compactness also diminishes metric perturbations and gravitational wave damping, leading to longer damping times. In contrast, hyperonic EoS are generally softer, producing NSs with smaller radii and higher mean densities. This strengthens the restoring force, resulting in higher $f$-mode frequencies. The increased compactness amplifies metric perturbations, shortening the damping time and enhancing gravitational wave emission~\cite{Lindblom:1983ps}.

\subsection{Gravitational Wave Asteroseismology}

     
\subsubsection{Universal Relations}

\begin{figure*}[t!]
  \centering
  \begin{minipage}[b]{0.47\linewidth}
    \centering
    \includegraphics[width=\linewidth]{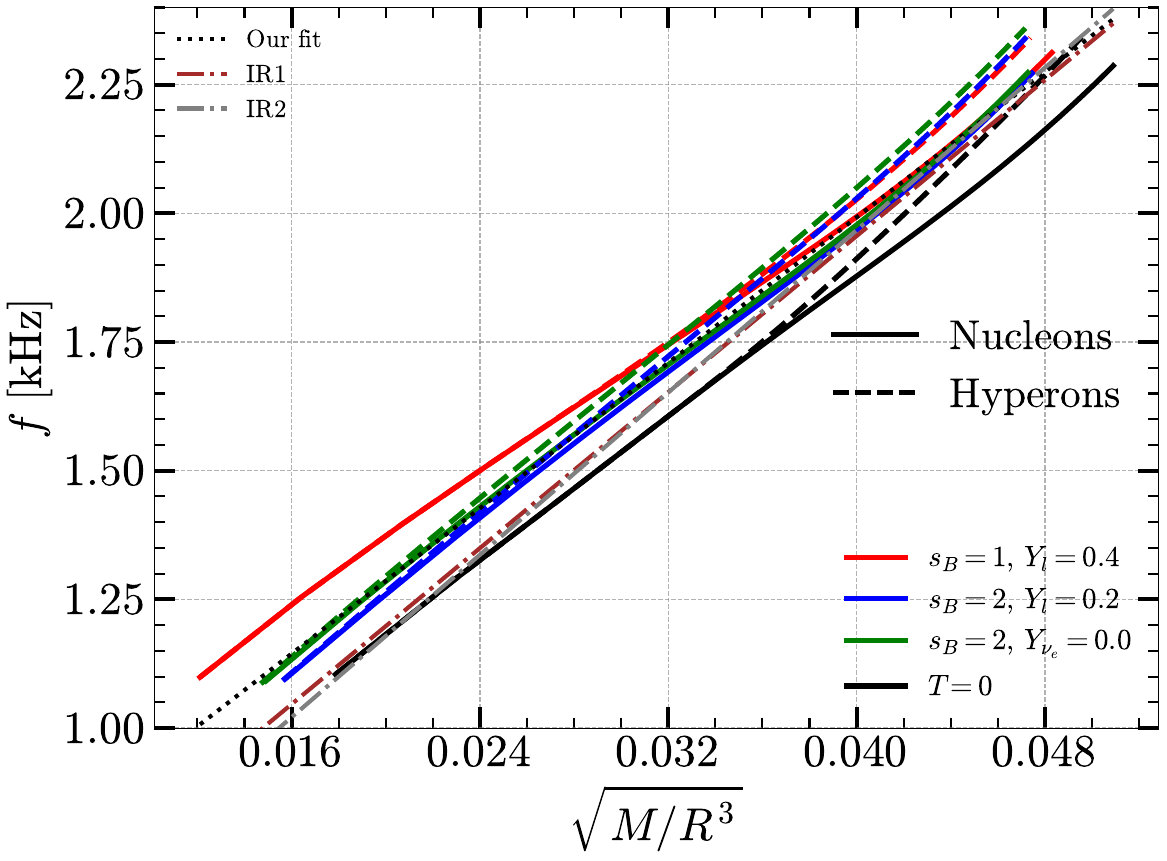}
  \end{minipage}\hspace{0.02\linewidth}%
  \begin{minipage}[b]{0.47\linewidth}
    \centering
    \includegraphics[width=\linewidth]{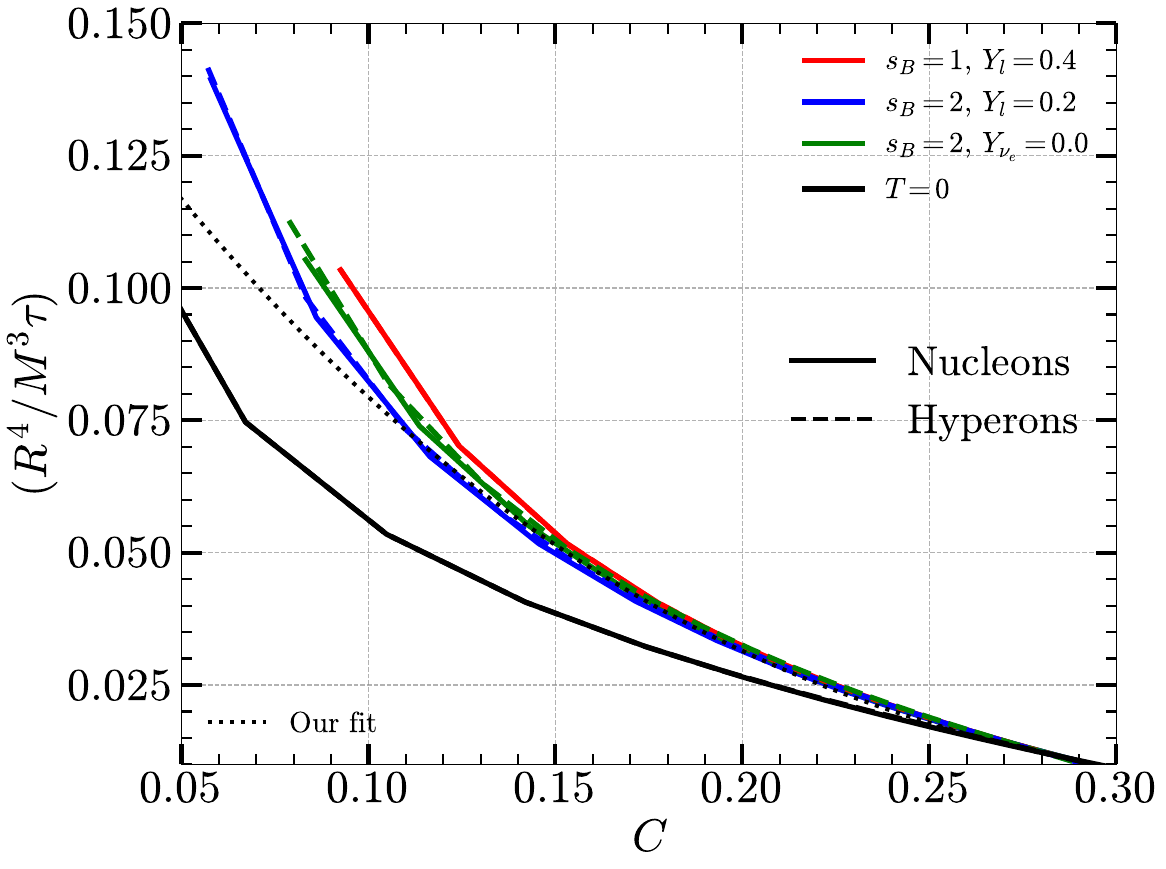}
  \end{minipage}
  \caption{Left: The $f$-mode frequency as a function of the mean stellar density. The IR1 (brown) and IR2 (gray) are the fits from the earlier study, represented by a dot-dashed line \cite{Rather:2024nry}, while "Our fit" (dotted line) represents the fit from the current work. Right: Normalized damping time of the $f$-mode as a function of the stellar compactness. The solid (dashed) lines correspond to the nucleonic (hyperonic) EoSs for different stages of the PNS evolution. }
  \label{frequency_mass}
\end{figure*}

In this section, we focus on universal relations, as they are independent of the underlying EoSs. Such relations are particularly valuable because determining NS properties directly from gravitational wave observations is complicated by the significant uncertainty in the EoS of dense matter. Universal relations offer a pathway to bypass this issue by connecting observable quantities in a manner that is largely insensitive to the internal composition of the star. Notable examples include the relation between the $f$-mode oscillation frequency and the mean stellar density, its dependence on the compactness of the star, and the spacetime ($w$) mode frequency and damping time as functions of compactness~\cite{Benhar:1998au,Andersson:1997rn,Chirenti:2015dda}

A representative example is shown in the left panel of \Cref{frequency_mass}, which illustrates the variation of the fundamental ($f$) mode oscillation frequency as a function of the square root of the mean density, $\sqrt{M/R^3}$, for NSs under different thermal and compositional conditions. This relation is motivated by the empirical observation that the $f$-mode frequency scales approximately linearly with the square root of the mean density, as originally proposed by Andersson \cite{Andersson:1997rn}, which is expressed as:
\begin{equation}
\frac{f}{\text{kHz}} = a + b \sqrt{\frac{M}{R^3}}.
\label{eq:f_mode_relation}
\end{equation}
where $a$ and $b$ are 0.22 and 32.16, respectively. The figure includes multiple curves corresponding to different $s_B$, $Y_l$, and the presence or absence of neutrino trapping. For instance, the red curve corresponds to hot, lepton-rich matter with $s_B = 1$ and $Y_l = 0.4$, while the blue and green curves show the cases with higher entropy and varying lepton fractions ($s_B = 2, Y_l = 0.2$ and $Y_{\nu_e} = 0$, respectively). The black line represents the cold NS configuration with $T = 0$. Each thermodynamic configuration includes both nucleonic (solid lines) and hyperonic (dashed lines) EoSs. 

Additionally, the figure also includes several model fits for comparison. The dotted black curve labeled ``Our fit" corresponds to the empirical relation we derived from our models covering different stages of PNS evolution. which is expressed as:
\begin{equation}
\frac{f}{\text{kHz}} \approx 0.57 + 35.32 \sqrt{\frac{M}{R^3}}.
\end{equation}

While this relation fits well across all the stellar evolutionary stages, it shows a significant deviation from the cold NS data. This indicates that the frequency-compactness relation is not truly universal, but instead depends on the thermodynamic state of the star. The underlying reason for this deviation lies in the internal structure. PNSs are hot, possess high entropy, and contain trapped neutrinos (first and second stages); all of these factors contribute to increased thermal pressure and a stiffer EoS. These factors modify the density and sound speed profiles within the star, generally resulting in lower $f$-mode frequencies for a given mass and radius. 

In contrast, cold NSs are more compact with no thermal or neutrino pressure support, which leads to higher $f$-mode frequencies. Therefore, a relation derived from PNS models cannot accurately capture the oscillation behavior of cold NSs, emphasizing the need for temperature-dependent relations in gravitational wave asteroseismology. Additionally, the brown and grey dot-dashed curves, labeled ``IR1'' and ``IR2,'' correspond to predictions from our previous work~\cite{Rather:2024nry} based on EoSs that include hyperons and $\Delta$ baryons. The ``IR1'' curve represents scenarios without phase transitions, with fit parameters $a = 0.44$ and $b = 37.90$, while ``IR2'' ($a = 0.39$, $b = 39.44$) reflects cases where the EoS undergoes a phase transition to quark matter. These curves align closely with the behavior of cold NSs, supporting their consistency with the $T = 0$ EoS scenarios.


The right panel of \Cref{frequency_mass} presents the normalized damping time of the $f$-mode, given by the dimensionless quantity $R^4 / (M^3 \tau)$, plotted as a function of the stellar compactness. This quantity characterizes how efficiently an NS loses energy through gravitational wave emission and is useful for exploring potential universal relations across different star configurations. The colored curves correspond to different thermal stages of PNS evolution, including $s_B = 1, Y_l = 0.4$ (red), $s_B = 2, Y_l = 0.2$ (blue), $s_B = 2, Y_{{\nu}_e} = 0$ (green), and cold NSs with $T = 0$ (black). As before, solid and dashed lines indicate nucleonic and hyperonic EoSs, respectively.

A clear inverse trend is observed: the normalized damping time, $R^4 / (M^3 \tau)$, decreases with increasing compactness $M/R$, consistent with the expectation that more compact stars radiate gravitational waves more efficiently. The black dotted curve corresponds to our empirical fit derived from thermal PNS models~\cite{Lioutas:2017xtn, PhysRevD.103.123015}, and is expressed as a cubic polynomial:
\begin{equation}
\frac{R^4}{M^3 \tau} = 0.16 - 1.08 \left(\frac{M}{R}\right) + 2.49 \left(\frac{M}{R}\right)^2 - 2.01 \left(\frac{M}{R}\right)^3.
\label{eq:our_fit_damping}
\end{equation}
This fit captures the overall trend and aligns well with the PNS data, demonstrating a quasi-universal behavior when thermal effects are included.

However, a key finding from the figure is that this fit significantly deviates for cold NSs. The damping times for cold stars fall well below the PNS trend, especially at low compactness. This indicates that the damping time--compactness relation is sensitive to the star's thermal state, and that cold stars cannot be accurately described by a fit calibrated solely on hot stellar configurations. In other words, the apparent universality breaks down once cold NSs are considered.  
Overall, the figure demonstrates that universal relations constructed for cold NSs are not fully consistent with those for PNSs. The inclusion, cold NS configurations causes a noticeable deviation from the fit obtained using only thermal models, indicating that a single universal relation cannot simultaneously capture the damping behavior of both cold and hot NS phases. This highlights the need for temperature-dependent formulations when modeling gravitational wave damping times.

 \begin{figure}
    \centering
    \includegraphics[width=\linewidth]{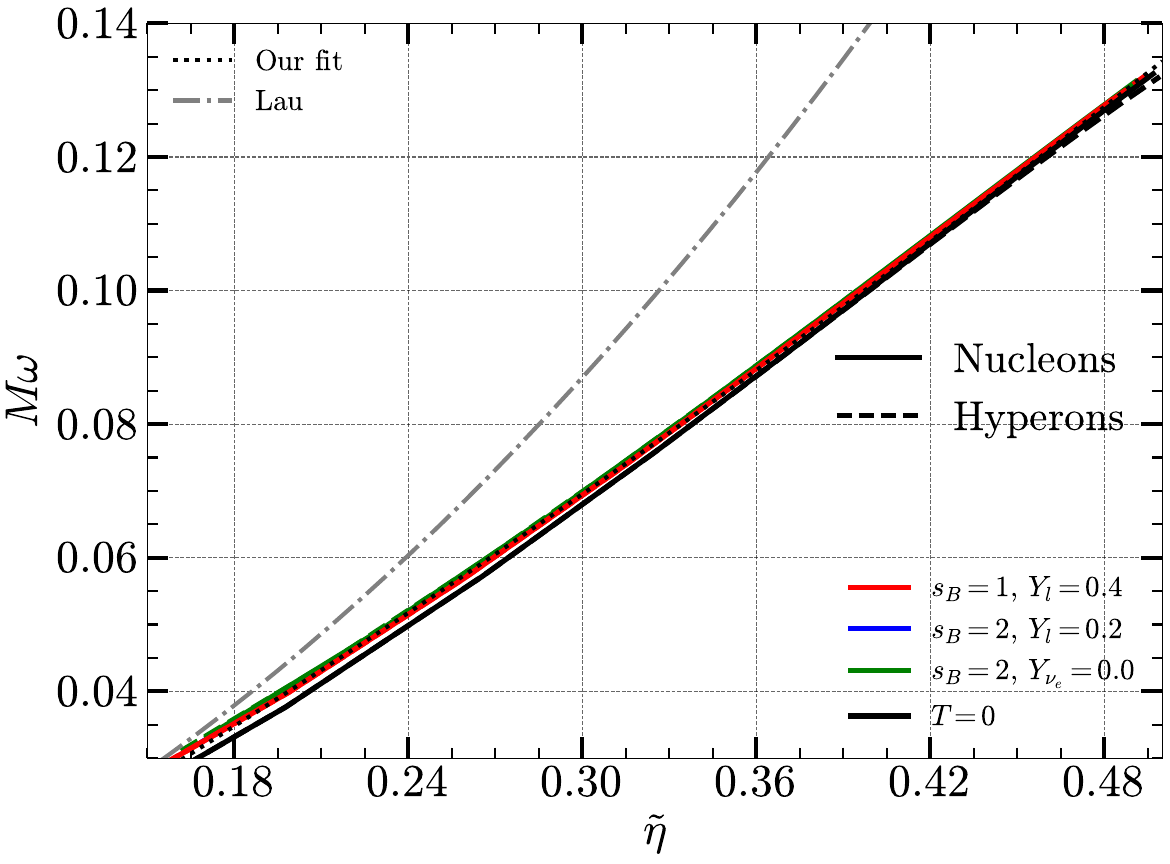}
   \caption{The dimensionless real part of the $f$-mode frequency, scaled by mass as $\omega_r = 2\pi f$, is plotted against the effective compactness $\tilde{\eta} = \sqrt{M^3/I}$. The dotted black line represents the fit obtained in this work, while the dot-dashed line corresponds to the fit proposed by Lau et al. \cite{Lau_2010}.
}
   \label{universal_MOI}
\end{figure}

\Cref{universal_MOI} shows the dimensionless form of the real part of the $f$-mode frequency, $M \omega_r = 2\pi M f $, plotted against the effective compactness, defined as $\tilde{\eta} = \sqrt{M^3/I}$,
where $M$ is the mass and $I$ is the moment of inertia of the NS. This representation is motivated by the search for universal relations that remain largely independent of the underlying EoS. The colored curves correspond to different stages of PNS evolution, including varying entropy and lepton fractions. Solid and dashed lines represent nucleonic and hyperonic EoSs, respectively. The consistency among these curves indicates that the scaled frequency correlates strongly with the effective compactness, regardless of thermal conditions or particle composition. 

The black dotted line labeled ``Our fit'' represents the empirical relation derived in this work, which captures the behavior of $M \omega$ as a function of effective compactness $\tilde{\eta}$. The best-fit linear relation is given by
\begin{equation}
M \omega = -0.011512 + 0.237550 \, \tilde{\eta} + 0.108332 \, \tilde{\eta}^2
\label{eq:our_fit_eta}
\end{equation}
and it shows excellent agreement with the data across all the stellar evolutionary stages. In contrast, the dot-dashed grey curve labeled ``Lau'' represents the universal relation proposed by Lau et al.~\cite{Lau_2010}, which shows a noticeable deviation from our fitted relation. Their corresponding fit parameters are $a = -0.0047$, $b = 0.133$, and $c = 0.0575$.

The discrepancy between our universal relation and that of Lau et al. \cite{Lau_2010} stems from the different physical conditions and EoS considered. Lau \textit{et al.} studied cold, catalyzed NSs with a barotropic EoS, where the moment of inertia $I$ closely tracks the mass distribution and largely determines the $f$-mode frequency, yielding an almost EoS-independent relation. In contrast, our work focuses on hot PNSs with non-barotropic EoSs, $P = P(n_B, s_B, Y_l)$, where entropy and lepton fraction strongly affect the pressure, altering the internal structure. Using a density-dependent relativistic mean-field model that includes hyperons, we demonstrate that these effects lead to the deviations in \Cref{universal_MOI} and signal the breakdown of cold NS universality. Our quadratic fit \Cref{eq:our_fit_eta} captures the $f$-mode behavior across thermal and compositional stages, reflecting both the quasi-universality of NS relations and the distinct influence of hot, non-barotropic matter.

The close clustering of all stellar profiles along our fit line underscores the robustness of this relation across different stellar compositions and evolutionary stages, supporting its applicability in realistic astrophysical scenarios. It also reinforces the idea that the effective compactness $\tilde{\eta}$ is a more suitable scaling variable than the mean density, especially when attempting to construct model-independent EoS relevant for temperature-dependent gravitational wave asteroseismology.

\subsection{$p_1$-mode}

\begin{figure*}[htbp]
  \centering
  \begin{minipage}[b]{0.47\linewidth}
    \centering
    \includegraphics[width=\linewidth]{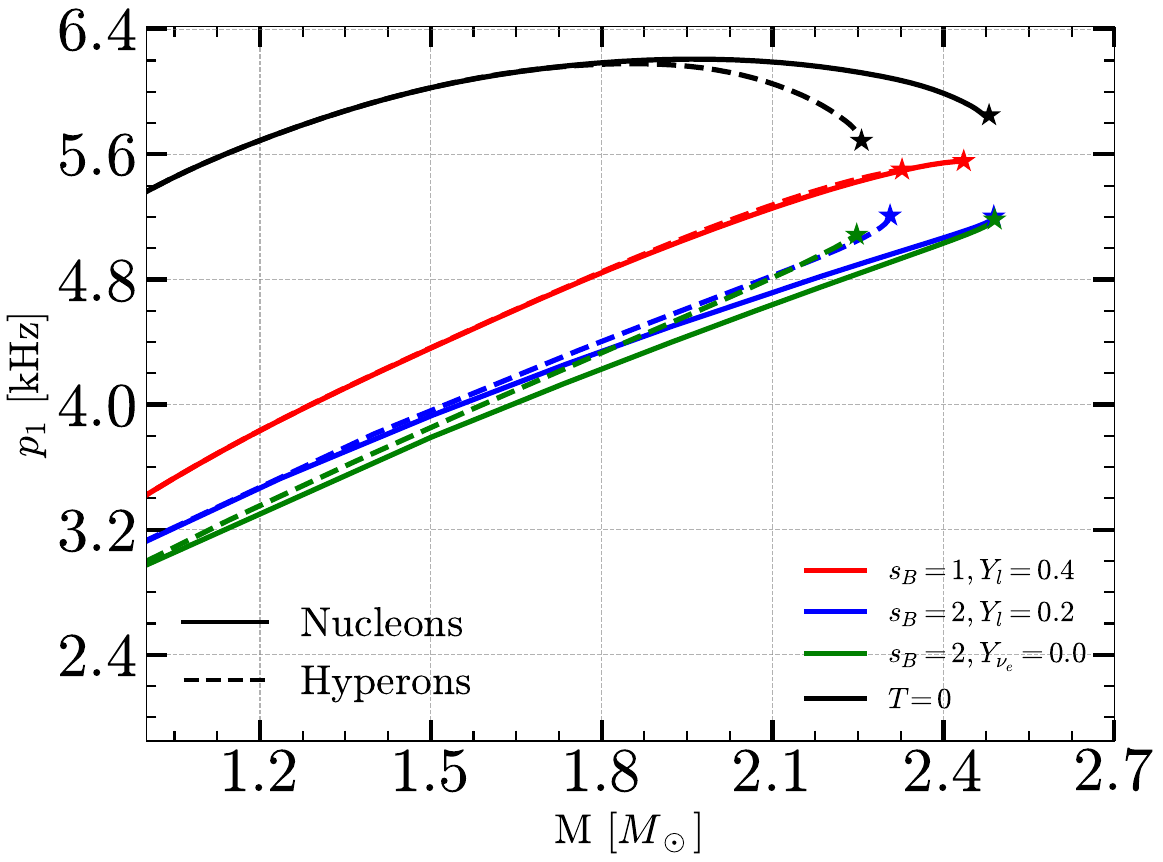}
  \end{minipage}\hspace{0.02\linewidth}%
  \begin{minipage}[b]{0.47\linewidth}
    \centering
    \includegraphics[width=\linewidth]{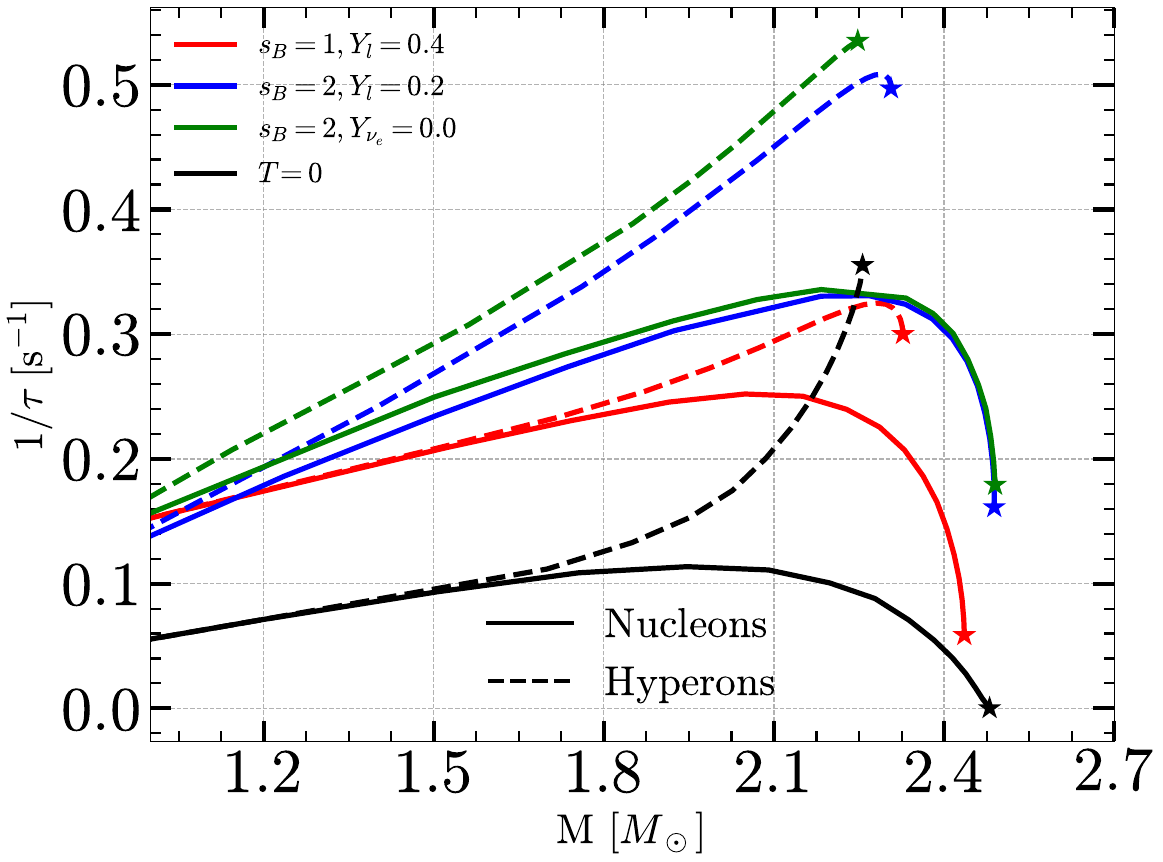}
  \end{minipage}
  \caption{Left: The $p_1$-mode frequency as a function of mass in the full GR framework. Right: Variation of the imaginary part $\omega$ of $p_1$-mode oscillations (inverse damping time) versus neutron star mass. }
  \label{p_frequency_mass}
\end{figure*}

Next, we focus on calculating the non-radial $p_1$-mode (unlike the radial oscillation, here $p_1$-modes refers to the pressure modes) for $\ell = 2$. The frequencies and damping times associated with the $p_1$-mode provide valuable insights into the internal structure and oscillation behavior of the star. The $p_1$-modes are primarily driven by pressure as the dominant restoring force. The $p_1$-mode frequency generally lies in the range of 5--7\,kHz~\cite{Kunjipurayil:2022zah,Rather:2024nry,Thakur:2024ijp}. However, due to their high frequencies, $p_1$-modes are generally not excited during NS mergers~\cite{Kokkotas:1999bd}. Interestingly, in the context of PNSs, the entropy and the lepton fraction cause a reduction in the $p_1$-mode frequencies, making their behavior particularly noteworthy — this aspect will be discussed below. \Cref{p_frequency_mass} presents the behavior of the $p_1$-mode in NSs, showing both the oscillation frequency (left panel) and the damping rate $1/\tau$ (right panel) illustrating the variation of the imaginary part of the oscillation frequency, which corresponds to the inverse damping time, $1/\tau$, computed within the full GR framework. Each curve corresponds to different internal conditions of the star, including temperature, composition, and lepton fraction, with a clear visual distinction made through color and linestyle. In both panels, solid lines represent stars composed of nucleonic matter, while dashed lines correspond to configurations that include hyperons. The black lines show results for cold NSs serving as a baseline. In contrast, colored lines correspond to PNS conditions at different $s_B$ and $Y_l$. Specifically, the red line denotes the first stage ($s_B = 1, Y_l = 0.4$), the blue line corresponds to the second stage ($s_B = 2, Y_l = 0.2$), and the green line represents the third stage  ($s_B = 2, Y_{\nu_e} = 0.0$) when all the neutrinos have escaped from the stellar core.

In the left panel, the $p_1$-mode frequency increases with stellar mass across all configurations. For cold NSs (shown in black), the frequency reaches significantly higher values, peaking around 6.4\,kHz, before slightly decreasing at higher masses. In contrast, hot PNS models exhibit noticeably lower frequencies. Among the configurations studied, those corresponding to the second and third evolutionary stages exhibit the lowest $p_1$-mode frequencies. This is primarily due to the expansion of the PNS during deleptonization, which reduces its compactness and weakens the restoring pressure forces that drive $p$-mode oscillations. Additionally, the adiabatic sound speed within the star decreases with increasing $s_B$, as a higher $s_B$ is associated with enhanced temperatures. This thermal softening of the EoS further contributes to the lowering of oscillation frequencies. These effects are observed both in purely nucleonic matter and in matter that includes hyperonic components. Comprehensive discussions on how stellar composition, thermal effects, and neutrino trapping influence the stellar structure and its oscillatory properties can be found in Refs.~\cite{Ferrari:2002ut, Pons:1998mm}. 
This observed trend is further corroborated by the data presented in \Cref{tab:combined_nucleon}, where the column labeled $p_1$ (GR) lists the $p_1$-mode frequencies calculated for a $1.4\,M_{\odot}$ configuration. Specifically, the frequencies for the second and third stages for nucleonic matter are 3.77\,kHz and 3.63\,kHz, respectively, affirming the reduction in frequency during deleptonization.

Another notable observation from \Cref{tab:combined_nucleon} is that the percentage error between the Cowling approximation and full GR is smaller for the first and last stages. Whereas, it increases in the higher $s_B$ cases for the second and the third stages 
contrary to the $f$-modes frequencies in \Cref{tab:model_frequencies_with_cowling}, where the Cowling approximation performs better at higher entropy second and third stages. Furthermore, the damping time of the $p_1$-mode, $\tau_{p_1}$, for cold NSs is significantly longer than that of the $f$-mode, consistent with the findings of Refs.~\cite{Kunjipurayil:2022zah, Thakur:2024ijp}. However, this damping time decreases substantially in the intermediate stages of the star's evolution, as shown in \Cref{tab:combined_nucleon}. Similar to the behavior of the $p_1$-mode frequency, the presence of hyperons has a minimal impact on the damping time.

\begin{table*}[ht]
  \centering
  \scriptsize
  \caption{Non-radial $p_1$-mode frequency with Cowling approximation (Cow), full (GR) in kHz, percentage error difference P. E., and damping time $\tau$ in sec., at 1.4\,$M_{\odot}$ for both nucleonic and hyperonic compositions, computed within the full GR framework. The PE column indicates the percentage by which the Cowling approximation overestimates the $p_1$-mode frequency.}
  \setlength{\tabcolsep}{4pt} 
  \renewcommand{\arraystretch}{1.5} 
  \begin{tabular}{ccccc}
    \hline
    Model & $p_1$ (Cow) & $p_1$ (GR) & P. E. (\%) & $\tau$ (sec) \\
    \hline
    \hline
    \multicolumn{5}{c}{Nucleons(Hyperons)} \\
    \hline
    \(s_B=1;\; Y_l=0.4\)     & 4.47 (4.48)   & 4.19 (4.19) & 6.64 (6.81)  & 5.205 (5.173) \\
    \(s_B=2;\; Y_l=0.2\)     & 4.19 (4.23)   & 3.77 (3.81) & 10.97 (11.02) & 4.693 (4.141)\\
    \(s_B=2;\; Y_{\nu_e}=0.0\) & 4.04 (4.10) & 3.63 (3.69) & 11.39 (11.11) & 4.511 (3.741) \\
    \(T=0\)                      & 6.32 (6.32) & 5.93 (5.93) & 6.51 (6.51)  & 11.856 (11.997) \\
    \hline
  \end{tabular}
  \label{tab:combined_nucleon}
\end{table*}

\subsection{Radial Profiles}

\begin{figure*}[t!]
		\begin{minipage}[t]{0.47\textwidth}		 		
  \includegraphics[width=\textwidth]{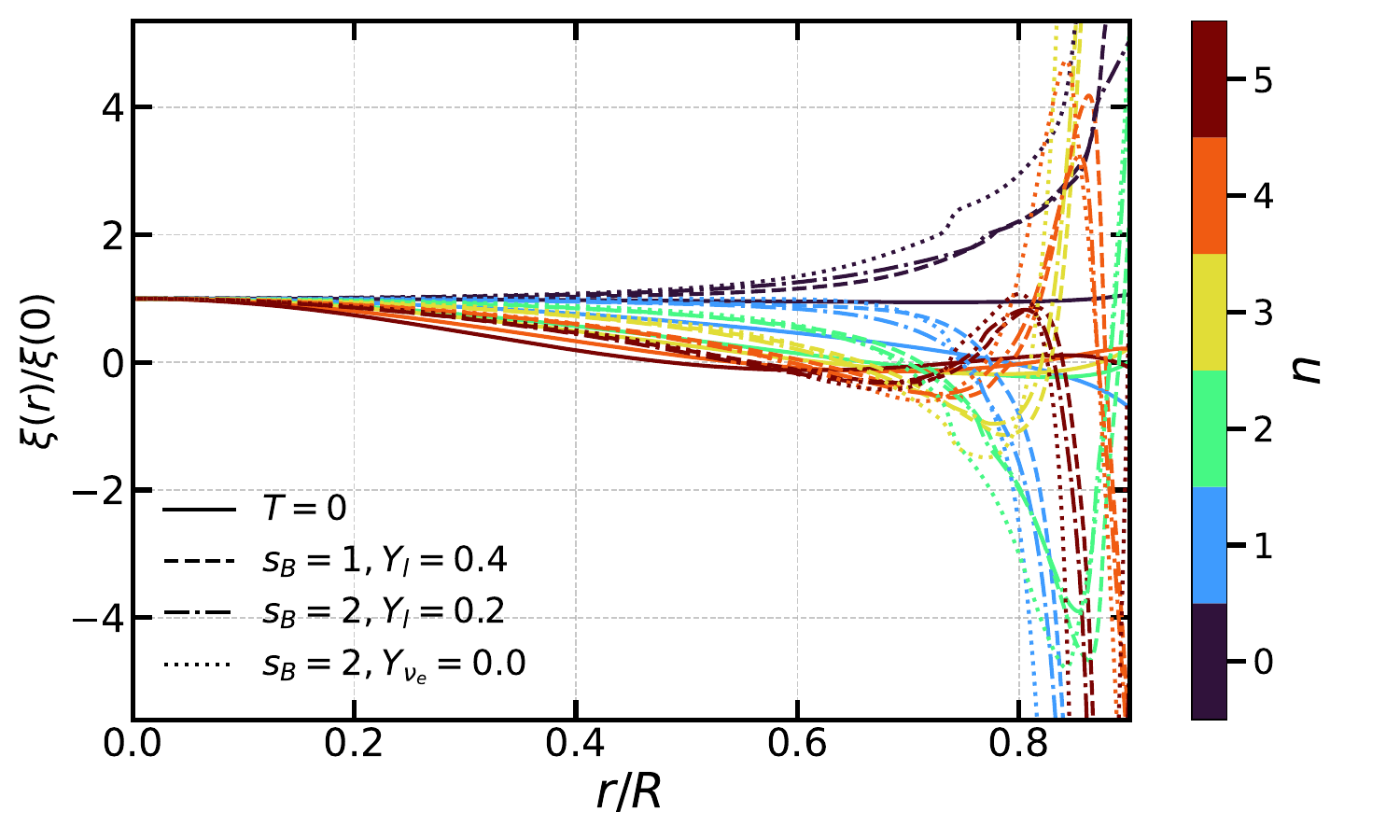}
			 	\end{minipage}
		 		\begin{minipage}[t]{0.47\textwidth}
			 		\includegraphics[width=\textwidth]{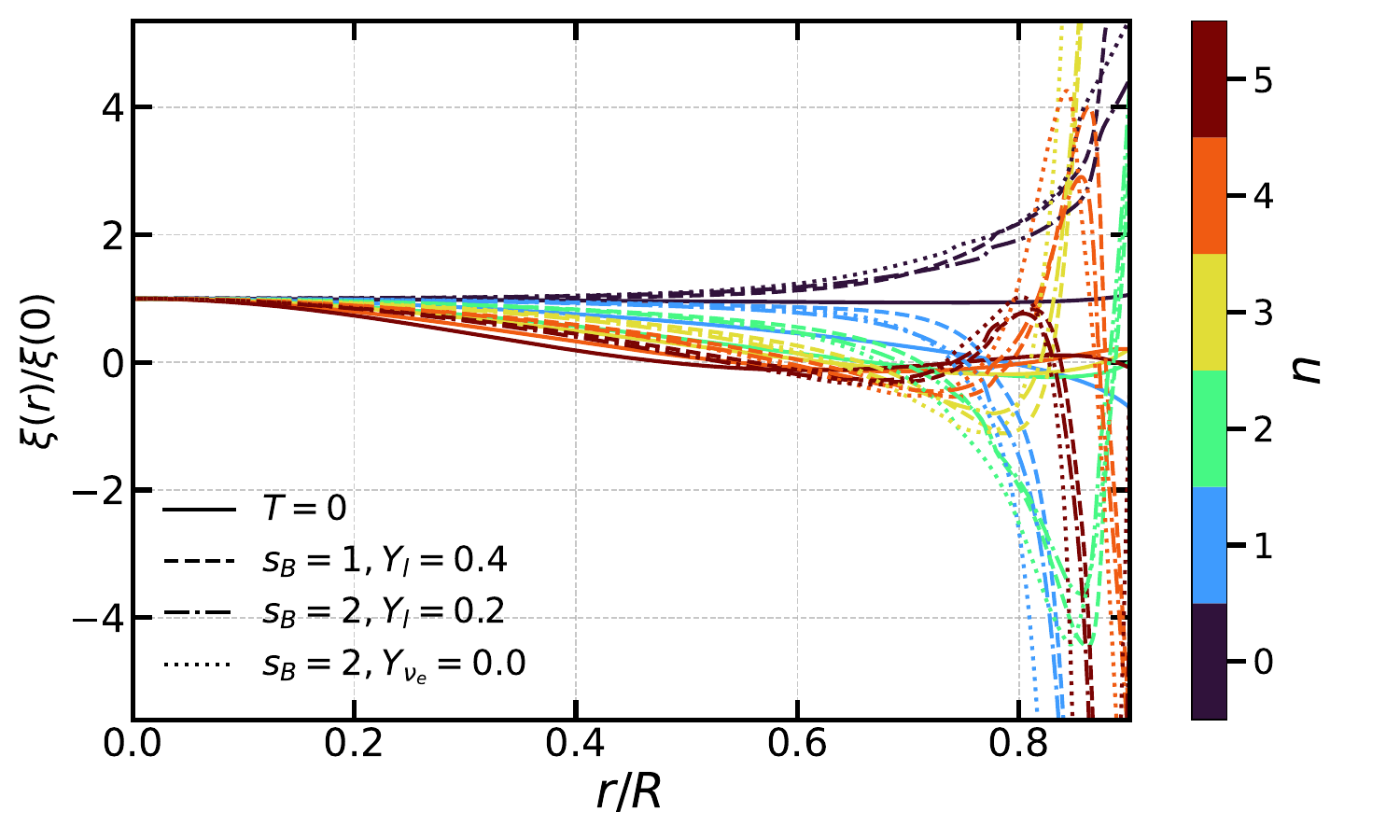}
			 	\end{minipage}
			 			\caption{Left: The radial displacement perturbation {$\xi(r)/\xi(0)$ = ${\Delta r/r}$} as a function of dimensionless radius distance ${r/R}$ for different stages of the PNS evolution. The colorbar represents the modes: ${f}$-mode ($n = 0$) and lower-order ${p}$-modes ($n = 1-5$) for pure nucleonic EoS. Right: Same as left plot, but for Hyperonic EoSs. All the modes are calculated at 1.4\,${M_{\odot}}$. }
		\label{fig:ksi}	 	
     \end{figure*}

\Cref{fig:ksi} presents the normalized radial displacement eigenfunctions $\xi(r)/\xi(0)$ for $1.4\,M_\odot$ NS models, illustrating how these stars oscillate in their various radial modes. The left plot depicts a purely nucleonic NS, while the second incorporates hyperons alongside nucleons in the stellar composition. Both plots display the dimensionless displacement function against normalized radius ($r/R$) for different combinations of $s_B$ and $Y_l$, as well as the cold and catalyzed NS. This perspective clearly shows the structural differences between the modes. The fundamental $f$-mode ($n=0$, dark purple) is nodeless, decreasing monotonically from 1 at the center to a positive value at the surface. The higher-order $p$-modes ($n=1, 2, \ldots$) are oscillatory, exhibiting an increasing number of nodes (zero-crossings) as $n$ increases. These nodes represent shells within the star that move in the opposite direction to their neighbors. 

A key thermal feature is immediately apparent: for the fundamental mode ($n=0$), the two intermediate-hot stages ($s_B=2$, $Y_l=0.2$ and $s_B=2$, $Y_{\nu_e}=0.0$), for the pure nucleonic case, are nearly indistinguishable, showing almost perfect overlap. The initial, relativel cooler stage ($s_B=1$, $Y_l=0.4$) and the final cold, catalyzed stage ($T=0$) are, however, distinct from the intermediate pair and from each other. This highlights that the $f$-mode's shape is highly sensitive to the star's thermal state and compactness.   
Since the intermediate stages are characterized by neutrino diffusion, elevated temperatures, and an expanded stellar radius, we can infer that the $\xi(r)/\xi(0)$ is sensitive to both the thermal state and the compactness of the star. Apart from the variations in thermal and pressure profiles driven by the changes in $s_B$  and $Y_l$, the speed of sound also evolves throughout the star. This evolution directly influences $\xi(r)$ through modifications in the internal structure and stratification of the stellar matter \cite{Gondek:1997fd}. Higher-order $p$-modes ($n= 1-5$) show decreasingly oscillatory behavior with additional nodes, effectively dividing the star into alternating regions of expansion and contraction. 

Comparing the nucleonic (left) and hyperonic (right) plots, the normalized $f$-modes ($n=0$) are similar. The EoS softening from hyperons, while affecting the star's global properties (like mass and radius), does not dramatically alter the shape of the fundamental displacement eigenfunction. The most notable differences appear in the higher-order $p$-modes ($n \ge 1$), where node positions and amplitudes near the surface are subtly shifted.

\begin{figure*}[t!]
		\begin{minipage}[t]{0.47\textwidth}		 		
  \includegraphics[width=\textwidth]{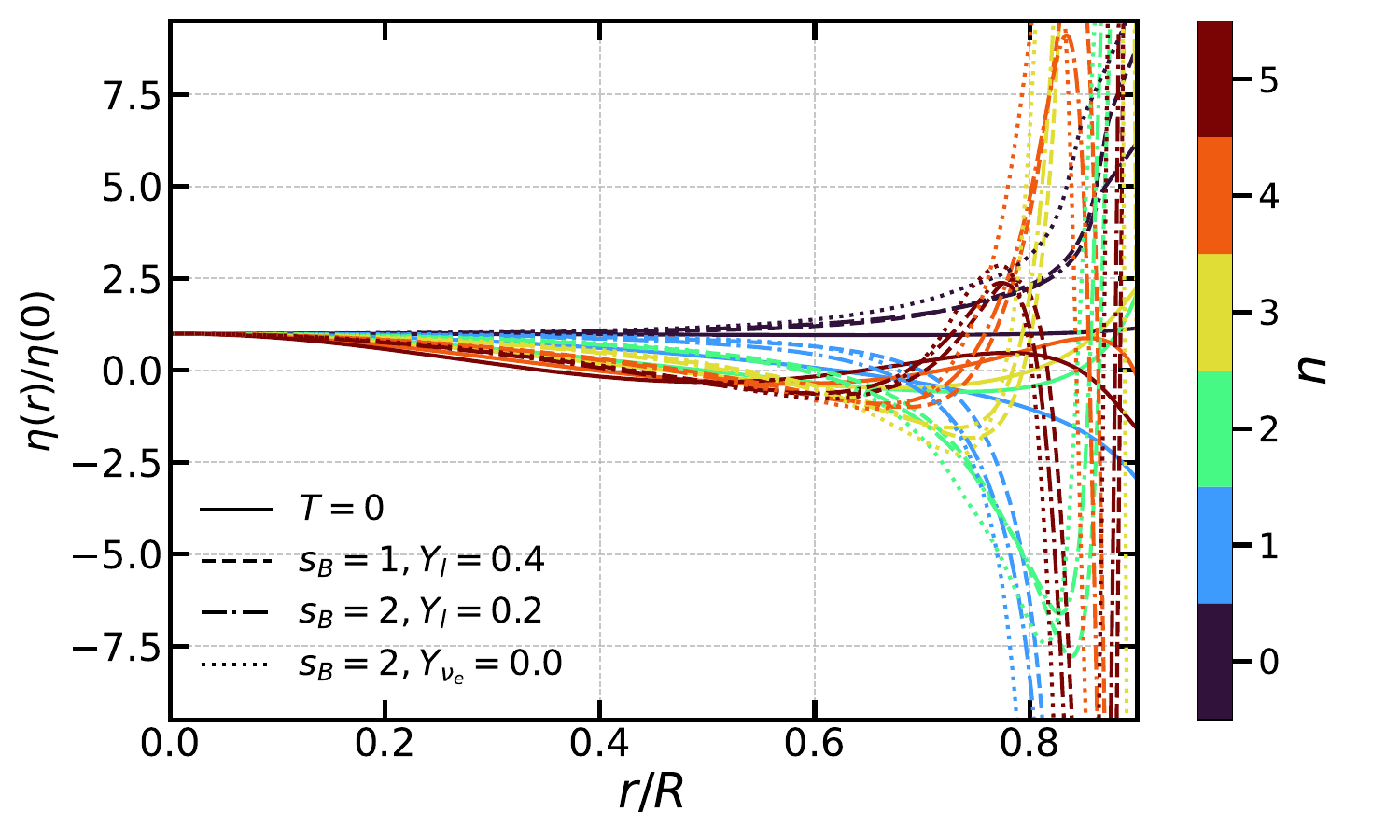}
			 	\end{minipage}
		 		\begin{minipage}[t]{0.47\textwidth}
			 		\includegraphics[width=\textwidth]{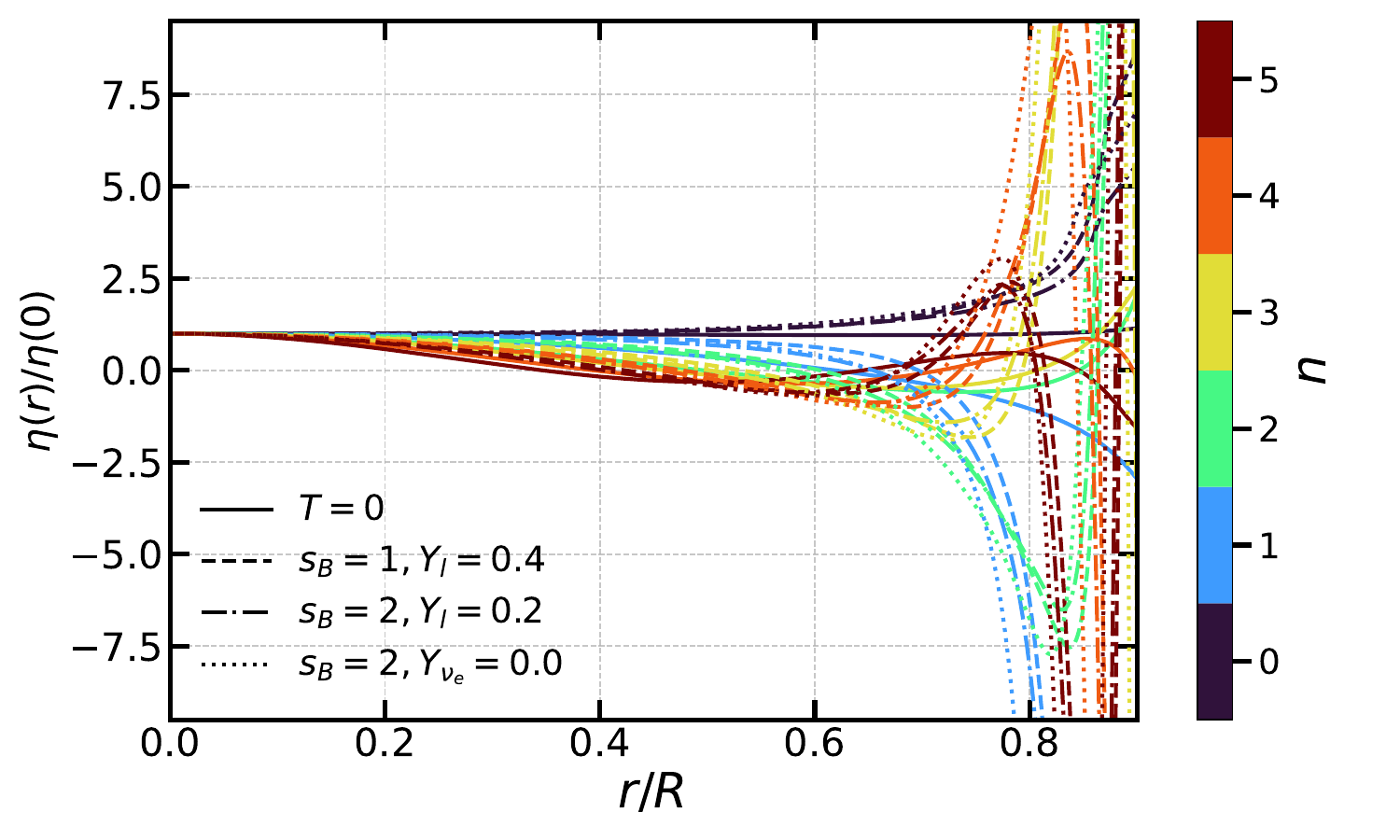}
			 	\end{minipage}
			 			\caption{Left: Same as Figure \ref{fig:ksi}, but for the radial pressure perturbation {${\eta(r)/\eta(0)}$ = ${\Delta P/P}$} as a function of dimensionless radius distance ${r/R}$.} 
		\label{fig:eta}	 	
     \end{figure*}

\Cref{fig:eta} presents another crucial aspect of NS oscillations through the normalized Lagrangian pressure perturbation, $\eta(r)/\eta(0)$, plotted against normalized radius ($r/R$) for $1.4\,M_\odot$ NS models. While the previously discussed $\xi(r)/\xi(0)$ function represented radial displacement, $\eta(r)$ characterizes the Lagrangian pressure perturbation, offering complementary insights into the stellar pulsation mechanics. The left plot depicts the $\eta(r)/\eta(0)$ profiles for a purely nucleonic NS model, while the right plot shows the corresponding profiles for an NS incorporating hyperons alongside nucleons. The amplitude for the $s_B = 2, Y_{\nu_e} =0.0$ case is always higher for both nucleons as well as hyperons compared to the other stages of the PNS evolution. For intermediate stages of $s_B = 2, Y_l =0.2$ and $s_B = 2, Y_{\nu_e} =0.0$, we can see that the fundamental $f$-mode ($n=0$) overlaps for pure nucleonic cases, similar to the $\xi(r)/\xi(0)$ as seen in \Cref{fig:ksi}. The $\eta(r)$ profiles display distinctive features compared to the $\xi(r)/\xi(0)$ functions, with more pronounced negative amplitudes near the stellar core and complex oscillatory behavior that intensifies with increasing mode order. The introduction of hyperons (right plot) again has only a very subtle impact on the shapes of the normalized eigenfunctions, especially for the dominant $f$-mode.

 For comparison of how the profiles change for different stages of PNSs at different fixed masses, we also computed the radial perturbations at $2.0\,M_\odot$. The eigenfunctions $\xi(r)$ are more compact and exhibit lower amplitudes compared to the $1.4\,M_\odot$ case, reflecting stronger gravitational binding. The effect of hyperons is more pronounced at higher mass, further damping the oscillations, especially in higher-order modes. Node positions shift slightly outward, and differences between thermal stages become less distinct, indicating that increased compactness dominates over thermal effects in shaping radial oscillations. Compared to the $1.4\,M_\odot$ models, the $2.0\,M_\odot$ NSs show more pronounced oscillatory behavior in $\eta(r)$, likely due to increased compactness and stronger gravitational binding. While the core amplitude range remains nearly unchanged, the overall profiles display enhanced structure and more complex mode features. The cold, catalyzed state continues to exhibit the highest amplitudes, and the overlap of fundamental modes in intermediate stages persists, indicating robust behavior across mass ranges. Hyperonic effects remain significant, though potentially more pronounced at higher central densities.

\begin{figure}[t]	 
\centering
  \includegraphics[width=0.45\textwidth]{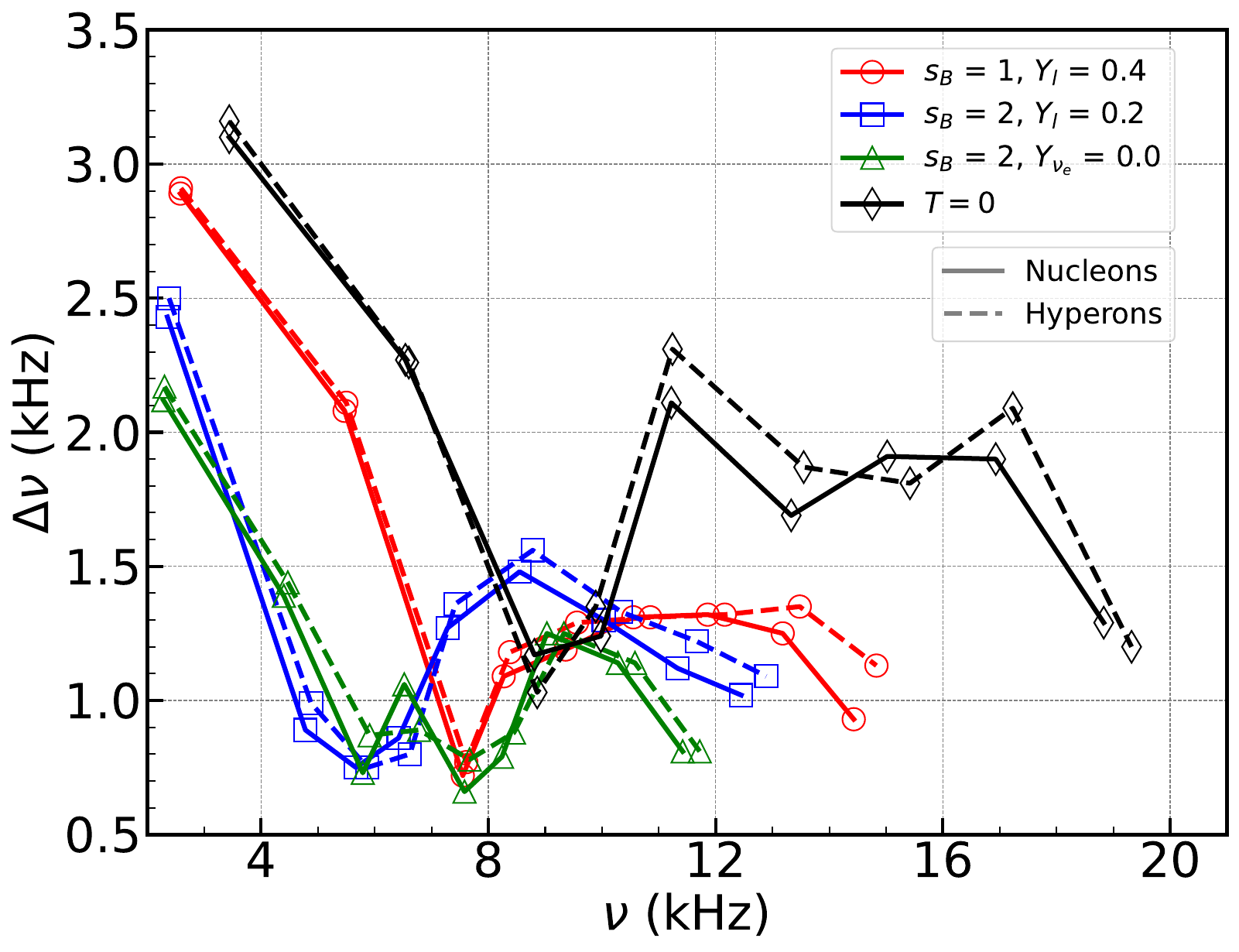}
			 			\caption{Frequency difference ${\Delta \nu_n}$ = ${\nu_{n+1}}$ - ${\nu_n}$ versus ${\nu_n}$ in kHz for several configurations considered in the text. Solid (Dasehd) lines correspond to the EoS with nucleons (hyperons) only.  }
		\label{fig:f_sep}	 	
     \end{figure} 

The eigenfunctions shown in Figures \Cref{fig:ksi} and \ref{fig:eta} satisfies the boundary condition that the Lagrangian pressure perturbation, $\Delta P$, vanishes at the stellar surface ($r=R$). Near the surface, the amplitudes of successive modes $\eta_n(r)$ and $\eta_{n+1}(r)$ grow, but they oscillate with opposite signs. Their cancellation ensures the boundary condition holds. As a consequence, the differences $\eta_{n+1}-\eta_n$ and $\xi_{n+1}-\xi_n$ are most responsive to variations in the stellar core. This heightened sensitivity implies that the frequency gap $\Delta\nu_n = \nu_{n+1} - \nu_n$ can act as a key observational marker of the star’s innermost layers and its crust-core transition.

\Cref{fig:f_sep} shows the frequency separation, $\Delta\nu_{n} = \nu_{n+1} - \nu_{n}$, versus the mode frequency $\nu_{n}$ ($n=0-9$) for the $1.4~M_{\odot}$ configurations. The cold, catalyzed star ($T=0$) has a large, clearly distinct $\Delta\nu_n$ at low frequencies (starting $\Delta\nu \approx 3.1$ kHz). The hot PNS stages are clustered at lower values. Among these, the hottest, most expanded stages ($s_B=2$) show the lowest $\Delta\nu_n$, while the relatively cooler, neutrino-trapped stage ($s_B=1$) is intermediate. This shows that $\Delta\nu_n$ is a sensitive probe of the star's thermal profile and compactness during its evolution. The nucleonic models (solid lines) display a relatively smooth, quasi-periodic variation in $\Delta\nu_n$. This pattern is characteristic of a standard star, with the wiggles arising from the sound wave resonating in the crust and core.

In sharp contrast, the hyperon-containing scenario (dashed lines) introduces a fundamentally different landscape, characterized by pronounced non-linear behaviors and unexpected curvatures. This is a clear signature of EoS softening. These strong ``avoided crossings'' are caused by an enhanced interaction, or resonance, between the core-dominated $p$-modes and modes trapped in the stellar crust. The softer hyperonic EoS alters the density profile and sound speed, strengthening this crust-core coupling and producing the distinct fluctuating pattern \cite{Sagun:2020qvc, Rather:2023dom, Rather:2024hmo, Sun:2024vdt, Routaray:2022utr}.

The observed variations in the frequency separation $\Delta \nu$ as $\nu$ increases can be mainly attributed to the transition between the crust and the core of the NS, as discussed in Refs.~\cite{Rather:2023dom,Rather:2024hmo,Sen:2022kva,Routaray:2022utr, Glendenning:1997wn, haensel2006neutron, Sagun:2020qvc, Thakur:2025zhi}. In the crust, where the density is relatively low and an inhomogeneous mixture of nuclei and free neutrons prevails, this composite structure significantly influences the oscillation modes. Once the density exceeds a critical threshold, matter becomes uniform in the core, and the resulting crust–core boundary gives rise to mode coupling or conversion effects that lead to fluctuations in the oscillation frequencies. The extent of these fluctuations is strongly dependent on the assumed crust–core transition density, which sets the effective boundary conditions for the modes. At finite temperature, the transition remains smooth, with the main change being a shift in the transition density rather than the qualitative nature of the boundary. These effects are visible between each separation for all the cases in Figure~\ref{fig:f_sep}. Softer EoSs, which generate steeper density gradients and yield more compact stars, tend to amplify the interaction between crustal and global modes, thereby producing more pronounced fluctuations in $\Delta \nu$. Conversely, stiffer EoSs produce smoother transitions in density and support larger masses and radii, which shifts the oscillation modes and typically results in a reduced frequency separation. For comparison, EoSs constructed without an explicit crustal part exhibit a smooth trend in $\Delta \nu$, highlighting the role of the crust in shaping the oscillation modes, as discussed in Refs.~\cite{Sagun:2020qvc,Panotopoulos:2017eig, Routaray:2022utr, Rather:2023tly}.

PNSs undergo a series of evolutionary stages that further influence these oscillation characteristics. Each stage, from the high-entropy, neutrino-dominated state to the cooler, neutrino-transparent configuration, introduces changes in the density gradient and compositional profile, thereby affecting the coupling between crustal and core oscillation modes. Consequently, the frequency separation $\Delta \nu$ not only reflects the static properties of the stellar interior but also serves as a dynamic diagnostic tool for probing the evolving structure of the PNS during its early life.
     
\begin{figure}[t]
\centering
  \includegraphics[width=0.45\textwidth]{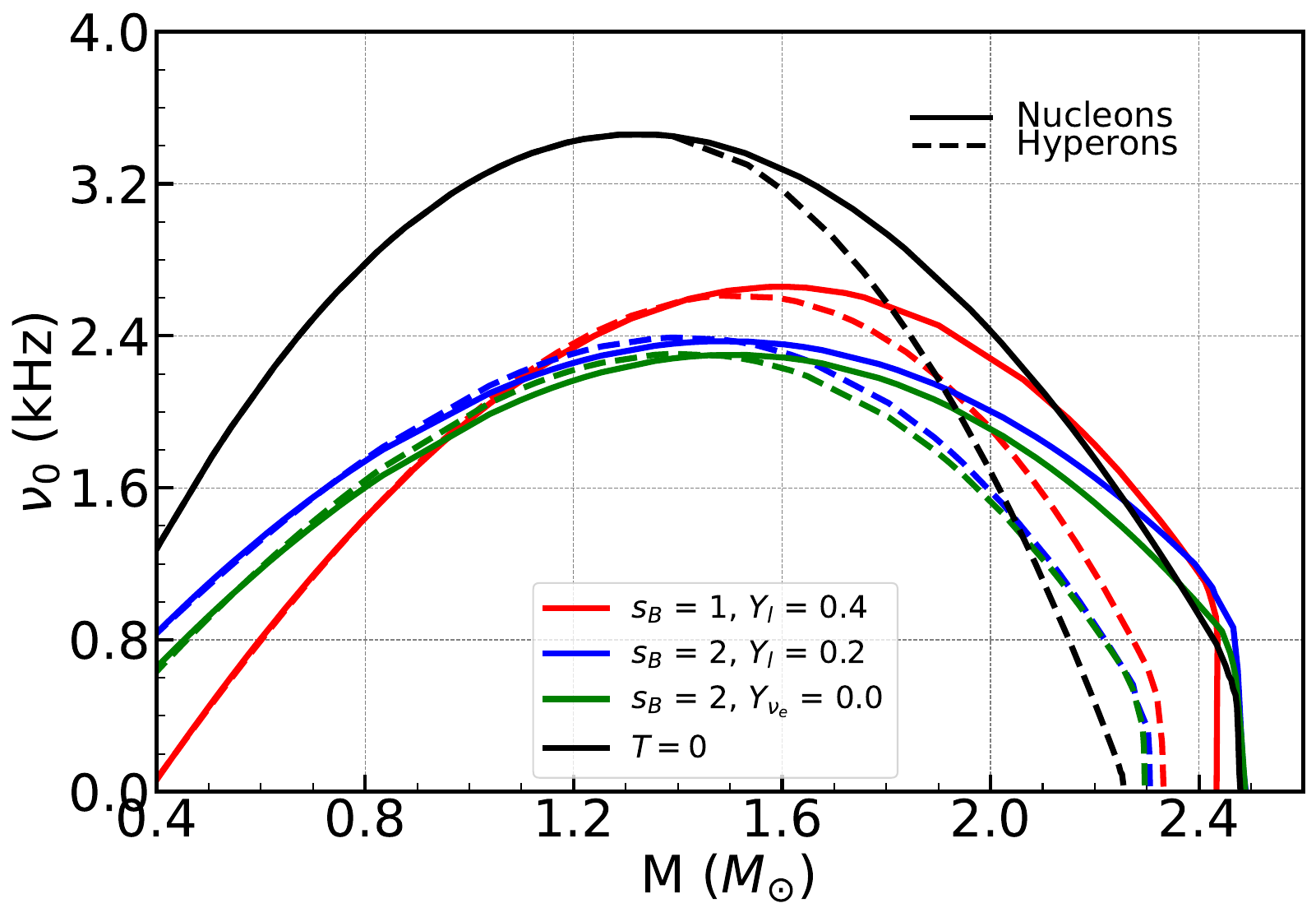}
			 			\caption{Radial $f$-mode ($n = 0$) frequency as a function of mass for different stages of PNS evolution. The solid (dashed) lines correspond to the pure nucleonic (hyperonic) EoSs. }
		\label{fig:fM}	 	
     \end{figure}   
\Cref{fig:fM} illustrates how the fundamental radial oscillation frequency, $\nu_0$, of the PNSs varies with stellar mass under different evolutionary conditions. Three distinct thermal stages are represented, corresponding to specific combinations of $s_B$ and $Y_l$, and are compared to the final cold, catalyzed ($T=0$) NS configuration. The fundamental frequency, which is sensitive to the star’s internal structure and the underlying EoS, varies with mass as well as the star's thermal and compositional profiles.

The solid and dashed lines in each pair correspond to the pure nucleonic and hyperonic EoSs, respectively. At low stellar masses (e.g., $M \lesssim 0.8~M_{\odot}$ for $s_B = 2, Y_l =0.2$ and $s_B = 2, Y_\nu =0.0$), the curves for both compositions are coincident for each thermodynamic scenario. This is the expected physical behavior, as the central densities of these stars are too low to trigger the onset of hyperons.

As the stellar mass increases, the curves for the hyperonic models (dashed lines) begin to diverge from their nucleonic counterparts (solid lines). This divergence occurs precisely when the central density becomes high enough for hyperons to appear, which softens the EoS. This softening leads to a more compact stellar configuration (a smaller radius $R$ for a given mass $M$). Because the fundamental frequency scales with the average density ($\nu_0 \propto \sqrt{M/R^3}$), this increased compactness results in slightly \textit{higher} fundamental radial oscillation frequencies for the hyperonic models compared to the purely nucleonic ones at the same mass. This effect becomes more pronounced at higher masses where the hyperon fraction is larger, highlighting the significant impact of hyperons on the star's structure.

Finally, as demonstrated in earlier investigations \cite{Sun:2021cez,Routaray:2022utr,Rather:2024hmo, Gondek:1997fd, Thakur:2025zhi}, the $f$-mode frequency $\nu_0$ for every model rapidly approaches zero near the point where the maximum numerical NS mass is reached. This behavior is in agreement with the stability criterion, $\partial M/\partial \varepsilon_c > 0$, where $\varepsilon_c$ is the central energy density. The vanishing of the fundamental frequency correctly identifies the onset of radial instability.

\begin{table}[ht]
\centering
\scriptsize
		\caption{10 lowest-order radial oscillation frequencies, ${\nu}$ in (kHz) for different compositions with nucleons (hyperons). For each EoS, the frequencies are calculated at 1.4\,${M_{\odot}}$.\label{table:radialf} }
\begin{tabular}{ ccccc }
 \hline
Order  $n$
  & $T = 0$ & $s_B = 1;\;  \newline Y_l = 0.4$ & $s_B = 2;\; \newline  Y_l = 0.2$ & $s_B = 2;\; \newline  Y_{\nu_e} = 0.0$ \\
\hline 
0   & 3.44 (3.44) & 2.58 (2.59) & 2.35 (2.38) & 2.28 (2.30)\\
1	&6.54 (6.60) & 5.47 (5.50) & 4.78 (4.88) & 4.40 (4.47)\\
2	&8.81 (8.86) & 7.55 (7.61) & 5.67 (5.87) & 5.79 (5.91)\\
3	&9.98 (9.89) & 8.27 (8.38) & 6.42 (6.62) & 6.52 (6.78)\\
4	&11.22 (11.24) & 9.36 (9.56) & 7.28 (7.42) & 7.58 (7.67)\\
5	&13.33 (13.55) & 10.55 (10.85) & 8.55 (8.78) & 8.24 (8.45)\\
6	&15.02 (15.42) & 11.86 (12.16) & 10.03 (10.34) & 9.03 (9.33)\\
7	&16.93 (17.23) & 13.18 (13.48) & 11.33 (11.67) & 10.28 (10.58)\\
8	&18.83 (19.32) & 14.43 (14.83) & 12.45 (12.89) & 11.42 (11.72)\\
9	&20.12 (20.52) & 15.36 (15.96) & 13.47 (13.98) & 12.23 (12.53)\\
\hline 
 \hline
\end{tabular}
\end{table}


\begin{table*}
\centering
\scriptsize

\caption{Gravitational wave energy estimates for various evolutionary stages of a PNS with a mass of $1.4\,M_{\odot}$ are presented at different source distances. The associated oscillation frequencies (in~kHz) and damping times (in~sec) are also provided for a range of spectral noise densities.}
\begin{tabular}{ccccc|ccc}
\hline
\multirow{2}{*}{Distance} & \multirow{2}{*}{Model} & \multicolumn{3}{c|}{$S_n = 2 \times 10^{-23}~\mathrm{Hz}^{-1}$} & \multicolumn{3}{c}{$S_n = 1 \times 10^{-24}~\mathrm{Hz}^{-1}$} \\
\cline{3-8}
& & $f$ (kHz) & $\tau$ (s) & $E_{\text{GW}}/(M_\odot c^2)$ & $f$ (kHz) & $\tau$ (s) & $E_{\text{GW}}/(M_\odot c^2)$ \\
\hline
\hline
\multicolumn{8}{c}{$f$-mode with Nucleons} \\
\hline
\hline 
\multirow{4}{*}{10 kpc}  
& $(s_B = 1;\; Y_l = 0.4)$   & 1.57 & 0.28 & $8.5 \times 10^{-8}$ & 1.57 & 0.28 & $2.13 \times 10^{-10}$ \\
&  $(s_B = 2;\; Y_l = 0.2)$   & 1.45 & 0.33 & $7.2 \times 10^{-8}$ & 1.45 & 0.33 & $1.82 \times 10^{-10}$ \\
&  $(s_B = 2;\; Y_{\nu_e} = 0.0)$ & 1.46 & 0.33 & $7.3 \times 10^{-8}$ & 1.46 & 0.33 & $1.84 \times 10^{-10}$ \\
&  $(T = 0)$                      & 1.60 & 0.28 & $8.8 \times 10^{-8}$ & 1.60 & 0.28 & $2.22 \times 10^{-10}$ \\
\hline

\multirow{4}{*}{100 kpc}  
& $(s_B = 1;\; Y_l = 0.4)$   & 1.57 & 0.28 & $8.5 \times 10^{-6}$ & 1.57 & 0.28 & $2.13 \times 10^{-8}$ \\
&  $(s_B = 2;\; Y_l = 0.2)$   & 1.45 & 0.33 & $7.2 \times 10^{-6}$ & 1.45 & 0.33 & $1.82 \times 10^{-8}$ \\
&  $(s_B = 2;\; Y_{\nu_e} = 0.0)$ & 1.46 & 0.33 & $7.3 \times 10^{-6}$ & 1.46 & 0.33 & $1.84 \times 10^{-8}$ \\
&  $(T = 0)$                      & 1.60 & 0.28 & $8.8 \times 10^{-6}$ & 1.60 & 0.28 & $2.22 \times 10^{-8}$ \\
\hline

\multirow{4}{*}{15 Mpc}  
& $(s_B = 1;\; Y_l = 0.4)$   & 1.57 & 0.28 & $1.92 \times 10^{-1}$ & 1.57 & 0.28 & $4.81 \times 10^{-4}$ \\
&  $(s_B = 2;\; Y_l = 0.2)$   & 1.45 & 0.33 & $1.64 \times 10^{-1}$ & 1.45 & 0.33 & $4.10 \times 10^{-4}$ \\
&  $(s_B = 2;\; Y_{\nu_e} = 0.0)$ & 1.46 & 0.33 & $1.66 \times 10^{-1}$ & 1.46 & 0.33 & $4.16 \times 10^{-4}$ \\
&  $(T = 0)$                      & 1.60 & 0.28 & $1.99 \times 10^{-1}$ & 1.60 & 0.28 & $4.99 \times 10^{-4}$ \\
\hline
\hline
\multicolumn{8}{c}{$p_1$-mode with Nucleons} \\
\hline
\hline 
\multirow{4}{*}{10 kpc}  
& $(s_B = 1;\; Y_l = 0.4)$   & 4.19 & 5.20  & $6.0 \times 10^{-7}$ & 4.19 & 5.20  & $1.52 \times 10^{-9}$  \\
&  $(s_B = 2;\; Y_l = 0.2)$   &  3.77 & 4.69  & $5.0 \times 10^{-7}$ &  3.77 & 4.69  & $1.23 \times 10^{-9}$ \\
&  $(s_B = 2;\; Y_{\nu_e} = 0.0)$ &  3.62 & 4.51  & $4.5 \times 10^{-7}$ & 3.62 & 4.51  & $1.13 \times 10^{-9}$ \\
&  $(T = 0)$                      & 5.93 & 11.85 & $1.2 \times 10^{-6}$ & 5.93 & 11.85 & $3.05 \times 10^{-9}$ \\
\hline

\multirow{4}{*}{100 kpc}  
& $(s_B = 1;\; Y_l = 0.4)$   & 4.19 & 5.20  & $6.0 \times 10^{-5}$ &  4.19 & 5.20  & $1.52 \times 10^{-7}$ \\
&  $(s_B = 2;\; Y_l = 0.2)$ & 3.77 & 4.69  & $5.0 \times 10^{-5}$ & 3.77 & 4.69  & $1.23 \times 10^{-7}$ \\
&  $(s_B = 2;\; Y_{\nu_e} = 0.0)$ & 3.62 & 4.51  & $4.5 \times 10^{-5}$ & 3.62 & 4.51  & $1.13 \times 10^{-7}$ \\
&  $(T = 0)$    & 5.93 & 11.85 & $1.2 \times 10^{-4}$ & 5.93 & 11.85 & $3.05 \times 10^{-7}$                   \\
\hline

\multirow{4}{*}{15 Mpc}  
& $(s_B = 1;\; Y_l = 0.4)$   & 4.19 & 5.20  & $1.37$ & 4.19 & 5.20  & $3.42 \times 10^{-3}$\\
&  $(s_B = 2;\; Y_l = 0.2)$   & 3.77 & 4.69  & $1.11$ & 3.77 & 4.69  & $2.77 \times 10^{-3}$\\
&  $(s_B = 2;\; Y_{\nu_e} = 0.0)$ & 3.62 & 4.51  & $1.02$ & 3.62 & 4.51  & $2.55 \times 10^{-3}$\\
&  $(T = 0)$                      & 5.93 & 11.85 & $2.74$ & 5.93 & 11.85 & $6.80 \times 10^{-3}$\\
\hline 
\end{tabular}
\label{CombinedTable}
\end{table*}

\subsection{Prospects of GW detection}

We extend our investigation to evaluate the gravitational wave (GW) energy emission in both the fundamental ($f$) and first pressure ($p_1$) modes for 1.4 $M_{\odot}$ for the different stages of the star's evolution. The GW energy is determined from the expression:
\begin{equation}
\begin{split}
\frac{E_{\text{GW}}}{M_\odot c^2} = 3.471 \times 10^{36} 
\left( \frac{S}{N} \right)^2 
\left( \frac{1 + 4Q^2}{Q^2} \right)
\left( \frac{D}{10\,\text{kpc}} \right)^2 \\
\times 
\left( \frac{f}{1\,\text{kHz}} \right)^2 
\left( \frac{S_n}{1\,\text{Hz}^{-1}} \right)
\end{split} \label{eq_EGW},
\end{equation}
 where $S/N$ signal-to-noise ratio, $E_{GW}$ is the total energy emitted in gravitational waves, ${M_\odot c^2}$ is the solar mass energy equivalent, $Q$ is the quality factor of the oscillation mode related to the dampping time, $D$ is the distance to the source in kiloparsecs (kpc), $f$ is the mode frequency, and $S_n$ is the spectral noise density. Using the formalism in Eq.~(\ref{eq_EGW}), we estimate the minimum energy required to achieve $S/N$ greater than $5$ for various configurations of $s_B$ and $Y_l$. Frequencies and damping times for each mode are extracted and used to compute the required GW energy for two representative spectral noise densities: $S_n = 2 \times 10^{-23}~\mathrm{Hz}^{-1}$, corresponding to the current sensitivity of Advanced LIGO/Virgo \cite{LIGOScientific:2016wof}, and $S_n = 1 \times 10^{-24}~\mathrm{Hz}^{-1}$, representing the capabilities of third-generation detectors like the Einstein Telescope \cite{Punturo:2010zz}, which are shown in \Cref{CombinedTable}.

For the $f$-mode, the oscillation frequencies range from 1.45 to 1.60 kHz, with damping times spanning 0.28 to 0.33 seconds across the different thermal configurations. At a distance of 10 kpc, the estimated GW energy required for detection ranges from $7.2 \times 10^{-8}$ to $8.8 \times 10^{-8}\,M_\odot c^2$ for $S_n = 2 \times 10^{-23}$, and drops significantly to values between $1.82 \times 10^{-10}$ and $2.22 \times 10^{-10}\,M_\odot c^2$ for $S_n = 1 \times 10^{-24}$. These energy levels are well within the range of expected GW emission from core-collapse supernova, indicating that $f$-mode signals from PNSs in our Galaxy are detectable with current and future detectors. For instance, in \cite{Andersen:2021vzo}, the authors found that the GW energy emitted during core collapse ranges from approximately \(10^{-9}\) to \(10^{-8} \, M_\odot c^2\), based on axisymmetric simulations and depending on the stiffness of the EoS. Similarly, in \cite{Mueller:2012sv}, the predicted GW energy falls within the range \(10^{-10}\) to \(10^{-8} \, M_\odot c^2\), depending on the mass of the progenitor star. Other studies reporting estimates within this range include \cite{Sotani:2017ubz, Andresen:2018aom} and references therein. At 100 kpc, the required energy increases by four orders of magnitude, reaching approximately $7.2 \times 10^{-6}$ to $8.8 \times 10^{-6}\,M_\odot c^2$ for current detectors, and $1.82 \times 10^{-8}$ to $2.22 \times 10^{-8}\,M_\odot c^2$ for third-generation detectors. Although this represents a more challenging detection scenario, the lower thresholds associated with advanced detectors may still permit observability under favorable conditions. At 15 Mpc, the energy thresholds become significantly more demanding: values of $1.64 \times 10^{-1}$ to $1.99 \times 10^{-1}\,M_\odot c^2$ are required for Advanced LIGO, which are prohibitively high. However, with the improved sensitivity of the Einstein Telescope, the required energy drops to between $4.10 \times 10^{-4}$ and $4.99 \times 10^{-4}\,M_\odot c^2$. While still high, such values are marginally within reach for extremely energetic or repeated events, suggesting that the $f$-mode from extragalactic sources may be detected by future observatories under exceptional circumstances.

The $p_1$ mode presents higher frequencies, ranging from 3.62 to 5.93 kHz, and significantly longer damping times between 4.51 and 11.85 seconds. As a result, the energy required for detection is generally higher. At 10 kpc, the required energy ranges from $4.5 \times 10^{-7}$ to $1.2 \times 10^{-6}\,M_\odot c^2$ for $S_n = 2 \times 10^{-23}$ and from $1.13 \times 10^{-9}$ to $3.05 \times 10^{-9}\,M_\odot c^2$ for $S_n = 1 \times 10^{-24}$. These values are close to the expected energy output from PNS oscillations following a supernova, making the detection of $p_1$-modes feasible, especially with third-generation detectors \cite{Andersen:2021vzo, Sotani:2017ubz, Andresen:2016pdt}. At 100 kpc, the required GW energy increases to between $4.5 \times 10^{-5}$ and $1.2 \times 10^{-4}\,M_\odot c^2$ for Advanced LIGO, and $1.13 \times 10^{-7}$ to $3.05 \times 10^{-7}\,M_\odot c^2$ for the Einstein Telescope. While detection becomes challenging at this distance with current instruments, advanced detectors may still succeed in capturing such signals from moderately distant sources. The scenario at 15 Mpc remains extremely demanding. Advanced LIGO's required energy ranges from $1.02$ to $2.74\,M_\odot c^2$, far exceeding realistic emission levels. For the Einstein Telescope, the energy requirement is reduced to $2.55 \times 10^{-3}$ to $6.80 \times 10^{-3}\,M_\odot c^2$, which, though still large, brings the possibility of detection marginally closer, particularly in the case of strong excitation mechanisms or multiple sources contributing to the signal.

In summary, the $f$ and $p_1$ mode oscillations of PNSs determined are within the detection limits of current GW observatories for Galactic sources, and future detectors like the Einstein Telescope are expected to considerably extend this reach, potentially up to the outskirts of the Local Group. However, for extragalactic distances such as 15 Mpc, only extremely energetic events or stacked signals from multiple PNSs might yield detectable gravitational wave signatures. These results reinforce the importance of continuous monitoring by highly sensitive instruments and highlight the diagnostic potential of GW observations in probing the thermal and compositional evolution of NSs shortly after their birth.


 \section{Summary and Outlook}
 \label{summary}

 The present work represents a natural extension of Ref.~\cite{Barman:2024zuo}, which investigated non-radial oscillations during similar stages of PNS evolution using fully GR calculations. However, the study did not incorporate neutrino effects and was limited to nucleonic matter, neglecting the presence of hyperons. Building upon this foundation, this study investigates the radial and non-radial oscillations during the evolution of a PNS, from its birth as a neutrino-rich object to its transition into a neutrino-transparent, cold, catalyzed NS. Our results show that the presence of trapped neutrinos at low $s_B$ increases the stellar compactness, as seen in the early post-bounce phase. As deleptonization proceeds and $s_B$ rises, the internal temperature increases, leading to an expanded stellar radius and reduced compactness, characteristic of the intermediate evolutionary stages. Once the neutrinos have fully escaped from the core, the star begins to contract, ultimately forming a cold NS at $T = 0$.

The structural evolution of PNSs into cold, catalyzed NSs, compared with observational mass, radius confidence contours, reveals that, beyond thermal effects, the inclusion of hyperons leads to a reduction in both stellar mass and radius, as expected \cite{Bednarek_2012}. Additionally, hyperons decrease the temperature distribution within the stellar matter by lowering the thermal energy per baryon due to their impact on the entire composition and thermodynamics of the EoS \cite{Pons:1998mm}.

By analyzing non-radial oscillations through the $f$- and $p_1$-modes (where $p_1$-mode refers to pressure mode) using both the Cowling approximation and the full GR approach, we observed that the Percentage Error (P.E.) in the $f$-mode frequencies is relatively minimal during the deleptonization phase, when the star is hot and expanded, thus favoring the Cowling approximation. While the Cowling approximation exhibits substantial systematic errors (16–20\%) across the evolutionary stages, particularly for $2.0\,\rm M_\odot$ stars, it remains valuable for identifying qualitative seismic trends. Its computational efficiency makes it useful for exploring parameter space, although full general relativity is required for accurate quantitative analyses. The decline in the P.E. in the intermediate stages suggests that gravitational perturbations are relatively insignificant during neutrino diffusion when the stellar matter is heated. As the star cools and reaches a catalyzed configuration, gravitational effects become more pronounced, leading to a higher P.E., since the Cowling approximation neglects these perturbations. In Ref.~\cite{Messios:2001br, Thapa:2023grg}, the authors provide an extensive comparison between the full GR framework and the Cowling approximation across various oscillation modes, offering valuable insights into the behavior and magnitude of the P.E. Additional studies on stellar oscillations, including both theoretical foundations and mode classifications, can be found in Ref.~\cite{Andersson:1997rn} and references therein. The results obtained in these references qualitatively agree with our findings.

In contrast, the analysis of the $p_1$-mode frequencies reveals a different trend. The P.E. is highest during the intermediate stages of the star’s evolution, while the lowest P.E. occurs when the star reaches the cold, catalyzed configuration. This discrepancy arises because $p_1$-modes are more sensitive to pressure gradients within the star. During the intermediate stages, the balance of temperature, pressure, and density becomes more complex as the star undergoes deleptonization, making it harder for both the Cowling and full GR approaches to accurately capture the density and pressure variations \cite{Kruger:2014pva}. As the star stabilizes and cools, the pressure and density gradients become more straightforward, allowing both approaches to more accurately describe the oscillations, resulting in a reduction in the P.E.. In Ref.~\cite{Ferrari:2002ut}, the authors demonstrate that thermal pressure and the stellar composition significantly alter the non-radial frequencies during deleptonization, in agreement with our findings.  Comparison between $f$- and $p_1$-modes for PNSs can also be found in \cite{Thapa:2023grg}. 

While the inclusion of hyperons modifies the EoS by softening it, leading to more compact stellar configurations, our analysis shows that the resulting changes in the $f$- and $p_1$-mode frequencies are relatively modest. This indicates that, within the mass range and thermal conditions considered in studying the stellar evolution, these modes are not strongly sensitive to variations in core compositions. The similarities observed between nucleonic and hyperonic models suggest that, although hyperons influence the overall structure, their effect on the non-radial oscillation spectrum remains limited. We note, however, that larger deviations may arise in more massive stars or in modes such as $g$-modes, which are more sensitive to composition gradients and thermal stratification—an investigation that lies beyond the scope of the present study.

We tested our EoSs against several universal relations to assess the influence of entropy and lepton fraction. From the behavior of the $f$-mode frequencies and damping times, we found that no existing universal relation accurately describes both PNSs through to their transition into cold, catalyzed NSs simultaneously. This underscores the need to incorporate temperature and neutrino effects when modeling gravitational-wave damping time. Motivated by this, we derived a new, more robust universal relation, presented in \Cref{eq:our_fit_eta} and illustrated in \Cref{universal_MOI}, by fitting to PNS data. This result accounts for the thermal effects and the neutron presence associated with PNSs and their evolution. Although this relation remains model-dependent until tested against a wider range of PNS models, it offers a more realistic description than those derived solely from cold NS configurations.

By analyzing the radial displacement $\xi(r)$ and pressure perturbation $\eta(r)$ eigenfunctions of radial profiles, we identified distinct signatures in the oscillation patterns associated with the thermal and compositional evolution of the star. Our results show that the inclusion of hyperons leads to a more compact configuration and hence higher oscillation frequencies. The evolution of the fundamental $f$-mode and its separation from higher $p$-modes reveal key structural changes, particularly near the {crust-core} interface, and provide a sensitive probe of the internal stellar dynamics.

The frequency separation $\Delta\nu$ exhibits non-linear behavior, especially in hyperonic models, offering insights into the softness/stiffness of the EoS and the crust–core coupling during PNS evolution. Furthermore, the correlation of the fundamental mode frequency $\nu_0$ with stellar mass highlights the stabilizing or destabilizing effects of thermal and compositional changes, with the $f$-mode frequency vanishing near the maximum mass limit, consistent with the stability criterion.

Finally, we estimate the GW energy required for detection within our model by combining the $f$- and $p_1$-mode non-radial frequencies of each stellar configuration with the spectral noise densities of Advanced LIGO/Virgo \cite{LIGOScientific:2016wof} and the next-generation Einstein Telescope \cite{Punturo:2010zz}, to evaluate the sensitivity of these detectors. The results indicate that the $f$- and $p_1$-mode oscillations of PNSs lie within the detection range of current GW observatories for galactic events, while next-generation detectors like the Einstein Telescope are expected to significantly extend this reach to extragalactic distances. 

As a future direction, extending this analysis to include more exotic phases of matter such as color-superconducting (CSC) quark phases \cite{Gholami:2024ety, Christian:2025dhe}, as well as the potential influence of dark matter \cite{Issifu:2024htq} on PNS oscillation spectra, would provide valuable insights into the internal dynamics and composition of NS. Exploring both radial and non-radial modes in hybrid stars and twin-star configurations will further enhance our understanding of the dense matter EoS and its astrophysical manifestations. These investigations may open new observational windows through gravitational wave asteroseismology and contribute to constraining the fundamental physics governing compact stars.


\section*{Acknowledgement}
A.I. acknowledges financial support from the São Paulo State Research Foundation (FAPESP), Grant No. 2023/09545-1. I.A.R. acknowledges support from the Alexander von Humboldt Foundation. T. F. also thanks the financial support from Improvement of Higher Education Personnel CAPES (Finance Code 001), the National Council for Scientific and Technological Development (CNPq) under Grants Nos. 306834/2022-7, and FAPESP (Grant 2019/07767-1). P.~Thakur and Y.~Lim is supported by the National Research Foundation of Korea (NRF) grant funded by the Korea government (MSIT) (No.~RS-2024-00457037). Y.~Lim is also supported by Global - Learning \& Academic research institution for Master's and PhD students, 
and Postdocs(LAMP) Program of the National Research Foundation of Korea(NRF) grant funded by the Ministry of Education(No.  RS-2024-00442483) and by the Yonsei University Research Fund of 2024-22-0121.

\bibliographystyle{apsrev4-112}
\bibliography{references}

\begin{thebibliography}{99}%
\makeatletter
\providecommand \@ifxundefined [1]{%
 \@ifx{#1\undefined}
}%
\providecommand \@ifnum [1]{%
 \ifnum #1\expandafter \@firstoftwo
 \else \expandafter \@secondoftwo
 \fi
}%
\providecommand \@ifx [1]{%
 \ifx #1\expandafter \@firstoftwo
 \else \expandafter \@secondoftwo
 \fi
}%
\providecommand \natexlab [1]{#1}%
\providecommand \enquote  [1]{``#1''}%
\providecommand \bibnamefont  [1]{#1}%
\providecommand \bibfnamefont [1]{#1}%
\providecommand \citenamefont [1]{#1}%
\providecommand \href@noop [0]{\@secondoftwo}%
\providecommand \href [0]{\begingroup \@sanitize@url \@href}%
\providecommand \@href[1]{\@@startlink{#1}\@@href}%
\providecommand \@@href[1]{\endgroup#1\@@endlink}%
\providecommand \@sanitize@url [0]{\catcode `\\12\catcode `\$12\catcode `\&12\catcode `\#12\catcode `\^12\catcode `\_12\catcode `\%12\relax}%
\providecommand \@@startlink[1]{}%
\providecommand \@@endlink[0]{}%
\providecommand \url  [0]{\begingroup\@sanitize@url \@url }%
\providecommand \@url [1]{\endgroup\@href {#1}{\urlprefix }}%
\providecommand \urlprefix  [0]{URL }%
\providecommand \Eprint [0]{\href }%
\providecommand \doibase [0]{http://dx.doi.org/}%
\providecommand \selectlanguage [0]{\@gobble}%
\providecommand \bibinfo  [0]{\@secondoftwo}%
\providecommand \bibfield  [0]{\@secondoftwo}%
\providecommand \translation [1]{[#1]}%
\providecommand \BibitemOpen [0]{}%
\providecommand \bibitemStop [0]{}%
\providecommand \bibitemNoStop [0]{.\EOS\space}%
\providecommand \EOS [0]{\spacefactor3000\relax}%
\providecommand \BibitemShut  [1]{\csname bibitem#1\endcsname}%
\let\auto@bib@innerbib\@empty
\bibitem [{\citenamefont {Janka}(2012)}]{Janka:2012wk}%
  \BibitemOpen
  \bibfield  {author} {\bibinfo {author} {\bibfnamefont {H.-T.}\ \bibnamefont {Janka}},\ }\href {\doibase 10.1146/annurev-nucl-102711-094901} {\bibfield  {journal} {\bibinfo  {journal} {Ann. Rev. Nucl. Part. Sci.}\ }\textbf {\bibinfo {volume} {62}},\ \bibinfo {pages} {407--451} (\bibinfo {year} {2012})},\ \Eprint {http://arxiv.org/abs/1206.2503} {arXiv:1206.2503 [astro-ph.SR]} \BibitemShut {NoStop}%
\bibitem [{\citenamefont {Kotake}\ \emph {et~al.}(2006)\citenamefont {Kotake}, \citenamefont {Sato},\ and\ \citenamefont {Takahashi}}]{Kotake:2005zn}%
  \BibitemOpen
  \bibfield  {author} {\bibinfo {author} {\bibfnamefont {K.}~\bibnamefont {Kotake}}, \bibinfo {author} {\bibfnamefont {K.}~\bibnamefont {Sato}}, \ and\ \bibinfo {author} {\bibfnamefont {K.}~\bibnamefont {Takahashi}},\ }\href {\doibase 10.1088/0034-4885/69/4/R03} {\bibfield  {journal} {\bibinfo  {journal} {Rept. Prog. Phys.}\ }\textbf {\bibinfo {volume} {69}},\ \bibinfo {pages} {971--1144} (\bibinfo {year} {2006})},\ \Eprint {http://arxiv.org/abs/astro-ph/0509456} {arXiv:astro-ph/0509456} \BibitemShut {NoStop}%
\bibitem [{\citenamefont {Raffelt}(1990)}]{Raffelt:1990yz}%
  \BibitemOpen
  \bibfield  {author} {\bibinfo {author} {\bibfnamefont {G.~G.}\ \bibnamefont {Raffelt}},\ }\href {\doibase 10.1016/0370-1573(90)90054-6} {\bibfield  {journal} {\bibinfo  {journal} {Phys. Rept.}\ }\textbf {\bibinfo {volume} {198}},\ \bibinfo {pages} {1--113} (\bibinfo {year} {1990})}\BibitemShut {NoStop}%
\bibitem [{\citenamefont {Raffelt}\ and\ \citenamefont {Seckel}(1988)}]{Raffelt:1987yt}%
  \BibitemOpen
  \bibfield  {author} {\bibinfo {author} {\bibfnamefont {G.}~\bibnamefont {Raffelt}}\ and\ \bibinfo {author} {\bibfnamefont {D.}~\bibnamefont {Seckel}},\ }\href {\doibase 10.1103/PhysRevLett.60.1793} {\bibfield  {journal} {\bibinfo  {journal} {Phys. Rev. Lett.}\ }\textbf {\bibinfo {volume} {60}},\ \bibinfo {pages} {1793} (\bibinfo {year} {1988})}\BibitemShut {NoStop}%
\bibitem [{\citenamefont {Hirata}\ \emph {et~al.}(1987)\citenamefont {Hirata} \emph {et~al.}}]{Kamiokande-II:1987idp}%
  \BibitemOpen
  \bibfield  {author} {\bibinfo {author} {\bibfnamefont {K.}~\bibnamefont {Hirata}} \emph {et~al.} (\bibinfo {collaboration} {Kamiokande-II}),\ }\href {\doibase 10.1103/PhysRevLett.58.1490} {\bibfield  {journal} {\bibinfo  {journal} {Phys. Rev. Lett.}\ }\textbf {\bibinfo {volume} {58}},\ \bibinfo {pages} {1490--1493} (\bibinfo {year} {1987})}\BibitemShut {NoStop}%
\bibitem [{\citenamefont {Hirata}\ \emph {et~al.}(1988)\citenamefont {Hirata} \emph {et~al.}}]{Hirata:1988ad}%
  \BibitemOpen
  \bibfield  {author} {\bibinfo {author} {\bibfnamefont {K.~S.}\ \bibnamefont {Hirata}} \emph {et~al.},\ }\href {\doibase 10.1103/PhysRevD.38.448} {\bibfield  {journal} {\bibinfo  {journal} {Phys. Rev. D}\ }\textbf {\bibinfo {volume} {38}},\ \bibinfo {pages} {448--458} (\bibinfo {year} {1988})}\BibitemShut {NoStop}%
\bibitem [{\citenamefont {Pons}\ \emph {et~al.}(1999)\citenamefont {Pons}, \citenamefont {Reddy}, \citenamefont {Prakash}, \citenamefont {Lattimer},\ and\ \citenamefont {Miralles}}]{Pons:1998mm}%
  \BibitemOpen
  \bibfield  {author} {\bibinfo {author} {\bibfnamefont {J.~A.}\ \bibnamefont {Pons}}, \bibinfo {author} {\bibfnamefont {S.}~\bibnamefont {Reddy}}, \bibinfo {author} {\bibfnamefont {M.}~\bibnamefont {Prakash}}, \bibinfo {author} {\bibfnamefont {J.~M.}\ \bibnamefont {Lattimer}}, \ and\ \bibinfo {author} {\bibfnamefont {J.~A.}\ \bibnamefont {Miralles}},\ }\href {\doibase 10.1086/306889} {\bibfield  {journal} {\bibinfo  {journal} {Astrophys. J.}\ }\textbf {\bibinfo {volume} {513}},\ \bibinfo {pages} {780} (\bibinfo {year} {1999})},\ \Eprint {http://arxiv.org/abs/astro-ph/9807040} {arXiv:astro-ph/9807040} \BibitemShut {NoStop}%
\bibitem [{\citenamefont {Prakash}\ \emph {et~al.}(1997)\citenamefont {Prakash}, \citenamefont {Bombaci}, \citenamefont {Prakash}, \citenamefont {Ellis}, \citenamefont {Lattimer},\ and\ \citenamefont {Knorren}}]{Prakash:1996xs}%
  \BibitemOpen
  \bibfield  {author} {\bibinfo {author} {\bibfnamefont {M.}~\bibnamefont {Prakash}}, \bibinfo {author} {\bibfnamefont {I.}~\bibnamefont {Bombaci}}, \bibinfo {author} {\bibfnamefont {M.}~\bibnamefont {Prakash}}, \bibinfo {author} {\bibfnamefont {P.~J.}\ \bibnamefont {Ellis}}, \bibinfo {author} {\bibfnamefont {J.~M.}\ \bibnamefont {Lattimer}}, \ and\ \bibinfo {author} {\bibfnamefont {R.}~\bibnamefont {Knorren}},\ }\href {\doibase 10.1016/S0370-1573(96)00023-3} {\bibfield  {journal} {\bibinfo  {journal} {Phys. Rept.}\ }\textbf {\bibinfo {volume} {280}},\ \bibinfo {pages} {1--77} (\bibinfo {year} {1997})},\ \Eprint {http://arxiv.org/abs/nucl-th/9603042} {arXiv:nucl-th/9603042} \BibitemShut {NoStop}%
\bibitem [{\citenamefont {Vartanyan}\ \emph {et~al.}(2019)\citenamefont {Vartanyan}, \citenamefont {Burrows}, \citenamefont {Radice}, \citenamefont {Skinner},\ and\ \citenamefont {Dolence}}]{Vartanyan:2018iah}%
  \BibitemOpen
  \bibfield  {author} {\bibinfo {author} {\bibfnamefont {D.}~\bibnamefont {Vartanyan}}, \bibinfo {author} {\bibfnamefont {A.}~\bibnamefont {Burrows}}, \bibinfo {author} {\bibfnamefont {D.}~\bibnamefont {Radice}}, \bibinfo {author} {\bibfnamefont {A.~M.}\ \bibnamefont {Skinner}}, \ and\ \bibinfo {author} {\bibfnamefont {J.}~\bibnamefont {Dolence}},\ }\href {\doibase 10.1093/mnras/sty2585} {\bibfield  {journal} {\bibinfo  {journal} {Mon. Not. Roy. Astron. Soc.}\ }\textbf {\bibinfo {volume} {482}},\ \bibinfo {pages} {351--369} (\bibinfo {year} {2019})},\ \Eprint {http://arxiv.org/abs/1809.05106} {arXiv:1809.05106 [astro-ph.HE]} \BibitemShut {NoStop}%
\bibitem [{\citenamefont {Camelio}\ \emph {et~al.}(2017)\citenamefont {Camelio}, \citenamefont {Lovato}, \citenamefont {Gualtieri}, \citenamefont {Benhar}, \citenamefont {Pons},\ and\ \citenamefont {Ferrari}}]{Camelio:2017nka}%
  \BibitemOpen
  \bibfield  {author} {\bibinfo {author} {\bibfnamefont {G.}~\bibnamefont {Camelio}}, \bibinfo {author} {\bibfnamefont {A.}~\bibnamefont {Lovato}}, \bibinfo {author} {\bibfnamefont {L.}~\bibnamefont {Gualtieri}}, \bibinfo {author} {\bibfnamefont {O.}~\bibnamefont {Benhar}}, \bibinfo {author} {\bibfnamefont {J.~A.}\ \bibnamefont {Pons}}, \ and\ \bibinfo {author} {\bibfnamefont {V.}~\bibnamefont {Ferrari}},\ }\href {\doibase 10.1103/PhysRevD.96.043015} {\bibfield  {journal} {\bibinfo  {journal} {Phys. Rev. D}\ }\textbf {\bibinfo {volume} {96}},\ \bibinfo {pages} {043015} (\bibinfo {year} {2017})},\ \Eprint {http://arxiv.org/abs/1704.01923} {arXiv:1704.01923 [astro-ph.HE]} \BibitemShut {NoStop}%
\bibitem [{\citenamefont {Nakazato}\ and\ \citenamefont {Suzuki}(2019)}]{Nakazato:2019ojk}%
  \BibitemOpen
  \bibfield  {author} {\bibinfo {author} {\bibfnamefont {K.}~\bibnamefont {Nakazato}}\ and\ \bibinfo {author} {\bibfnamefont {H.}~\bibnamefont {Suzuki}},\ }\href {\doibase 10.3847/1538-4357/ab1d4b} {\bibfield  {journal} {\bibinfo  {journal} {Astrophys. J.}\ }\textbf {\bibinfo {volume} {878}},\ \bibinfo {pages} {25} (\bibinfo {year} {2019})},\ \Eprint {http://arxiv.org/abs/1905.00014} {arXiv:1905.00014 [astro-ph.HE]} \BibitemShut {NoStop}%
\bibitem [{\citenamefont {Lattimer}\ and\ \citenamefont {Prakash}(2016)}]{Lattimer:2015nhk}%
  \BibitemOpen
  \bibfield  {author} {\bibinfo {author} {\bibfnamefont {J.~M.}\ \bibnamefont {Lattimer}}\ and\ \bibinfo {author} {\bibfnamefont {M.}~\bibnamefont {Prakash}},\ }\href {\doibase 10.1016/j.physrep.2015.12.005} {\bibfield  {journal} {\bibinfo  {journal} {Phys. Rep.}\ }\textbf {\bibinfo {volume} {\textbf{621}}},\ \bibinfo {pages} {127} (\bibinfo {year} {2016})}\BibitemShut {NoStop}%
\bibitem [{\citenamefont {Oertel}\ \emph {et~al.}(2017)\citenamefont {Oertel}, \citenamefont {Hempel}, \citenamefont {Kl\"ahn},\ and\ \citenamefont {Typel}}]{Oertel:2016bki}%
  \BibitemOpen
  \bibfield  {author} {\bibinfo {author} {\bibfnamefont {M.}~\bibnamefont {Oertel}}, \bibinfo {author} {\bibfnamefont {M.}~\bibnamefont {Hempel}}, \bibinfo {author} {\bibfnamefont {T.}~\bibnamefont {Kl\"ahn}}, \ and\ \bibinfo {author} {\bibfnamefont {S.}~\bibnamefont {Typel}},\ }\href {\doibase 10.1103/RevModPhys.89.015007} {\bibfield  {journal} {\bibinfo  {journal} {Rev. Mod. Phys.}\ }\textbf {\bibinfo {volume} {89}},\ \bibinfo {pages} {015007} (\bibinfo {year} {2017})}\BibitemShut {NoStop}%
\bibitem [{\citenamefont {Issifu}\ \emph {et~al.}(2023)\citenamefont {Issifu}, \citenamefont {Marquez}, \citenamefont {Pelicer},\ and\ \citenamefont {Menezes}}]{Issifu:2023qyi}%
  \BibitemOpen
  \bibfield  {author} {\bibinfo {author} {\bibfnamefont {A.}~\bibnamefont {Issifu}}, \bibinfo {author} {\bibfnamefont {K.~D.}\ \bibnamefont {Marquez}}, \bibinfo {author} {\bibfnamefont {M.~R.}\ \bibnamefont {Pelicer}}, \ and\ \bibinfo {author} {\bibfnamefont {D.~P.}\ \bibnamefont {Menezes}},\ }\href {\doibase 10.1093/mnras/stad1198} {\bibfield  {journal} {\bibinfo  {journal} {Mon. Not. Roy. Astron. Soc.}\ }\textbf {\bibinfo {volume} {522}},\ \bibinfo {pages} {3263--3270} (\bibinfo {year} {2023})},\ \Eprint {http://arxiv.org/abs/2302.04364} {arXiv:2302.04364 [nucl-th]} \BibitemShut {NoStop}%
\bibitem [{\citenamefont {Issifu}\ \emph {et~al.}(2025{\natexlab{a}})\citenamefont {Issifu}, \citenamefont {Menezes}, \citenamefont {Rezaei},\ and\ \citenamefont {Frederico}}]{Issifu:2024fuw}%
  \BibitemOpen
  \bibfield  {author} {\bibinfo {author} {\bibfnamefont {A.}~\bibnamefont {Issifu}}, \bibinfo {author} {\bibfnamefont {D.~P.}\ \bibnamefont {Menezes}}, \bibinfo {author} {\bibfnamefont {Z.}~\bibnamefont {Rezaei}}, \ and\ \bibinfo {author} {\bibfnamefont {T.}~\bibnamefont {Frederico}},\ }\href {\doibase 10.1088/1475-7516/2025/01/024} {\bibfield  {journal} {\bibinfo  {journal} {JCAP}\ }\textbf {\bibinfo {volume} {01}},\ \bibinfo {pages} {024} (\bibinfo {year} {2025}{\natexlab{a}})},\ \Eprint {http://arxiv.org/abs/2405.10386} {arXiv:2405.10386 [nucl-th]} \BibitemShut {NoStop}%
\bibitem [{\citenamefont {Raduta}\ \emph {et~al.}(2020)\citenamefont {Raduta}, \citenamefont {Oertel},\ and\ \citenamefont {Sedrakian}}]{Raduta:2020fdn}%
  \BibitemOpen
  \bibfield  {author} {\bibinfo {author} {\bibfnamefont {A.~R.}\ \bibnamefont {Raduta}}, \bibinfo {author} {\bibfnamefont {M.}~\bibnamefont {Oertel}}, \ and\ \bibinfo {author} {\bibfnamefont {A.}~\bibnamefont {Sedrakian}},\ }\href {\doibase 10.1093/mnras/staa2491} {\bibfield  {journal} {\bibinfo  {journal} {Mon. Not. Roy. Astron. Soc.}\ }\textbf {\bibinfo {volume} {499}},\ \bibinfo {pages} {914--931} (\bibinfo {year} {2020})},\ \Eprint {http://arxiv.org/abs/2008.00213} {arXiv:2008.00213 [nucl-th]} \BibitemShut {NoStop}%
\bibitem [{\citenamefont {Kunkel}\ \emph {et~al.}(2025)\citenamefont {Kunkel}, \citenamefont {Wystub},\ and\ \citenamefont {Schaffner-Bielich}}]{Kunkel:2024otq}%
  \BibitemOpen
  \bibfield  {author} {\bibinfo {author} {\bibfnamefont {S.}~\bibnamefont {Kunkel}}, \bibinfo {author} {\bibfnamefont {S.}~\bibnamefont {Wystub}}, \ and\ \bibinfo {author} {\bibfnamefont {J.}~\bibnamefont {Schaffner-Bielich}},\ }\href {\doibase 10.1103/PhysRevC.111.035807} {\bibfield  {journal} {\bibinfo  {journal} {Phys. Rev. C}\ }\textbf {\bibinfo {volume} {111}},\ \bibinfo {pages} {035807} (\bibinfo {year} {2025})},\ \Eprint {http://arxiv.org/abs/2411.14930} {arXiv:2411.14930 [nucl-th]} \BibitemShut {NoStop}%
\bibitem [{\citenamefont {Issifu}\ \emph {et~al.}(2025{\natexlab{b}})\citenamefont {Issifu}, \citenamefont {Thakur}, \citenamefont {da~Silva}, \citenamefont {Marquez}, \citenamefont {Menezes}, \citenamefont {Dutra}, \citenamefont {Louren\c{c}o},\ and\ \citenamefont {Frederico}}]{Issifu:2024htq}%
  \BibitemOpen
  \bibfield  {author} {\bibinfo {author} {\bibfnamefont {A.}~\bibnamefont {Issifu}}, \bibinfo {author} {\bibfnamefont {P.}~\bibnamefont {Thakur}}, \bibinfo {author} {\bibfnamefont {F.~M.}\ \bibnamefont {da~Silva}}, \bibinfo {author} {\bibfnamefont {K.~D.}\ \bibnamefont {Marquez}}, \bibinfo {author} {\bibfnamefont {D.~P.}\ \bibnamefont {Menezes}}, \bibinfo {author} {\bibfnamefont {M.}~\bibnamefont {Dutra}}, \bibinfo {author} {\bibfnamefont {O.}~\bibnamefont {Louren\c{c}o}}, \ and\ \bibinfo {author} {\bibfnamefont {T.}~\bibnamefont {Frederico}},\ }\href {\doibase 10.1103/PhysRevD.111.083026} {\bibfield  {journal} {\bibinfo  {journal} {Phys. Rev. D}\ }\textbf {\bibinfo {volume} {111}},\ \bibinfo {pages} {083026} (\bibinfo {year} {2025}{\natexlab{b}})},\ \Eprint {http://arxiv.org/abs/2412.17946} {arXiv:2412.17946 [hep-ph]} \BibitemShut {NoStop}%
\bibitem [{\citenamefont {Sun}\ \emph {et~al.}(2025)\citenamefont {Sun}, \citenamefont {Zheng}, \citenamefont {Chen}, \citenamefont {Wei}, \citenamefont {Burgio},\ and\ \citenamefont {Schulze}}]{Sun:2024vdt}%
  \BibitemOpen
  \bibfield  {author} {\bibinfo {author} {\bibfnamefont {T.~T.}\ \bibnamefont {Sun}}, \bibinfo {author} {\bibfnamefont {Z.~Y.}\ \bibnamefont {Zheng}}, \bibinfo {author} {\bibfnamefont {H.}~\bibnamefont {Chen}}, \bibinfo {author} {\bibfnamefont {J.~B.}\ \bibnamefont {Wei}}, \bibinfo {author} {\bibfnamefont {G.~F.}\ \bibnamefont {Burgio}}, \ and\ \bibinfo {author} {\bibfnamefont {H.~J.}\ \bibnamefont {Schulze}},\ }\href {\doibase 10.1103/PhysRevD.111.043008} {\bibfield  {journal} {\bibinfo  {journal} {Phys. Rev. D}\ }\textbf {\bibinfo {volume} {111}},\ \bibinfo {pages} {043008} (\bibinfo {year} {2025})},\ \Eprint {http://arxiv.org/abs/2408.06599} {arXiv:2408.06599 [nucl-th]} \BibitemShut {NoStop}%
\bibitem [{\citenamefont {Gondek}\ \emph {et~al.}(1997)\citenamefont {Gondek}, \citenamefont {Haensel},\ and\ \citenamefont {Zdunik}}]{Gondek:1997fd}%
  \BibitemOpen
  \bibfield  {author} {\bibinfo {author} {\bibfnamefont {D.}~\bibnamefont {Gondek}}, \bibinfo {author} {\bibfnamefont {P.}~\bibnamefont {Haensel}}, \ and\ \bibinfo {author} {\bibfnamefont {J.~L.}\ \bibnamefont {Zdunik}},\ }\href@noop {} {\bibfield  {journal} {\bibinfo  {journal} {Astron. Astrophys.}\ }\textbf {\bibinfo {volume} {325}},\ \bibinfo {pages} {217--227} (\bibinfo {year} {1997})},\ \Eprint {http://arxiv.org/abs/astro-ph/9705157} {arXiv:astro-ph/9705157} \BibitemShut {NoStop}%
\bibitem [{\citenamefont {Kokkotas}\ and\ \citenamefont {Ruoff}(2001)}]{Kokkotas:2000up}%
  \BibitemOpen
  \bibfield  {author} {\bibinfo {author} {\bibfnamefont {K.~D.}\ \bibnamefont {Kokkotas}}\ and\ \bibinfo {author} {\bibfnamefont {J.}~\bibnamefont {Ruoff}},\ }\href {\doibase 10.1051/0004-6361:20000216} {\bibfield  {journal} {\bibinfo  {journal} {Astron. Astrophys.}\ }\textbf {\bibinfo {volume} {366}},\ \bibinfo {pages} {565} (\bibinfo {year} {2001})},\ \Eprint {http://arxiv.org/abs/gr-qc/0011093} {arXiv:gr-qc/0011093} \BibitemShut {NoStop}%
\bibitem [{\citenamefont {Sotani}\ and\ \citenamefont {Sumiyoshi}(2021)}]{Sotani:2021kvj}%
  \BibitemOpen
  \bibfield  {author} {\bibinfo {author} {\bibfnamefont {H.}~\bibnamefont {Sotani}}\ and\ \bibinfo {author} {\bibfnamefont {K.}~\bibnamefont {Sumiyoshi}},\ }\href {\doibase 10.1093/mnras/stab2301} {\bibfield  {journal} {\bibinfo  {journal} {Mon. Not. Roy. Astron. Soc.}\ }\textbf {\bibinfo {volume} {507}},\ \bibinfo {pages} {2766--2776} (\bibinfo {year} {2021})},\ \Eprint {http://arxiv.org/abs/2108.02484} {arXiv:2108.02484 [astro-ph.HE]} \BibitemShut {NoStop}%
\bibitem [{\citenamefont {Rodriguez}\ \emph {et~al.}(2023)\citenamefont {Rodriguez}, \citenamefont {Ranea-Sandoval}, \citenamefont {Chirenti},\ and\ \citenamefont {Radice}}]{Rodriguez:2023nay}%
  \BibitemOpen
  \bibfield  {author} {\bibinfo {author} {\bibfnamefont {M.~C.}\ \bibnamefont {Rodriguez}}, \bibinfo {author} {\bibfnamefont {I.~F.}\ \bibnamefont {Ranea-Sandoval}}, \bibinfo {author} {\bibfnamefont {C.}~\bibnamefont {Chirenti}}, \ and\ \bibinfo {author} {\bibfnamefont {D.}~\bibnamefont {Radice}},\ }\href {\doibase 10.1093/mnras/stad1459} {\bibfield  {journal} {\bibinfo  {journal} {Mon. Not. Roy. Astron. Soc.}\ }\textbf {\bibinfo {volume} {523}},\ \bibinfo {pages} {2236--2246} (\bibinfo {year} {2023})},\ \Eprint {http://arxiv.org/abs/2304.00033} {arXiv:2304.00033 [astro-ph.HE]} \BibitemShut {NoStop}%
\bibitem [{\citenamefont {Barman}\ \emph {et~al.}(2025)\citenamefont {Barman}, \citenamefont {Pradhan},\ and\ \citenamefont {Chatterjee}}]{Barman:2024zuo}%
  \BibitemOpen
  \bibfield  {author} {\bibinfo {author} {\bibfnamefont {N.}~\bibnamefont {Barman}}, \bibinfo {author} {\bibfnamefont {B.~K.}\ \bibnamefont {Pradhan}}, \ and\ \bibinfo {author} {\bibfnamefont {D.}~\bibnamefont {Chatterjee}},\ }\href {\doibase 10.1103/PhysRevD.111.023017} {\bibfield  {journal} {\bibinfo  {journal} {Phys. Rev. D}\ }\textbf {\bibinfo {volume} {111}},\ \bibinfo {pages} {023017} (\bibinfo {year} {2025})},\ \Eprint {http://arxiv.org/abs/2408.00739} {arXiv:2408.00739 [astro-ph.HE]} \BibitemShut {NoStop}%
\bibitem [{\citenamefont {Pradhan}\ \emph {et~al.}(2022)\citenamefont {Pradhan}, \citenamefont {Chatterjee}, \citenamefont {Lanoye},\ and\ \citenamefont {Jaikumar}}]{Pradhan:2022vdf}%
  \BibitemOpen
  \bibfield  {author} {\bibinfo {author} {\bibfnamefont {B.~K.}\ \bibnamefont {Pradhan}}, \bibinfo {author} {\bibfnamefont {D.}~\bibnamefont {Chatterjee}}, \bibinfo {author} {\bibfnamefont {M.}~\bibnamefont {Lanoye}}, \ and\ \bibinfo {author} {\bibfnamefont {P.}~\bibnamefont {Jaikumar}},\ }\href {\doibase 10.1103/PhysRevC.106.015805} {\bibfield  {journal} {\bibinfo  {journal} {Phys. Rev. C}\ }\textbf {\bibinfo {volume} {106}},\ \bibinfo {pages} {015805} (\bibinfo {year} {2022})},\ \Eprint {http://arxiv.org/abs/2203.03141} {arXiv:2203.03141 [astro-ph.HE]} \BibitemShut {NoStop}%
\bibitem [{\citenamefont {Kunjipurayil}\ \emph {et~al.}(2022)\citenamefont {Kunjipurayil}, \citenamefont {Zhao}, \citenamefont {Kumar}, \citenamefont {Agrawal},\ and\ \citenamefont {Prakash}}]{Kunjipurayil:2022zah}%
  \BibitemOpen
  \bibfield  {author} {\bibinfo {author} {\bibfnamefont {A.}~\bibnamefont {Kunjipurayil}}, \bibinfo {author} {\bibfnamefont {T.}~\bibnamefont {Zhao}}, \bibinfo {author} {\bibfnamefont {B.}~\bibnamefont {Kumar}}, \bibinfo {author} {\bibfnamefont {B.~K.}\ \bibnamefont {Agrawal}}, \ and\ \bibinfo {author} {\bibfnamefont {M.}~\bibnamefont {Prakash}},\ }\href {\doibase 10.1103/PhysRevD.106.063005} {\bibfield  {journal} {\bibinfo  {journal} {Phys. Rev. D}\ }\textbf {\bibinfo {volume} {106}},\ \bibinfo {pages} {063005} (\bibinfo {year} {2022})},\ \Eprint {http://arxiv.org/abs/2205.02081} {arXiv:2205.02081 [nucl-th]} \BibitemShut {NoStop}%
\bibitem [{\citenamefont {Roy}\ \emph {et~al.}(2024)\citenamefont {Roy}, \citenamefont {Malik}, \citenamefont {Bhattacharya},\ and\ \citenamefont {Banik}}]{Roy:2023gzi}%
  \BibitemOpen
  \bibfield  {author} {\bibinfo {author} {\bibfnamefont {D.~G.}\ \bibnamefont {Roy}}, \bibinfo {author} {\bibfnamefont {T.}~\bibnamefont {Malik}}, \bibinfo {author} {\bibfnamefont {S.}~\bibnamefont {Bhattacharya}}, \ and\ \bibinfo {author} {\bibfnamefont {S.}~\bibnamefont {Banik}},\ }\href {\doibase 10.3847/1538-4357/ad43e6} {\bibfield  {journal} {\bibinfo  {journal} {Astrophys. J.}\ }\textbf {\bibinfo {volume} {968}},\ \bibinfo {pages} {124} (\bibinfo {year} {2024})},\ \Eprint {http://arxiv.org/abs/2312.02061} {arXiv:2312.02061 [astro-ph.HE]} \BibitemShut {NoStop}%
\bibitem [{\citenamefont {Zheng}\ \emph {et~al.}(2025)\citenamefont {Zheng}, \citenamefont {Sun}, \citenamefont {Chen}, \citenamefont {Wei}, \citenamefont {Zheng}, \citenamefont {Schulze},\ and\ \citenamefont {Burgio}}]{Zheng:2025xlr}%
  \BibitemOpen
  \bibfield  {author} {\bibinfo {author} {\bibfnamefont {Z.-Y.}\ \bibnamefont {Zheng}}, \bibinfo {author} {\bibfnamefont {T.-t.}\ \bibnamefont {Sun}}, \bibinfo {author} {\bibfnamefont {H.}~\bibnamefont {Chen}}, \bibinfo {author} {\bibfnamefont {J.-B.}\ \bibnamefont {Wei}}, \bibinfo {author} {\bibfnamefont {X.-P.}\ \bibnamefont {Zheng}}, \bibinfo {author} {\bibfnamefont {H.~J.}\ \bibnamefont {Schulze}}, \ and\ \bibinfo {author} {\bibfnamefont {G.~F.}\ \bibnamefont {Burgio}},\ }\href@noop {} {\  (\bibinfo {year} {2025})},\ \Eprint {http://arxiv.org/abs/2505.10133} {arXiv:2505.10133 [astro-ph.HE]} \BibitemShut {NoStop}%
\bibitem [{\citenamefont {Kumar}\ \emph {et~al.}(2025)\citenamefont {Kumar}, \citenamefont {Karan}, \citenamefont {Verma}, \citenamefont {Mishra},\ and\ \citenamefont {Mallick}}]{Kumar:2024bvd}%
  \BibitemOpen
  \bibfield  {author} {\bibinfo {author} {\bibfnamefont {D.}~\bibnamefont {Kumar}}, \bibinfo {author} {\bibfnamefont {A.}~\bibnamefont {Karan}}, \bibinfo {author} {\bibfnamefont {A.}~\bibnamefont {Verma}}, \bibinfo {author} {\bibfnamefont {H.}~\bibnamefont {Mishra}}, \ and\ \bibinfo {author} {\bibfnamefont {R.}~\bibnamefont {Mallick}},\ }\href {\doibase 10.1103/PhysRevC.111.055805} {\bibfield  {journal} {\bibinfo  {journal} {Phys. Rev. C}\ }\textbf {\bibinfo {volume} {111}},\ \bibinfo {pages} {055805} (\bibinfo {year} {2025})},\ \Eprint {http://arxiv.org/abs/2409.01785} {arXiv:2409.01785 [astro-ph.HE]} \BibitemShut {NoStop}%
\bibitem [{\citenamefont {Ferrari}\ \emph {et~al.}(2003)\citenamefont {Ferrari}, \citenamefont {Miniutti},\ and\ \citenamefont {Pons}}]{Ferrari:2002ut}%
  \BibitemOpen
  \bibfield  {author} {\bibinfo {author} {\bibfnamefont {V.}~\bibnamefont {Ferrari}}, \bibinfo {author} {\bibfnamefont {G.}~\bibnamefont {Miniutti}}, \ and\ \bibinfo {author} {\bibfnamefont {J.~A.}\ \bibnamefont {Pons}},\ }\href {\doibase 10.1046/j.1365-8711.2003.06580.x} {\bibfield  {journal} {\bibinfo  {journal} {Mon. Not. Roy. Astron. Soc.}\ }\textbf {\bibinfo {volume} {342}},\ \bibinfo {pages} {629} (\bibinfo {year} {2003})},\ \Eprint {http://arxiv.org/abs/astro-ph/0210581} {arXiv:astro-ph/0210581} \BibitemShut {NoStop}%
\bibitem [{\citenamefont {Thapa}\ \emph {et~al.}(2023)\citenamefont {Thapa}, \citenamefont {Beznogov}, \citenamefont {Raduta},\ and\ \citenamefont {Thakur}}]{Thapa:2023grg}%
  \BibitemOpen
  \bibfield  {author} {\bibinfo {author} {\bibfnamefont {V.~B.}\ \bibnamefont {Thapa}}, \bibinfo {author} {\bibfnamefont {M.~V.}\ \bibnamefont {Beznogov}}, \bibinfo {author} {\bibfnamefont {A.~R.}\ \bibnamefont {Raduta}}, \ and\ \bibinfo {author} {\bibfnamefont {P.}~\bibnamefont {Thakur}},\ }\href {\doibase 10.1103/PhysRevD.107.103054} {\bibfield  {journal} {\bibinfo  {journal} {Phys. Rev. D}\ }\textbf {\bibinfo {volume} {107}},\ \bibinfo {pages} {103054} (\bibinfo {year} {2023})},\ \Eprint {http://arxiv.org/abs/2302.11469} {arXiv:2302.11469 [nucl-th]} \BibitemShut {NoStop}%
\bibitem [{\citenamefont {Kumar}\ \emph {et~al.}(2024)\citenamefont {Kumar}, \citenamefont {Thakur},\ and\ \citenamefont {Sinha}}]{Kumar:2024jky}%
  \BibitemOpen
  \bibfield  {author} {\bibinfo {author} {\bibfnamefont {A.}~\bibnamefont {Kumar}}, \bibinfo {author} {\bibfnamefont {P.}~\bibnamefont {Thakur}}, \ and\ \bibinfo {author} {\bibfnamefont {M.}~\bibnamefont {Sinha}},\ }\href {\doibase 10.1093/mnras/stae834} {\bibfield  {journal} {\bibinfo  {journal} {Mon. Not. Roy. Astron. Soc.}\ }\textbf {\bibinfo {volume} {530}},\ \bibinfo {pages} {501--513} (\bibinfo {year} {2024})},\ \Eprint {http://arxiv.org/abs/2404.01252} {arXiv:2404.01252 [astro-ph.HE]} \BibitemShut {NoStop}%
\bibitem [{\citenamefont {Serot}\ and\ \citenamefont {Walecka}(1997)}]{Serot:1997xg}%
  \BibitemOpen
  \bibfield  {author} {\bibinfo {author} {\bibfnamefont {B.~D.}\ \bibnamefont {Serot}}\ and\ \bibinfo {author} {\bibfnamefont {J.~D.}\ \bibnamefont {Walecka}},\ }\href {\doibase 10.1142/S0218301397000299} {\bibfield  {journal} {\bibinfo  {journal} {Int. J. Mod. Phys. E}\ }\textbf {\bibinfo {volume} {6}},\ \bibinfo {pages} {515--631} (\bibinfo {year} {1997})},\ \Eprint {http://arxiv.org/abs/nucl-th/9701058} {arXiv:nucl-th/9701058} \BibitemShut {NoStop}%
\bibitem [{\citenamefont {Menezes}(2021)}]{Menezes:2021jmw}%
  \BibitemOpen
  \bibfield  {author} {\bibinfo {author} {\bibfnamefont {D.~P.}\ \bibnamefont {Menezes}},\ }\href {\doibase 10.3390/universe7080267} {\bibfield  {journal} {\bibinfo  {journal} {Universe}\ }\textbf {\bibinfo {volume} {7}},\ \bibinfo {pages} {267} (\bibinfo {year} {2021})},\ \Eprint {http://arxiv.org/abs/2106.09515} {arXiv:2106.09515 [astro-ph.HE]} \BibitemShut {NoStop}%
\bibitem [{\citenamefont {Bombaci}(2017)}]{Bombaci:2016xzl}%
  \BibitemOpen
  \bibfield  {author} {\bibinfo {author} {\bibfnamefont {I.}~\bibnamefont {Bombaci}},\ }\href {\doibase 10.7566/JPSCP.17.101002} {\bibfield  {journal} {\bibinfo  {journal} {JPS Conf. Proc.}\ }\textbf {\bibinfo {volume} {17}},\ \bibinfo {pages} {101002} (\bibinfo {year} {2017})},\ \Eprint {http://arxiv.org/abs/1601.05339} {arXiv:1601.05339 [nucl-th]} \BibitemShut {NoStop}%
\bibitem [{\citenamefont {Lalazissis}\ \emph {et~al.}(2005)\citenamefont {Lalazissis}, \citenamefont {Nik\ifmmode \check{s}\else \v{s}\fi{}i\ifmmode~\acute{c}\else \'{c}\fi{}}, \citenamefont {Vretenar},\ and\ \citenamefont {Ring}}]{ddme2}%
  \BibitemOpen
  \bibfield  {author} {\bibinfo {author} {\bibfnamefont {G.~A.}\ \bibnamefont {Lalazissis}}, \bibinfo {author} {\bibfnamefont {T.}~\bibnamefont {Nik\ifmmode \check{s}\else \v{s}\fi{}i\ifmmode~\acute{c}\else \'{c}\fi{}}}, \bibinfo {author} {\bibfnamefont {D.}~\bibnamefont {Vretenar}}, \ and\ \bibinfo {author} {\bibfnamefont {P.}~\bibnamefont {Ring}},\ }\href {\doibase 10.1103/PhysRevC.71.024312} {\bibfield  {journal} {\bibinfo  {journal} {Phys. Rev. C}\ }\textbf {\bibinfo {volume} {71}},\ \bibinfo {pages} {024312} (\bibinfo {year} {2005})}\BibitemShut {NoStop}%
\bibitem [{\citenamefont {Dutra}\ \emph {et~al.}(2012)\citenamefont {Dutra}, \citenamefont {Louren\ifmmode~\mbox{\c{c}}\else \c{c}\fi{}o}, \citenamefont {S\'a~Martins}, \citenamefont {Delfino}, \citenamefont {Stone},\ and\ \citenamefont {Stevenson}}]{Dutra}%
  \BibitemOpen
  \bibfield  {author} {\bibinfo {author} {\bibfnamefont {M.}~\bibnamefont {Dutra}}, \bibinfo {author} {\bibfnamefont {O.}~\bibnamefont {Louren\ifmmode~\mbox{\c{c}}\else \c{c}\fi{}o}}, \bibinfo {author} {\bibfnamefont {J.~S.}\ \bibnamefont {S\'a~Martins}}, \bibinfo {author} {\bibfnamefont {A.}~\bibnamefont {Delfino}}, \bibinfo {author} {\bibfnamefont {J.~R.}\ \bibnamefont {Stone}}, \ and\ \bibinfo {author} {\bibfnamefont {P.~D.}\ \bibnamefont {Stevenson}},\ }\href {\doibase 10.1103/PhysRevC.85.035201} {\bibfield  {journal} {\bibinfo  {journal} {Phys. Rev. C}\ }\textbf {\bibinfo {volume} {\textbf{85}}},\ \bibinfo {pages} {035201} (\bibinfo {year} {2012})}\BibitemShut {NoStop}%
\bibitem [{\citenamefont {Lattimer}\ and\ \citenamefont {Lim}(2013)}]{Lattimer:2012xj}%
  \BibitemOpen
  \bibfield  {author} {\bibinfo {author} {\bibfnamefont {J.~M.}\ \bibnamefont {Lattimer}}\ and\ \bibinfo {author} {\bibfnamefont {Y.}~\bibnamefont {Lim}},\ }\href {\doibase 10.1088/0004-637X/771/1/51} {\bibfield  {journal} {\bibinfo  {journal} {Astrophys. J.}\ }\textbf {\bibinfo {volume} {771}},\ \bibinfo {pages} {51} (\bibinfo {year} {2013})},\ \Eprint {http://arxiv.org/abs/1203.4286} {arXiv:1203.4286 [nucl-th]} \BibitemShut {NoStop}%
\bibitem [{\citenamefont {Reed}\ \emph {et~al.}(2021)\citenamefont {Reed}, \citenamefont {Fattoyev}, \citenamefont {Horowitz},\ and\ \citenamefont {Piekarewicz}}]{reed2021}%
  \BibitemOpen
  \bibfield  {author} {\bibinfo {author} {\bibfnamefont {B.~T.}\ \bibnamefont {Reed}}, \bibinfo {author} {\bibfnamefont {F.~J.}\ \bibnamefont {Fattoyev}}, \bibinfo {author} {\bibfnamefont {C.~J.}\ \bibnamefont {Horowitz}}, \ and\ \bibinfo {author} {\bibfnamefont {J.}~\bibnamefont {Piekarewicz}},\ }\href {\doibase 10.1103/PhysRevLett.126.172503} {\bibfield  {journal} {\bibinfo  {journal} {Phys. Rev. Lett.}\ }\textbf {\bibinfo {volume} {126}},\ \bibinfo {pages} {172503} (\bibinfo {year} {2021})}\BibitemShut {NoStop}%
\bibitem [{\citenamefont {Glendenning}\ and\ \citenamefont {Moszkowski}(1991)}]{Glendenning:1991es}%
  \BibitemOpen
  \bibfield  {author} {\bibinfo {author} {\bibfnamefont {N.~K.}\ \bibnamefont {Glendenning}}\ and\ \bibinfo {author} {\bibfnamefont {S.~A.}\ \bibnamefont {Moszkowski}},\ }\href {\doibase 10.1103/PhysRevLett.67.2414} {\bibfield  {journal} {\bibinfo  {journal} {Phys. Rev. Lett.}\ }\textbf {\bibinfo {volume} {67}},\ \bibinfo {pages} {2414--2417} (\bibinfo {year} {1991})}\BibitemShut {NoStop}%
\bibitem [{\citenamefont {Weissenborn}\ \emph {et~al.}(2012)\citenamefont {Weissenborn}, \citenamefont {Chatterjee},\ and\ \citenamefont {Schaffner-Bielich}}]{Weissenborn:2011kb}%
  \BibitemOpen
  \bibfield  {author} {\bibinfo {author} {\bibfnamefont {S.}~\bibnamefont {Weissenborn}}, \bibinfo {author} {\bibfnamefont {D.}~\bibnamefont {Chatterjee}}, \ and\ \bibinfo {author} {\bibfnamefont {J.}~\bibnamefont {Schaffner-Bielich}},\ }\href {\doibase 10.1016/j.nuclphysa.2012.02.012} {\bibfield  {journal} {\bibinfo  {journal} {Nucl. Phys. A}\ }\textbf {\bibinfo {volume} {881}},\ \bibinfo {pages} {62--77} (\bibinfo {year} {2012})},\ \Eprint {http://arxiv.org/abs/1111.6049} {arXiv:1111.6049 [astro-ph.HE]} \BibitemShut {NoStop}%
\bibitem [{\citenamefont {Lopes}\ \emph {et~al.}(2023)\citenamefont {Lopes}, \citenamefont {Marquez},\ and\ \citenamefont {Menezes}}]{Lopes:2022vjx}%
  \BibitemOpen
  \bibfield  {author} {\bibinfo {author} {\bibfnamefont {L.~L.}\ \bibnamefont {Lopes}}, \bibinfo {author} {\bibfnamefont {K.~D.}\ \bibnamefont {Marquez}}, \ and\ \bibinfo {author} {\bibfnamefont {D.~P.}\ \bibnamefont {Menezes}},\ }\href {\doibase 10.1103/PhysRevD.107.036011} {\bibfield  {journal} {\bibinfo  {journal} {Phys. Rev. D}\ }\textbf {\bibinfo {volume} {107}},\ \bibinfo {pages} {036011} (\bibinfo {year} {2023})},\ \Eprint {http://arxiv.org/abs/2211.17153} {arXiv:2211.17153 [hep-ph]} \BibitemShut {NoStop}%
\bibitem [{\citenamefont {Glendenning}(1985)}]{Glendenning1985}%
  \BibitemOpen
  \bibfield  {author} {\bibinfo {author} {\bibfnamefont {N.~K.}\ \bibnamefont {Glendenning}},\ }\href {\doibase 10.1086/163253} {\bibfield  {journal} {\bibinfo  {journal} {Astrophysical Journal}\ }\textbf {\bibinfo {volume} {293}},\ \bibinfo {pages} {470--493} (\bibinfo {year} {1985})},\ \bibinfo {note} {provided by the SAO/NASA Astrophysics Data System}\BibitemShut {NoStop}%
\bibitem [{\citenamefont {Fortin}\ \emph {et~al.}(2016)\citenamefont {Fortin}, \citenamefont {Providencia}, \citenamefont {Raduta}, \citenamefont {Gulminelli}, \citenamefont {Zdunik}, \citenamefont {Haensel},\ and\ \citenamefont {Bejger}}]{Fortin:2016hny}%
  \BibitemOpen
  \bibfield  {author} {\bibinfo {author} {\bibfnamefont {M.}~\bibnamefont {Fortin}}, \bibinfo {author} {\bibfnamefont {C.}~\bibnamefont {Providencia}}, \bibinfo {author} {\bibfnamefont {A.~R.}\ \bibnamefont {Raduta}}, \bibinfo {author} {\bibfnamefont {F.}~\bibnamefont {Gulminelli}}, \bibinfo {author} {\bibfnamefont {J.~L.}\ \bibnamefont {Zdunik}}, \bibinfo {author} {\bibfnamefont {P.}~\bibnamefont {Haensel}}, \ and\ \bibinfo {author} {\bibfnamefont {M.}~\bibnamefont {Bejger}},\ }\href {\doibase 10.1103/PhysRevC.94.035804} {\bibfield  {journal} {\bibinfo  {journal} {Phys. Rev. C}\ }\textbf {\bibinfo {volume} {94}},\ \bibinfo {pages} {035804} (\bibinfo {year} {2016})},\ \Eprint {http://arxiv.org/abs/1604.01944} {arXiv:1604.01944 [astro-ph.SR]} \BibitemShut {NoStop}%
\bibitem [{\citenamefont {Balberg}\ and\ \citenamefont {Gal}(1997)}]{Balberg:1997yw}%
  \BibitemOpen
  \bibfield  {author} {\bibinfo {author} {\bibfnamefont {S.}~\bibnamefont {Balberg}}\ and\ \bibinfo {author} {\bibfnamefont {A.}~\bibnamefont {Gal}},\ }\href {\doibase 10.1016/S0375-9474(97)81465-0} {\bibfield  {journal} {\bibinfo  {journal} {Nucl. Phys. A}\ }\textbf {\bibinfo {volume} {625}},\ \bibinfo {pages} {435--472} (\bibinfo {year} {1997})},\ \Eprint {http://arxiv.org/abs/nucl-th/9704013} {arXiv:nucl-th/9704013} \BibitemShut {NoStop}%
\bibitem [{\citenamefont {Fortin}\ \emph {et~al.}(2018)\citenamefont {Fortin}, \citenamefont {Oertel},\ and\ \citenamefont {Provid{\^e}ncia}}]{Fortin:2017dsj}%
  \BibitemOpen
  \bibfield  {author} {\bibinfo {author} {\bibfnamefont {M.}~\bibnamefont {Fortin}}, \bibinfo {author} {\bibfnamefont {M.}~\bibnamefont {Oertel}}, \ and\ \bibinfo {author} {\bibfnamefont {C.}~\bibnamefont {Provid{\^e}ncia}},\ }\href {\doibase 10.1017/pasa.2018.32} {\bibfield  {journal} {\bibinfo  {journal} {Publ. Astron. Soc. Austral.}\ }\textbf {\bibinfo {volume} {35}},\ \bibinfo {pages} {44} (\bibinfo {year} {2018})},\ \Eprint {http://arxiv.org/abs/1711.09427} {arXiv:1711.09427 [astro-ph.HE]} \BibitemShut {NoStop}%
\bibitem [{\citenamefont {Bednarek}\ \emph {et~al.}(2012{\natexlab{a}})\citenamefont {Bednarek}, \citenamefont {Haensel}, \citenamefont {Zdunik}, \citenamefont {Bejger},\ and\ \citenamefont {Ma{\'n}ka}}]{Bednarek2012}%
  \BibitemOpen
  \bibfield  {author} {\bibinfo {author} {\bibfnamefont {I.}~\bibnamefont {Bednarek}}, \bibinfo {author} {\bibfnamefont {P.}~\bibnamefont {Haensel}}, \bibinfo {author} {\bibfnamefont {J.~L.}\ \bibnamefont {Zdunik}}, \bibinfo {author} {\bibfnamefont {M.}~\bibnamefont {Bejger}}, \ and\ \bibinfo {author} {\bibfnamefont {R.}~\bibnamefont {Ma{\'n}ka}},\ }\href {\doibase 10.1051/0004-6361/201118560} {\bibfield  {journal} {\bibinfo  {journal} {Astronomy \& Astrophysics}\ }\textbf {\bibinfo {volume} {543}},\ \bibinfo {pages} {A157} (\bibinfo {year} {2012}{\natexlab{a}})}\BibitemShut {NoStop}%
\bibitem [{\citenamefont {Djapo}\ \emph {et~al.}(2010)\citenamefont {Djapo}, \citenamefont {Schaefer},\ and\ \citenamefont {Wambach}}]{Djapo:2008au}%
  \BibitemOpen
  \bibfield  {author} {\bibinfo {author} {\bibfnamefont {H.}~\bibnamefont {Djapo}}, \bibinfo {author} {\bibfnamefont {B.-J.}\ \bibnamefont {Schaefer}}, \ and\ \bibinfo {author} {\bibfnamefont {J.}~\bibnamefont {Wambach}},\ }\href {\doibase 10.1103/PhysRevC.81.035803} {\bibfield  {journal} {\bibinfo  {journal} {Phys. Rev. C}\ }\textbf {\bibinfo {volume} {81}},\ \bibinfo {pages} {035803} (\bibinfo {year} {2010})},\ \Eprint {http://arxiv.org/abs/0811.2939} {arXiv:0811.2939 [nucl-th]} \BibitemShut {NoStop}%
\bibitem [{\citenamefont {Abbott}\ \emph {et~al.}(2017{\natexlab{a}})\citenamefont {Abbott} \emph {et~al.}}]{LIGOScientific:2017vwq}%
  \BibitemOpen
  \bibfield  {author} {\bibinfo {author} {\bibfnamefont {B.~P.}\ \bibnamefont {Abbott}} \emph {et~al.} (\bibinfo {collaboration} {LIGO Scientific, Virgo}),\ }\href {\doibase 10.1103/PhysRevLett.119.161101} {\bibfield  {journal} {\bibinfo  {journal} {Phys. Rev. Lett.}\ }\textbf {\bibinfo {volume} {119}},\ \bibinfo {pages} {161101} (\bibinfo {year} {2017}{\natexlab{a}})},\ \Eprint {http://arxiv.org/abs/1710.05832} {arXiv:1710.05832 [gr-qc]} \BibitemShut {NoStop}%
\bibitem [{\citenamefont {Abbott}\ \emph {et~al.}(2018)\citenamefont {Abbott} \emph {et~al.}}]{LIGOScientific:2018cki}%
  \BibitemOpen
  \bibfield  {author} {\bibinfo {author} {\bibfnamefont {B.~P.}\ \bibnamefont {Abbott}} \emph {et~al.} (\bibinfo {collaboration} {LIGO Scientific, Virgo}),\ }\href {\doibase 10.1103/PhysRevLett.121.161101} {\bibfield  {journal} {\bibinfo  {journal} {Phys. Rev. Lett.}\ }\textbf {\bibinfo {volume} {121}},\ \bibinfo {pages} {161101} (\bibinfo {year} {2018})},\ \Eprint {http://arxiv.org/abs/1805.11581} {arXiv:1805.11581 [gr-qc]} \BibitemShut {NoStop}%
\bibitem [{\citenamefont {Riley}\ \emph {et~al.}(2021{\natexlab{a}})\citenamefont {Riley} \emph {et~al.}}]{riley2021}%
  \BibitemOpen
  \bibfield  {author} {\bibinfo {author} {\bibfnamefont {T.~E.}\ \bibnamefont {Riley}} \emph {et~al.},\ }\href {\doibase 10.3847/2041-8213/ac0a81} {\bibfield  {journal} {\bibinfo  {journal} {Astrophys. J. Lett.}\ }\textbf {\bibinfo {volume} {918}},\ \bibinfo {pages} {L27} (\bibinfo {year} {2021}{\natexlab{a}})},\ \Eprint {http://arxiv.org/abs/2105.06980} {arXiv:2105.06980 [astro-ph.HE]} \BibitemShut {NoStop}%
\bibitem [{\citenamefont {Miller}\ \emph {et~al.}(2021)\citenamefont {Miller} \emph {et~al.}}]{Miller:2021qha}%
  \BibitemOpen
  \bibfield  {author} {\bibinfo {author} {\bibfnamefont {M.~C.}\ \bibnamefont {Miller}} \emph {et~al.},\ }\href {\doibase 10.3847/2041-8213/ac089b} {\bibfield  {journal} {\bibinfo  {journal} {Astrophys. J. Lett.}\ }\textbf {\bibinfo {volume} {918}},\ \bibinfo {pages} {L28} (\bibinfo {year} {2021})},\ \Eprint {http://arxiv.org/abs/2105.06979} {arXiv:2105.06979 [astro-ph.HE]} \BibitemShut {NoStop}%
\bibitem [{\citenamefont {Riley}\ \emph {et~al.}(2019{\natexlab{a}})\citenamefont {Riley} \emph {et~al.}}]{riley2019}%
  \BibitemOpen
  \bibfield  {author} {\bibinfo {author} {\bibfnamefont {T.~E.}\ \bibnamefont {Riley}} \emph {et~al.},\ }\href {\doibase 10.3847/2041-8213/ab481c} {\bibfield  {journal} {\bibinfo  {journal} {Astrophys. J. Lett.}\ }\textbf {\bibinfo {volume} {887}},\ \bibinfo {pages} {L21} (\bibinfo {year} {2019}{\natexlab{a}})},\ \Eprint {http://arxiv.org/abs/1912.05702} {arXiv:1912.05702 [astro-ph.HE]} \BibitemShut {NoStop}%
\bibitem [{\citenamefont {Miller}\ \emph {et~al.}(2019)\citenamefont {Miller} \emph {et~al.}}]{Miller:2019cac}%
  \BibitemOpen
  \bibfield  {author} {\bibinfo {author} {\bibfnamefont {M.}~\bibnamefont {Miller}} \emph {et~al.},\ }\href {\doibase 10.3847/2041-8213/ab50c5} {\bibfield  {journal} {\bibinfo  {journal} {Astrophys. J. Lett.}\ }\textbf {\bibinfo {volume} {887}},\ \bibinfo {pages} {L24} (\bibinfo {year} {2019})},\ \Eprint {http://arxiv.org/abs/1912.05705} {arXiv:1912.05705 [astro-ph.HE]} \BibitemShut {NoStop}%
\bibitem [{\citenamefont {Rather}\ \emph {et~al.}(2025)\citenamefont {Rather}, \citenamefont {Marquez}, \citenamefont {Thakur},\ and\ \citenamefont {Louren{\c{c}}o}}]{Rather:2024nry}%
  \BibitemOpen
  \bibfield  {author} {\bibinfo {author} {\bibfnamefont {I.~A.}\ \bibnamefont {Rather}}, \bibinfo {author} {\bibfnamefont {K.~D.}\ \bibnamefont {Marquez}}, \bibinfo {author} {\bibfnamefont {P.}~\bibnamefont {Thakur}}, \ and\ \bibinfo {author} {\bibfnamefont {O.}~\bibnamefont {Louren{\c{c}}o}},\ }\href {\doibase 10.1103/7qns-616m} {\bibfield  {journal} {\bibinfo  {journal} {Phys. Rev. D}\ }\textbf {\bibinfo {volume} {112}},\ \bibinfo {pages} {023013} (\bibinfo {year} {2025})},\ \Eprint {http://arxiv.org/abs/2412.12002} {arXiv:2412.12002 [astro-ph.HE]} \BibitemShut {NoStop}%
\bibitem [{\citenamefont {Zhao}\ \emph {et~al.}(2022)\citenamefont {Zhao}, \citenamefont {Constantinou}, \citenamefont {Jaikumar},\ and\ \citenamefont {Prakash}}]{Zhao:2022toc}%
  \BibitemOpen
  \bibfield  {author} {\bibinfo {author} {\bibfnamefont {T.}~\bibnamefont {Zhao}}, \bibinfo {author} {\bibfnamefont {C.}~\bibnamefont {Constantinou}}, \bibinfo {author} {\bibfnamefont {P.}~\bibnamefont {Jaikumar}}, \ and\ \bibinfo {author} {\bibfnamefont {M.}~\bibnamefont {Prakash}},\ }\href {\doibase 10.1103/PhysRevD.105.103025} {\bibfield  {journal} {\bibinfo  {journal} {Phys. Rev. D}\ }\textbf {\bibinfo {volume} {105}},\ \bibinfo {pages} {103025} (\bibinfo {year} {2022})},\ \Eprint {http://arxiv.org/abs/2202.01403} {arXiv:2202.01403 [gr-qc]} \BibitemShut {NoStop}%
\bibitem [{\citenamefont {{Bardeen}}\ \emph {et~al.}(1966)\citenamefont {{Bardeen}}, \citenamefont {{Thorne}},\ and\ \citenamefont {{Meltzer}}}]{1966ApJ...145..505B}%
  \BibitemOpen
  \bibfield  {author} {\bibinfo {author} {\bibfnamefont {J.~M.}\ \bibnamefont {{Bardeen}}}, \bibinfo {author} {\bibfnamefont {K.~S.}\ \bibnamefont {{Thorne}}}, \ and\ \bibinfo {author} {\bibfnamefont {D.~W.}\ \bibnamefont {{Meltzer}}},\ }\href {\doibase 10.1086/148791} {\bibfield  {journal} {\bibinfo  {journal} {Astrophys. J.}\ }\textbf {\bibinfo {volume} {145}},\ \bibinfo {pages} {505} (\bibinfo {year} {1966})}\BibitemShut {NoStop}%
\bibitem [{\citenamefont {Rather}\ \emph {et~al.}(2023{\natexlab{a}})\citenamefont {Rather}, \citenamefont {Marquez}, \citenamefont {Panotopoulos},\ and\ \citenamefont {Lopes}}]{Rather:2023dom}%
  \BibitemOpen
  \bibfield  {author} {\bibinfo {author} {\bibfnamefont {I.~A.}\ \bibnamefont {Rather}}, \bibinfo {author} {\bibfnamefont {K.~D.}\ \bibnamefont {Marquez}}, \bibinfo {author} {\bibfnamefont {G.}~\bibnamefont {Panotopoulos}}, \ and\ \bibinfo {author} {\bibfnamefont {I.}~\bibnamefont {Lopes}},\ }\href {\doibase 10.1103/PhysRevD.107.123022} {\bibfield  {journal} {\bibinfo  {journal} {Phys. Rev. D}\ }\textbf {\bibinfo {volume} {107}},\ \bibinfo {pages} {123022} (\bibinfo {year} {2023}{\natexlab{a}})},\ \Eprint {http://arxiv.org/abs/2303.11006} {arXiv:2303.11006 [nucl-th]} \BibitemShut {NoStop}%
\bibitem [{\citenamefont {Rather}\ \emph {et~al.}(2024)\citenamefont {Rather}, \citenamefont {Marquez}, \citenamefont {Backes}, \citenamefont {Panotopoulos},\ and\ \citenamefont {Lopes}}]{Rather:2024hmo}%
  \BibitemOpen
  \bibfield  {author} {\bibinfo {author} {\bibfnamefont {I.~A.}\ \bibnamefont {Rather}}, \bibinfo {author} {\bibfnamefont {K.~D.}\ \bibnamefont {Marquez}}, \bibinfo {author} {\bibfnamefont {B.~C.}\ \bibnamefont {Backes}}, \bibinfo {author} {\bibfnamefont {G.}~\bibnamefont {Panotopoulos}}, \ and\ \bibinfo {author} {\bibfnamefont {I.}~\bibnamefont {Lopes}},\ }\href {\doibase 10.1088/1475-7516/2024/05/130} {\bibfield  {journal} {\bibinfo  {journal} {JCAP}\ }\textbf {\bibinfo {volume} {05}},\ \bibinfo {pages} {130} (\bibinfo {year} {2024})},\ \Eprint {http://arxiv.org/abs/2401.07789} {arXiv:2401.07789 [nucl-th]} \BibitemShut {NoStop}%
\bibitem [{\citenamefont {Glendenning}(1997)}]{Glendenning:1997wn}%
  \BibitemOpen
  \bibfield  {author} {\bibinfo {author} {\bibfnamefont {N.~K.}\ \bibnamefont {Glendenning}},\ }\href@noop {} {\emph {\bibinfo {title} {{Compact stars: Nuclear physics, particle physics, and general relativity}}}}\ (\bibinfo {year} {1997})\BibitemShut {NoStop}%
\bibitem [{\citenamefont {Riley}\ \emph {et~al.}(2019{\natexlab{b}})\citenamefont {Riley} \emph {et~al.}}]{Riley:2019yda}%
  \BibitemOpen
  \bibfield  {author} {\bibinfo {author} {\bibfnamefont {T.~E.}\ \bibnamefont {Riley}} \emph {et~al.},\ }\href {\doibase 10.3847/2041-8213/ab481c} {\bibfield  {journal} {\bibinfo  {journal} {Astrophys. J. Lett.}\ }\textbf {\bibinfo {volume} {887}},\ \bibinfo {pages} {L21} (\bibinfo {year} {2019}{\natexlab{b}})},\ \Eprint {http://arxiv.org/abs/1912.05702} {arXiv:1912.05702 [astro-ph.HE]} \BibitemShut {NoStop}%
\bibitem [{\citenamefont {Riley}\ \emph {et~al.}(2021{\natexlab{b}})\citenamefont {Riley} \emph {et~al.}}]{Riley:2021pdl}%
  \BibitemOpen
  \bibfield  {author} {\bibinfo {author} {\bibfnamefont {T.~E.}\ \bibnamefont {Riley}} \emph {et~al.},\ }\href {\doibase 10.3847/2041-8213/ac0a81} {\bibfield  {journal} {\bibinfo  {journal} {Astrophys. J. Lett.}\ }\textbf {\bibinfo {volume} {918}},\ \bibinfo {pages} {L27} (\bibinfo {year} {2021}{\natexlab{b}})},\ \Eprint {http://arxiv.org/abs/2105.06980} {arXiv:2105.06980 [astro-ph.HE]} \BibitemShut {NoStop}%
\bibitem [{\citenamefont {Choudhury}\ \emph {et~al.}(2024)\citenamefont {Choudhury} \emph {et~al.}}]{Choudhury:2024xbk}%
  \BibitemOpen
  \bibfield  {author} {\bibinfo {author} {\bibfnamefont {D.}~\bibnamefont {Choudhury}} \emph {et~al.},\ }\href {\doibase 10.3847/2041-8213/ad5a6f} {\bibfield  {journal} {\bibinfo  {journal} {The Astrophysical Journal Letters}\ }\textbf {\bibinfo {volume} {971}},\ \bibinfo {pages} {L20} (\bibinfo {year} {2024})}\BibitemShut {NoStop}%
\bibitem [{\citenamefont {Oppenheimer}\ and\ \citenamefont {Volkoff}(1939{\natexlab{a}})}]{PhysRev.55.374}%
  \BibitemOpen
  \bibfield  {author} {\bibinfo {author} {\bibfnamefont {J.~R.}\ \bibnamefont {Oppenheimer}}\ and\ \bibinfo {author} {\bibfnamefont {G.~M.}\ \bibnamefont {Volkoff}},\ }\href {\doibase 10.1103/PhysRev.55.374} {\bibfield  {journal} {\bibinfo  {journal} {Phys. Rev.}\ }\textbf {\bibinfo {volume} {55}},\ \bibinfo {pages} {374--381} (\bibinfo {year} {1939}{\natexlab{a}})}\BibitemShut {NoStop}%
\bibitem [{\citenamefont {Oppenheimer}\ and\ \citenamefont {Volkoff}(1939{\natexlab{b}})}]{Oppenheimer:1939ne}%
  \BibitemOpen
  \bibfield  {author} {\bibinfo {author} {\bibfnamefont {J.~R.}\ \bibnamefont {Oppenheimer}}\ and\ \bibinfo {author} {\bibfnamefont {G.~M.}\ \bibnamefont {Volkoff}},\ }\href {\doibase 10.1103/PhysRev.55.374} {\bibfield  {journal} {\bibinfo  {journal} {Phys. Rev.}\ }\textbf {\bibinfo {volume} {55}},\ \bibinfo {pages} {374--381} (\bibinfo {year} {1939}{\natexlab{b}})}\BibitemShut {NoStop}%
\bibitem [{\citenamefont {Hempel}\ and\ \citenamefont {Schaffner-Bielich}(2010)}]{Hempel:2009mc}%
  \BibitemOpen
  \bibfield  {author} {\bibinfo {author} {\bibfnamefont {M.}~\bibnamefont {Hempel}}\ and\ \bibinfo {author} {\bibfnamefont {J.}~\bibnamefont {Schaffner-Bielich}},\ }\href {\doibase 10.1016/j.nuclphysa.2010.02.010} {\bibfield  {journal} {\bibinfo  {journal} {Nucl. Phys. A}\ }\textbf {\bibinfo {volume} {837}},\ \bibinfo {pages} {210--254} (\bibinfo {year} {2010})},\ \Eprint {http://arxiv.org/abs/0911.4073} {arXiv:0911.4073 [nucl-th]} \BibitemShut {NoStop}%
\bibitem [{\citenamefont {Rather}\ \emph {et~al.}(2021)\citenamefont {Rather}, \citenamefont {Usmani},\ and\ \citenamefont {Patra}}]{rather2020effect}%
  \BibitemOpen
  \bibfield  {author} {\bibinfo {author} {\bibfnamefont {I.~A.}\ \bibnamefont {Rather}}, \bibinfo {author} {\bibfnamefont {A.}~\bibnamefont {Usmani}}, \ and\ \bibinfo {author} {\bibfnamefont {S.}~\bibnamefont {Patra}},\ }\href {\doibase https://doi.org/10.1016/j.nuclphysa.2021.122189} {\bibfield  {journal} {\bibinfo  {journal} {Nuclear Physics A}\ }\textbf {\bibinfo {volume} {1010}},\ \bibinfo {pages} {122189} (\bibinfo {year} {2021})}\BibitemShut {NoStop}%
\bibitem [{\citenamefont {Sotani}\ and\ \citenamefont {Takiwaki}(2020)}]{Sotani:2020mwc}%
  \BibitemOpen
  \bibfield  {author} {\bibinfo {author} {\bibfnamefont {H.}~\bibnamefont {Sotani}}\ and\ \bibinfo {author} {\bibfnamefont {T.}~\bibnamefont {Takiwaki}},\ }\href {\doibase 10.1103/PhysRevD.102.063025} {\bibfield  {journal} {\bibinfo  {journal} {Phys. Rev. D}\ }\textbf {\bibinfo {volume} {102}},\ \bibinfo {pages} {063025} (\bibinfo {year} {2020})},\ \Eprint {http://arxiv.org/abs/2009.05206} {arXiv:2009.05206 [astro-ph.HE]} \BibitemShut {NoStop}%
\bibitem [{\citenamefont {Zhao}\ and\ \citenamefont {Lattimer}(2022)}]{Zhao:2022tcw}%
  \BibitemOpen
  \bibfield  {author} {\bibinfo {author} {\bibfnamefont {T.}~\bibnamefont {Zhao}}\ and\ \bibinfo {author} {\bibfnamefont {J.~M.}\ \bibnamefont {Lattimer}},\ }\href {\doibase 10.1103/PhysRevD.106.123002} {\bibfield  {journal} {\bibinfo  {journal} {Phys. Rev. D}\ }\textbf {\bibinfo {volume} {106}},\ \bibinfo {pages} {123002} (\bibinfo {year} {2022})},\ \Eprint {http://arxiv.org/abs/2204.03037} {arXiv:2204.03037 [astro-ph.HE]} \BibitemShut {NoStop}%
\bibitem [{\citenamefont {Kokkotas}\ and\ \citenamefont {Schmidt}(1999)}]{Kokkotas:1999bd}%
  \BibitemOpen
  \bibfield  {author} {\bibinfo {author} {\bibfnamefont {K.~D.}\ \bibnamefont {Kokkotas}}\ and\ \bibinfo {author} {\bibfnamefont {B.~G.}\ \bibnamefont {Schmidt}},\ }\href {\doibase 10.12942/lrr-1999-2} {\bibfield  {journal} {\bibinfo  {journal} {Living Rev. Rel.}\ }\textbf {\bibinfo {volume} {2}},\ \bibinfo {pages} {2} (\bibinfo {year} {1999})},\ \Eprint {http://arxiv.org/abs/gr-qc/9909058} {arXiv:gr-qc/9909058} \BibitemShut {NoStop}%
\bibitem [{\citenamefont {Pradhan}\ and\ \citenamefont {Chatterjee}(2021)}]{Pradhan:2020amo}%
  \BibitemOpen
  \bibfield  {author} {\bibinfo {author} {\bibfnamefont {B.~K.}\ \bibnamefont {Pradhan}}\ and\ \bibinfo {author} {\bibfnamefont {D.}~\bibnamefont {Chatterjee}},\ }\href {\doibase 10.1103/PhysRevC.103.035810} {\bibfield  {journal} {\bibinfo  {journal} {Phys. Rev. C}\ }\textbf {\bibinfo {volume} {103}},\ \bibinfo {pages} {035810} (\bibinfo {year} {2021})},\ \Eprint {http://arxiv.org/abs/2011.02204} {arXiv:2011.02204 [astro-ph.HE]} \BibitemShut {NoStop}%
\bibitem [{\citenamefont {Lindblom}\ and\ \citenamefont {Detweiler}(1983)}]{Lindblom:1983ps}%
  \BibitemOpen
  \bibfield  {author} {\bibinfo {author} {\bibfnamefont {L.}~\bibnamefont {Lindblom}}\ and\ \bibinfo {author} {\bibfnamefont {S.~L.}\ \bibnamefont {Detweiler}},\ }\href {\doibase 10.1086/190884} {\bibfield  {journal} {\bibinfo  {journal} {Astrophys. J. Suppl.}\ }\textbf {\bibinfo {volume} {53}},\ \bibinfo {pages} {73--92} (\bibinfo {year} {1983})}\BibitemShut {NoStop}%
\bibitem [{\citenamefont {Benhar}\ \emph {et~al.}(1999)\citenamefont {Benhar}, \citenamefont {Berti},\ and\ \citenamefont {Ferrari}}]{Benhar:1998au}%
  \BibitemOpen
  \bibfield  {author} {\bibinfo {author} {\bibfnamefont {O.}~\bibnamefont {Benhar}}, \bibinfo {author} {\bibfnamefont {E.}~\bibnamefont {Berti}}, \ and\ \bibinfo {author} {\bibfnamefont {V.}~\bibnamefont {Ferrari}},\ }\href {\doibase 10.1046/j.1365-8711.1999.02983.x} {\bibfield  {journal} {\bibinfo  {journal} {Mon. Not. Roy. Astron. Soc.}\ }\textbf {\bibinfo {volume} {310}},\ \bibinfo {pages} {797--803} (\bibinfo {year} {1999})},\ \Eprint {http://arxiv.org/abs/gr-qc/9901037} {arXiv:gr-qc/9901037} \BibitemShut {NoStop}%
\bibitem [{\citenamefont {Andersson}\ and\ \citenamefont {Kokkotas}(1998)}]{Andersson:1997rn}%
  \BibitemOpen
  \bibfield  {author} {\bibinfo {author} {\bibfnamefont {N.}~\bibnamefont {Andersson}}\ and\ \bibinfo {author} {\bibfnamefont {K.~D.}\ \bibnamefont {Kokkotas}},\ }\href {\doibase 10.1046/j.1365-8711.1998.01840.x} {\bibfield  {journal} {\bibinfo  {journal} {Mon. Not. Roy. Astron. Soc.}\ }\textbf {\bibinfo {volume} {299}},\ \bibinfo {pages} {1059--1068} (\bibinfo {year} {1998})},\ \Eprint {http://arxiv.org/abs/gr-qc/9711088} {arXiv:gr-qc/9711088} \BibitemShut {NoStop}%
\bibitem [{\citenamefont {Chirenti}\ \emph {et~al.}(2015)\citenamefont {Chirenti}, \citenamefont {de~Souza},\ and\ \citenamefont {Kastaun}}]{Chirenti:2015dda}%
  \BibitemOpen
  \bibfield  {author} {\bibinfo {author} {\bibfnamefont {C.}~\bibnamefont {Chirenti}}, \bibinfo {author} {\bibfnamefont {G.~H.}\ \bibnamefont {de~Souza}}, \ and\ \bibinfo {author} {\bibfnamefont {W.}~\bibnamefont {Kastaun}},\ }\href {\doibase 10.1103/PhysRevD.91.044034} {\bibfield  {journal} {\bibinfo  {journal} {Phys. Rev. D}\ }\textbf {\bibinfo {volume} {91}},\ \bibinfo {pages} {044034} (\bibinfo {year} {2015})},\ \Eprint {http://arxiv.org/abs/1501.02970} {arXiv:1501.02970 [gr-qc]} \BibitemShut {NoStop}%
\bibitem [{\citenamefont {Lioutas}\ and\ \citenamefont {Stergioulas}(2018)}]{Lioutas:2017xtn}%
  \BibitemOpen
  \bibfield  {author} {\bibinfo {author} {\bibfnamefont {G.}~\bibnamefont {Lioutas}}\ and\ \bibinfo {author} {\bibfnamefont {N.}~\bibnamefont {Stergioulas}},\ }\href {\doibase 10.1007/s10714-017-2331-7} {\bibfield  {journal} {\bibinfo  {journal} {Gen. Rel. Grav.}\ }\textbf {\bibinfo {volume} {50}},\ \bibinfo {pages} {12} (\bibinfo {year} {2018})},\ \Eprint {http://arxiv.org/abs/1709.10067} {arXiv:1709.10067 [gr-qc]} \BibitemShut {NoStop}%
\bibitem [{\citenamefont {Sotani}(2021)}]{PhysRevD.103.123015}%
  \BibitemOpen
  \bibfield  {author} {\bibinfo {author} {\bibfnamefont {H.}~\bibnamefont {Sotani}},\ }\href {\doibase 10.1103/PhysRevD.103.123015} {\bibfield  {journal} {\bibinfo  {journal} {Phys. Rev. D}\ }\textbf {\bibinfo {volume} {103}},\ \bibinfo {pages} {123015} (\bibinfo {year} {2021})}\BibitemShut {NoStop}%
\bibitem [{\citenamefont {Lau}\ \emph {et~al.}(2010)\citenamefont {Lau}, \citenamefont {Leung},\ and\ \citenamefont {Lin}}]{Lau_2010}%
  \BibitemOpen
  \bibfield  {author} {\bibinfo {author} {\bibfnamefont {H.~K.}\ \bibnamefont {Lau}}, \bibinfo {author} {\bibfnamefont {P.~T.}\ \bibnamefont {Leung}}, \ and\ \bibinfo {author} {\bibfnamefont {L.~M.}\ \bibnamefont {Lin}},\ }\href {\doibase 10.1088/0004-637X/714/2/1234} {\bibfield  {journal} {\bibinfo  {journal} {The Astrophysical Journal}\ }\textbf {\bibinfo {volume} {714}},\ \bibinfo {pages} {1234} (\bibinfo {year} {2010})}\BibitemShut {NoStop}%
\bibitem [{\citenamefont {Thakur}\ \emph {et~al.}(2024)\citenamefont {Thakur}, \citenamefont {Chatterjee}, \citenamefont {Nath},\ and\ \citenamefont {Mallick}}]{Thakur:2024ijp}%
  \BibitemOpen
  \bibfield  {author} {\bibinfo {author} {\bibfnamefont {P.}~\bibnamefont {Thakur}}, \bibinfo {author} {\bibfnamefont {S.}~\bibnamefont {Chatterjee}}, \bibinfo {author} {\bibfnamefont {K.~K.}\ \bibnamefont {Nath}}, \ and\ \bibinfo {author} {\bibfnamefont {R.}~\bibnamefont {Mallick}},\ }\href {\doibase 10.1103/PhysRevD.110.103045} {\bibfield  {journal} {\bibinfo  {journal} {Phys. Rev. D}\ }\textbf {\bibinfo {volume} {110}},\ \bibinfo {pages} {103045} (\bibinfo {year} {2024})},\ \Eprint {http://arxiv.org/abs/2407.12601} {arXiv:2407.12601 [gr-qc]} \BibitemShut {NoStop}%
\bibitem [{\citenamefont {Sagun}\ \emph {et~al.}(2020)\citenamefont {Sagun}, \citenamefont {Panotopoulos},\ and\ \citenamefont {Lopes}}]{Sagun:2020qvc}%
  \BibitemOpen
  \bibfield  {author} {\bibinfo {author} {\bibfnamefont {V.}~\bibnamefont {Sagun}}, \bibinfo {author} {\bibfnamefont {G.}~\bibnamefont {Panotopoulos}}, \ and\ \bibinfo {author} {\bibfnamefont {I.}~\bibnamefont {Lopes}},\ }\href {\doibase 10.1103/PhysRevD.101.063025} {\bibfield  {journal} {\bibinfo  {journal} {Phys. Rev. D}\ }\textbf {\bibinfo {volume} {101}},\ \bibinfo {pages} {063025} (\bibinfo {year} {2020})},\ \Eprint {http://arxiv.org/abs/2002.12209} {arXiv:2002.12209 [astro-ph.HE]} \BibitemShut {NoStop}%
\bibitem [{\citenamefont {Routaray}\ \emph {et~al.}(2023)\citenamefont {Routaray}, \citenamefont {Das}, \citenamefont {Sen}, \citenamefont {Kumar}, \citenamefont {Panotopoulos},\ and\ \citenamefont {Zhao}}]{Routaray:2022utr}%
  \BibitemOpen
  \bibfield  {author} {\bibinfo {author} {\bibfnamefont {P.}~\bibnamefont {Routaray}}, \bibinfo {author} {\bibfnamefont {H.~C.}\ \bibnamefont {Das}}, \bibinfo {author} {\bibfnamefont {S.}~\bibnamefont {Sen}}, \bibinfo {author} {\bibfnamefont {B.}~\bibnamefont {Kumar}}, \bibinfo {author} {\bibfnamefont {G.}~\bibnamefont {Panotopoulos}}, \ and\ \bibinfo {author} {\bibfnamefont {T.}~\bibnamefont {Zhao}},\ }\href {\doibase 10.1103/PhysRevD.107.103039} {\bibfield  {journal} {\bibinfo  {journal} {Phys. Rev. D}\ }\textbf {\bibinfo {volume} {107}},\ \bibinfo {pages} {103039} (\bibinfo {year} {2023})},\ \Eprint {http://arxiv.org/abs/2211.12808} {arXiv:2211.12808 [nucl-th]} \BibitemShut {NoStop}%
\bibitem [{\citenamefont {Sen}\ \emph {et~al.}(2023)\citenamefont {Sen}, \citenamefont {Kumar}, \citenamefont {Kunjipurayil}, \citenamefont {Routaray}, \citenamefont {Ghosh}, \citenamefont {Kalita}, \citenamefont {Zhao},\ and\ \citenamefont {Kumar}}]{Sen:2022kva}%
  \BibitemOpen
  \bibfield  {author} {\bibinfo {author} {\bibfnamefont {S.}~\bibnamefont {Sen}}, \bibinfo {author} {\bibfnamefont {S.}~\bibnamefont {Kumar}}, \bibinfo {author} {\bibfnamefont {A.}~\bibnamefont {Kunjipurayil}}, \bibinfo {author} {\bibfnamefont {P.}~\bibnamefont {Routaray}}, \bibinfo {author} {\bibfnamefont {S.}~\bibnamefont {Ghosh}}, \bibinfo {author} {\bibfnamefont {P.~J.}\ \bibnamefont {Kalita}}, \bibinfo {author} {\bibfnamefont {T.}~\bibnamefont {Zhao}}, \ and\ \bibinfo {author} {\bibfnamefont {B.}~\bibnamefont {Kumar}},\ }\href {\doibase 10.3390/galaxies11020060} {\bibfield  {journal} {\bibinfo  {journal} {Galaxies}\ }\textbf {\bibinfo {volume} {11}},\ \bibinfo {pages} {60} (\bibinfo {year} {2023})},\ \Eprint {http://arxiv.org/abs/2205.02076} {arXiv:2205.02076 [nucl-th]} \BibitemShut {NoStop}%
\bibitem [{\citenamefont {Haensel}\ \emph {et~al.}(2006)\citenamefont {Haensel}, \citenamefont {Potekhin},\ and\ \citenamefont {Yakovlev}}]{haensel2006neutron}%
  \BibitemOpen
  \bibfield  {author} {\bibinfo {author} {\bibfnamefont {P.}~\bibnamefont {Haensel}}, \bibinfo {author} {\bibfnamefont {A.}~\bibnamefont {Potekhin}}, \ and\ \bibinfo {author} {\bibfnamefont {D.}~\bibnamefont {Yakovlev}},\ }\href {https://books.google.fr/books?id=iIrj9nfHnesC} {\emph {\bibinfo {title} {Neutron Stars 1: Equation of State and Structure}}},\ Astrophysics and Space Science Library\ (\bibinfo  {publisher} {Springer New York},\ \bibinfo {year} {2006})\BibitemShut {NoStop}%
\bibitem [{\citenamefont {Thakur}\ \emph {et~al.}(2025)\citenamefont {Thakur}, \citenamefont {Rather},\ and\ \citenamefont {Lim}}]{Thakur:2025zhi}%
  \BibitemOpen
  \bibfield  {author} {\bibinfo {author} {\bibfnamefont {P.}~\bibnamefont {Thakur}}, \bibinfo {author} {\bibfnamefont {I.~A.}\ \bibnamefont {Rather}}, \ and\ \bibinfo {author} {\bibfnamefont {Y.}~\bibnamefont {Lim}},\ }\href {\doibase 10.1103/7tnx-s66h} {\bibfield  {journal} {\bibinfo  {journal} {Phys. Rev. D}\ }\textbf {\bibinfo {volume} {112}},\ \bibinfo {pages} {043017} (\bibinfo {year} {2025})},\ \Eprint {http://arxiv.org/abs/2507.13227} {arXiv:2507.13227 [astro-ph.HE]} \BibitemShut {NoStop}%
\bibitem [{\citenamefont {Panotopoulos}\ and\ \citenamefont {Lopes}(2017)}]{Panotopoulos:2017eig}%
  \BibitemOpen
  \bibfield  {author} {\bibinfo {author} {\bibfnamefont {G.}~\bibnamefont {Panotopoulos}}\ and\ \bibinfo {author} {\bibfnamefont {I.}~\bibnamefont {Lopes}},\ }\href {\doibase 10.1103/PhysRevD.96.083013} {\bibfield  {journal} {\bibinfo  {journal} {Phys. Rev. D}\ }\textbf {\bibinfo {volume} {96}},\ \bibinfo {pages} {083013} (\bibinfo {year} {2017})},\ \Eprint {http://arxiv.org/abs/1709.06643} {arXiv:1709.06643 [gr-qc]} \BibitemShut {NoStop}%
\bibitem [{\citenamefont {Rather}\ \emph {et~al.}(2023{\natexlab{b}})\citenamefont {Rather}, \citenamefont {Panotopoulos},\ and\ \citenamefont {Lopes}}]{Rather:2023tly}%
  \BibitemOpen
  \bibfield  {author} {\bibinfo {author} {\bibfnamefont {I.~A.}\ \bibnamefont {Rather}}, \bibinfo {author} {\bibfnamefont {G.}~\bibnamefont {Panotopoulos}}, \ and\ \bibinfo {author} {\bibfnamefont {I.}~\bibnamefont {Lopes}},\ }\href {\doibase 10.1140/epjc/s10052-023-12223-1} {\bibfield  {journal} {\bibinfo  {journal} {Eur. Phys. J. C}\ }\textbf {\bibinfo {volume} {83}},\ \bibinfo {pages} {1065} (\bibinfo {year} {2023}{\natexlab{b}})},\ \Eprint {http://arxiv.org/abs/2307.03703} {arXiv:2307.03703 [astro-ph.HE]} \BibitemShut {NoStop}%
\bibitem [{\citenamefont {Sun}\ \emph {et~al.}(2021)\citenamefont {Sun}, \citenamefont {Zheng}, \citenamefont {Chen}, \citenamefont {Burgio},\ and\ \citenamefont {Schulze}}]{Sun:2021cez}%
  \BibitemOpen
  \bibfield  {author} {\bibinfo {author} {\bibfnamefont {T.-T.}\ \bibnamefont {Sun}}, \bibinfo {author} {\bibfnamefont {Z.-Y.}\ \bibnamefont {Zheng}}, \bibinfo {author} {\bibfnamefont {H.}~\bibnamefont {Chen}}, \bibinfo {author} {\bibfnamefont {G.~F.}\ \bibnamefont {Burgio}}, \ and\ \bibinfo {author} {\bibfnamefont {H.-J.}\ \bibnamefont {Schulze}},\ }\href {\doibase 10.1103/PhysRevD.103.103003} {\bibfield  {journal} {\bibinfo  {journal} {Phys. Rev. D}\ }\textbf {\bibinfo {volume} {103}},\ \bibinfo {pages} {103003} (\bibinfo {year} {2021})},\ \Eprint {http://arxiv.org/abs/2101.07515} {arXiv:2101.07515 [nucl-th]} \BibitemShut {NoStop}%
\bibitem [{\citenamefont {Abbott}\ \emph {et~al.}(2017{\natexlab{b}})\citenamefont {Abbott} \emph {et~al.}}]{LIGOScientific:2016wof}%
  \BibitemOpen
  \bibfield  {author} {\bibinfo {author} {\bibfnamefont {B.~P.}\ \bibnamefont {Abbott}} \emph {et~al.} (\bibinfo {collaboration} {LIGO Scientific}),\ }\href {\doibase 10.1088/1361-6382/aa51f4} {\bibfield  {journal} {\bibinfo  {journal} {Class. Quant. Grav.}\ }\textbf {\bibinfo {volume} {34}},\ \bibinfo {pages} {044001} (\bibinfo {year} {2017}{\natexlab{b}})},\ \Eprint {http://arxiv.org/abs/1607.08697} {arXiv:1607.08697 [astro-ph.IM]} \BibitemShut {NoStop}%
\bibitem [{\citenamefont {Punturo}\ \emph {et~al.}(2010)\citenamefont {Punturo} \emph {et~al.}}]{Punturo:2010zz}%
  \BibitemOpen
  \bibfield  {author} {\bibinfo {author} {\bibfnamefont {M.}~\bibnamefont {Punturo}} \emph {et~al.},\ }\href {\doibase 10.1088/0264-9381/27/19/194002} {\bibfield  {journal} {\bibinfo  {journal} {Class. Quant. Grav.}\ }\textbf {\bibinfo {volume} {27}},\ \bibinfo {pages} {194002} (\bibinfo {year} {2010})}\BibitemShut {NoStop}%
\bibitem [{\citenamefont {Andersen}\ \emph {et~al.}(2021)\citenamefont {Andersen}, \citenamefont {Zha}, \citenamefont {da~Silva~Schneider}, \citenamefont {Betranhandy}, \citenamefont {Couch},\ and\ \citenamefont {O'Connor}}]{Andersen:2021vzo}%
  \BibitemOpen
  \bibfield  {author} {\bibinfo {author} {\bibfnamefont {O.~E.}\ \bibnamefont {Andersen}}, \bibinfo {author} {\bibfnamefont {S.}~\bibnamefont {Zha}}, \bibinfo {author} {\bibfnamefont {A.}~\bibnamefont {da~Silva~Schneider}}, \bibinfo {author} {\bibfnamefont {A.}~\bibnamefont {Betranhandy}}, \bibinfo {author} {\bibfnamefont {S.~M.}\ \bibnamefont {Couch}}, \ and\ \bibinfo {author} {\bibfnamefont {E.~P.}\ \bibnamefont {O'Connor}},\ }\href {\doibase 10.3847/1538-4357/ac294c} {\bibfield  {journal} {\bibinfo  {journal} {Astrophys. J.}\ }\textbf {\bibinfo {volume} {923}},\ \bibinfo {pages} {201} (\bibinfo {year} {2021})},\ \Eprint {http://arxiv.org/abs/2106.09734} {arXiv:2106.09734 [astro-ph.HE]} \BibitemShut {NoStop}%
\bibitem [{\citenamefont {Mueller}\ \emph {et~al.}(2013)\citenamefont {Mueller}, \citenamefont {Janka},\ and\ \citenamefont {Marek}}]{Mueller:2012sv}%
  \BibitemOpen
  \bibfield  {author} {\bibinfo {author} {\bibfnamefont {B.}~\bibnamefont {Mueller}}, \bibinfo {author} {\bibfnamefont {H.-T.}\ \bibnamefont {Janka}}, \ and\ \bibinfo {author} {\bibfnamefont {A.}~\bibnamefont {Marek}},\ }\href {\doibase 10.1088/0004-637X/766/1/43} {\bibfield  {journal} {\bibinfo  {journal} {Astrophys. J.}\ }\textbf {\bibinfo {volume} {766}},\ \bibinfo {pages} {43} (\bibinfo {year} {2013})},\ \Eprint {http://arxiv.org/abs/1210.6984} {arXiv:1210.6984 [astro-ph.SR]} \BibitemShut {NoStop}%
\bibitem [{\citenamefont {Sotani}\ \emph {et~al.}(2017)\citenamefont {Sotani}, \citenamefont {Kuroda}, \citenamefont {Takiwaki},\ and\ \citenamefont {Kotake}}]{Sotani:2017ubz}%
  \BibitemOpen
  \bibfield  {author} {\bibinfo {author} {\bibfnamefont {H.}~\bibnamefont {Sotani}}, \bibinfo {author} {\bibfnamefont {T.}~\bibnamefont {Kuroda}}, \bibinfo {author} {\bibfnamefont {T.}~\bibnamefont {Takiwaki}}, \ and\ \bibinfo {author} {\bibfnamefont {K.}~\bibnamefont {Kotake}},\ }\href {\doibase 10.1103/PhysRevD.96.063005} {\bibfield  {journal} {\bibinfo  {journal} {Phys. Rev. D}\ }\textbf {\bibinfo {volume} {96}},\ \bibinfo {pages} {063005} (\bibinfo {year} {2017})},\ \Eprint {http://arxiv.org/abs/1708.03738} {arXiv:1708.03738 [astro-ph.HE]} \BibitemShut {NoStop}%
\bibitem [{\citenamefont {Andresen}\ \emph {et~al.}(2019)\citenamefont {Andresen}, \citenamefont {M\"uller}, \citenamefont {Janka}, \citenamefont {Summa}, \citenamefont {Gill},\ and\ \citenamefont {Zanolin}}]{Andresen:2018aom}%
  \BibitemOpen
  \bibfield  {author} {\bibinfo {author} {\bibfnamefont {H.}~\bibnamefont {Andresen}}, \bibinfo {author} {\bibfnamefont {E.}~\bibnamefont {M\"uller}}, \bibinfo {author} {\bibfnamefont {H.~T.}\ \bibnamefont {Janka}}, \bibinfo {author} {\bibfnamefont {A.}~\bibnamefont {Summa}}, \bibinfo {author} {\bibfnamefont {K.}~\bibnamefont {Gill}}, \ and\ \bibinfo {author} {\bibfnamefont {M.}~\bibnamefont {Zanolin}},\ }\href {\doibase 10.1093/mnras/stz990} {\bibfield  {journal} {\bibinfo  {journal} {Mon. Not. Roy. Astron. Soc.}\ }\textbf {\bibinfo {volume} {486}},\ \bibinfo {pages} {2238--2253} (\bibinfo {year} {2019})},\ \Eprint {http://arxiv.org/abs/1810.07638} {arXiv:1810.07638 [astro-ph.HE]} \BibitemShut {NoStop}%
\bibitem [{\citenamefont {Andresen}\ \emph {et~al.}(2017)\citenamefont {Andresen}, \citenamefont {M\"uller}, \citenamefont {M\"uller},\ and\ \citenamefont {Janka}}]{Andresen:2016pdt}%
  \BibitemOpen
  \bibfield  {author} {\bibinfo {author} {\bibfnamefont {H.}~\bibnamefont {Andresen}}, \bibinfo {author} {\bibfnamefont {B.}~\bibnamefont {M\"uller}}, \bibinfo {author} {\bibfnamefont {E.}~\bibnamefont {M\"uller}}, \ and\ \bibinfo {author} {\bibfnamefont {H.-T.}\ \bibnamefont {Janka}},\ }\href {\doibase 10.1093/mnras/stx618} {\bibfield  {journal} {\bibinfo  {journal} {Mon. Not. Roy. Astron. Soc.}\ }\textbf {\bibinfo {volume} {468}},\ \bibinfo {pages} {2032--2051} (\bibinfo {year} {2017})},\ \Eprint {http://arxiv.org/abs/1607.05199} {arXiv:1607.05199 [astro-ph.HE]} \BibitemShut {NoStop}%
\bibitem [{\citenamefont {Bednarek}\ \emph {et~al.}(2012{\natexlab{b}})\citenamefont {Bednarek}, \citenamefont {Haensel}, \citenamefont {Zdunik}, \citenamefont {Bejger},\ and\ \citenamefont {Ma{\'n}ka}}]{Bednarek_2012}%
  \BibitemOpen
  \bibfield  {author} {\bibinfo {author} {\bibfnamefont {I.}~\bibnamefont {Bednarek}}, \bibinfo {author} {\bibfnamefont {P.}~\bibnamefont {Haensel}}, \bibinfo {author} {\bibfnamefont {J.}~\bibnamefont {Zdunik}}, \bibinfo {author} {\bibfnamefont {M.}~\bibnamefont {Bejger}}, \ and\ \bibinfo {author} {\bibfnamefont {R.}~\bibnamefont {Ma{\'n}ka}},\ }\href {\doibase 10.1051/0004-6361/201118560} {\bibfield  {journal} {\bibinfo  {journal} {Astronomy \& Astrophysics}\ }\textbf {\bibinfo {volume} {543}},\ \bibinfo {pages} {A157} (\bibinfo {year} {2012}{\natexlab{b}})}\BibitemShut {NoStop}%
\bibitem [{\citenamefont {Messios}\ \emph {et~al.}(2001)\citenamefont {Messios}, \citenamefont {Papadopoulos},\ and\ \citenamefont {Stergioulas}}]{Messios:2001br}%
  \BibitemOpen
  \bibfield  {author} {\bibinfo {author} {\bibfnamefont {N.}~\bibnamefont {Messios}}, \bibinfo {author} {\bibfnamefont {D.~B.}\ \bibnamefont {Papadopoulos}}, \ and\ \bibinfo {author} {\bibfnamefont {N.}~\bibnamefont {Stergioulas}},\ }\href {\doibase 10.1046/j.1365-8711.2001.04645.x} {\bibfield  {journal} {\bibinfo  {journal} {Mon. Not. Roy. Astron. Soc.}\ }\textbf {\bibinfo {volume} {328}},\ \bibinfo {pages} {1161} (\bibinfo {year} {2001})},\ \Eprint {http://arxiv.org/abs/astro-ph/0105175} {arXiv:astro-ph/0105175} \BibitemShut {NoStop}%
\bibitem [{\citenamefont {Kr\"uger}\ \emph {et~al.}(2015)\citenamefont {Kr\"uger}, \citenamefont {Ho},\ and\ \citenamefont {Andersson}}]{Kruger:2014pva}%
  \BibitemOpen
  \bibfield  {author} {\bibinfo {author} {\bibfnamefont {C.~J.}\ \bibnamefont {Kr\"uger}}, \bibinfo {author} {\bibfnamefont {W.~C.~G.}\ \bibnamefont {Ho}}, \ and\ \bibinfo {author} {\bibfnamefont {N.}~\bibnamefont {Andersson}},\ }\href {\doibase 10.1103/PhysRevD.92.063009} {\bibfield  {journal} {\bibinfo  {journal} {Phys. Rev. D}\ }\textbf {\bibinfo {volume} {92}},\ \bibinfo {pages} {063009} (\bibinfo {year} {2015})},\ \Eprint {http://arxiv.org/abs/1402.5656} {arXiv:1402.5656 [gr-qc]} \BibitemShut {NoStop}%
\bibitem [{\citenamefont {Gholami}\ \emph {et~al.}(2025)\citenamefont {Gholami}, \citenamefont {Rather}, \citenamefont {Hofmann}, \citenamefont {Buballa},\ and\ \citenamefont {Schaffner-Bielich}}]{Gholami:2024ety}%
  \BibitemOpen
  \bibfield  {author} {\bibinfo {author} {\bibfnamefont {H.}~\bibnamefont {Gholami}}, \bibinfo {author} {\bibfnamefont {I.~A.}\ \bibnamefont {Rather}}, \bibinfo {author} {\bibfnamefont {M.}~\bibnamefont {Hofmann}}, \bibinfo {author} {\bibfnamefont {M.}~\bibnamefont {Buballa}}, \ and\ \bibinfo {author} {\bibfnamefont {J.}~\bibnamefont {Schaffner-Bielich}},\ }\href {\doibase 10.1103/PhysRevD.111.103034} {\bibfield  {journal} {\bibinfo  {journal} {Phys. Rev. D}\ }\textbf {\bibinfo {volume} {111}},\ \bibinfo {pages} {103034} (\bibinfo {year} {2025})},\ \Eprint {http://arxiv.org/abs/2411.04064} {arXiv:2411.04064 [hep-ph]} \BibitemShut {NoStop}%
\bibitem [{\citenamefont {Christian}\ \emph {et~al.}(2025)\citenamefont {Christian}, \citenamefont {Rather}, \citenamefont {Gholami},\ and\ \citenamefont {Hofmann}}]{Christian:2025dhe}%
  \BibitemOpen
  \bibfield  {author} {\bibinfo {author} {\bibfnamefont {J.-E.}\ \bibnamefont {Christian}}, \bibinfo {author} {\bibfnamefont {I.~A.}\ \bibnamefont {Rather}}, \bibinfo {author} {\bibfnamefont {H.}~\bibnamefont {Gholami}}, \ and\ \bibinfo {author} {\bibfnamefont {M.}~\bibnamefont {Hofmann}},\ }\href {\doibase 10.1051/0004-6361/202555009} {\bibfield  {journal} {\bibinfo  {journal} {Astron. Astrophys.}\ }\textbf {\bibinfo {volume} {701}},\ \bibinfo {pages} {A145} (\bibinfo {year} {2025})},\ \Eprint {http://arxiv.org/abs/2503.13626} {arXiv:2503.13626 [astro-ph.HE]} \BibitemShut {NoStop}%
\end{thebibliography}%

\end{document}